\documentclass[aps,prb,reprint,twocolumn,floatfix,superscriptaddress]{revtex4-1}
\usepackage[colorlinks=true,linkcolor=blue,citecolor=blue,breaklinks]{hyperref}
\usepackage{amsmath}
\usepackage{amsthm}
\usepackage{amssymb}
\usepackage{graphicx}
\usepackage[usenames,dvipsnames]{color}
\usepackage{array}
\usepackage{paralist}
\usepackage{amsfonts}
\usepackage{mathtools}
\usepackage{times}
\usepackage{braket}
\usepackage{slashed}
\usepackage{longtable}
\usepackage{tikz}\usetikzlibrary{calc,3d,shapes}

\begin{document}


\title{Transition to a many-body localized regime in a two-dimensional disordered quantum dimer model}



\author{Hugo Th\'eveniaut}
\affiliation{Laboratoire de Physique Th\'eorique, IRSAMC, Universit\'e de Toulouse, CNRS, UPS, 31062 Toulouse, France}
\author{Zhihao Lan}
\affiliation{Centre for the Mathematics and Theoretical Physics of Quantum Non-equilibrium Systems and School of Physics and Astronomy, University of Nottingham, Nottingham NG7 2RD, United Kingdom}
\author{Gabriel Meyer}
\affiliation{Laboratoire de Physique Th\'eorique, IRSAMC, Universit\'e de Toulouse, CNRS, UPS, 31062 Toulouse, France}
\affiliation{\'Ecole Normale Sup\'erieure de Lyon, F-69342 Lyon, France}

\author{Fabien Alet}
\email[]{alet@irsamc.ups-tlse.fr}
\affiliation{Laboratoire de Physique Th\'eorique, IRSAMC, Universit\'e de Toulouse, CNRS, UPS, 31062 Toulouse, France}

\date{\today}

\begin{abstract}
Many-body localization is a unique physical phenomenon driven by interactions and disorder for which a quantum system can evade thermalization. While the existence of a many-body localized phase is now well-established in one-dimensional systems, its fate in higher dimension is an open question.  We present evidence for the occurrence of a transition to a many-body localized regime in a two-dimensional quantum dimer model with interactions and disorder. Our analysis is based on the results of large-scale simulations for static and dynamical properties of a consequent number of observables. Our results pave the way for a generic understanding of occurrence of a many-body localization transition in dimension larger than one, and highlight the unusual quantum dynamics that can be present in constrained systems.

\end{abstract}


\maketitle

\section{\label{sec:intro} Introduction}

The emergence of thermalization in isolated quantum systems, which are not in contact with a bath and solely follow unitary Hamiltonian dynamics, has long been a central issue in statistical mechanics~\cite{neumann_beweis_1929,goldstein_long-time_2010,gogolin_equilibration_2016}. This question has been revived thanks to an extraordinary increase in the degree of control and isolation reached in cold-atom experiments~\cite{bloch_many-body_2008,gross_quantum_2017}. The central concept of Eigenstate Thermalization Hypothesis (ETH)~\cite{deutsch_quantum_1991,srednicki_chaos_1994}, a conjecture on expectation values of local operators in the eigenstates of the isolated system, allows to provide a physical understanding and justification of this thermalization. The ETH is found to be valid for the generic majority of many-body systems~\cite{rigol_thermalization_2008,deutsch_eigenstate_2018}. Recent years have witnessed a vast body of research concentrating on a situation where the ETH fails, namely the phenomenon of many-body localization (MBL)~\cite{basko_metalinsulator_2006,gornyi_interacting_2005}, which occurs most noticeably in the presence of strong disorder.

In a many-body localized system, eigenstates at the same energy density do not form a statistical ensemble, and local expectation values strongly differ from one eigenstate to the other. MBL states sustain a low level of quantum entanglement in the form of an area law~\cite{bauer_area_2013} (in contrast with ETH states which obey a volume law for entanglement entropy), and have been found to be able to sustain long-range order in cases where equilibrium states are forbidden to~\cite{huse_localization-protected_2013,bahri_localization_2015,pekker_hilbert-glass_2014,kjall_many-body_2014,chandran_many-body_2014}. Dynamics in MBL systems markedly differ from those of thermalizing systems, with {\it e.g.} long-time memory of the initial state, absence of transport for conserved quantities (localization), and a slow propagation of quantum information with logarithmic growth of entanglement~\cite{znidaric_many-body_2008,bardarson_unbounded_2012}. The latter example allows to distinguish MBL from Anderson localization~\cite{anderson_absence_1958} (where entanglement is bounded) which occurs in non-interacting systems.

The existence of many-body localization in one dimension is now established and reasonably well-understood in one-dimensional (1d) quantum lattice systems with short-range interactions, thanks to a concerted effort of approaches, including numerical simulations~\cite{pal_many-body_2010,luitz_many-body_2015,wahl_efficient_2017}, phenomenological renormalization~\cite{vosk_theory_2015,potter_universal_2015,dumitrescu_scaling_2017,thiery_many-body_2018,zhang_universal_2018} approaches, rigorous results~\cite{imbrie_diagonalization_2016,imbrie_many-body_2016} and cold-atom experiments~\cite{schreiber_observation_2015,smith_many-body_2016,lukin_probing_2018}.  Crucial to this understanding is the provable~\cite{imbrie_diagonalization_2016} existence of local integral of motions~\cite{huse_phenomenology_2014,serbyn_local_2013}, denoted as {\it l-}bits (for localized bits): these emergent localized operators, which diagonalize the Hamiltonian of the system at strong disorder, explain most of the salient features of the MBL phase~\cite{imbrie_local_2017}. In the typical setting of lattice models, the MBL phase indeed emerges in the presence of strong enough disorder, and is separated from a ETH phase at low disorder by a many-body localization transition involving an exponential number of eigenstates~\cite{pal_many-body_2010}.  The reviews~\cite{nandkishore_many-body_2015,altman_universal_2015,abanin_recent_2017,parameswaran_eigenstate_2017,alet_many-body_2018,abanin_ergodicity_2019} provide an introduction and recent insights on this alive field of research.

\subsection{Many-body localization in higher dimensions}

An outstanding question which remains open is the fate of many-body localization in dimension higher than one. A basic observation in this case is that localization is more difficult to achieve as there are more transport channels than in one dimension. The Anderson localization survives in two dimensions (2d) for all disorder strengths, but with a much higher localization length, and needs an increasingly stronger critical disorder in higher dimensions~\cite{evers_anderson_2008}.

In the many-body case, analytical arguments were presented~\cite{de_roeck_stability_2017,de_roeck_many-body_2017,potirniche_exploration_2019} that suggest the intrinsic instability of a MBL phase in dimension higher than one. They are based on the inevitable existence of thermal `bubbles' (created by {\it e.g.} local regions of space with low disorder), which grow in time and which eventually thermalize any MBL sample. The thermalization conveyed in this scenario is asymptotic, meaning that it may require an asymptotically large time before signs of localization (for instance memory of the initial state) disappear. In these works, the MBL region is represented in terms of the {\it l-}bits, the exponentially localized integral of motions which form a complete set of operators spanning the Hilbert space.

In spite of these arguments, recent remarkable numerical simulations~\cite{wahl_signatures_2018,thomson_time_2018,de_tomasi_solving_2019,doggen2020slow} analyze possible signatures of a MBL phase in two-dimensional lattice models of spins or bosons, in either static~\cite{wahl_signatures_2018} or dynamical quantities~\cite{thomson_time_2018,de_tomasi_solving_2019,doggen2020slow}. While results of Refs.~\onlinecite{wahl_signatures_2018,thomson_time_2018,de_tomasi_solving_2019} are in favor of the existence of a true thermodynamic MBL phase at finite disorder strenght in 2d, Ref.~\onlinecite{doggen2020slow} concluded opppositely (albeit evidencing slow dynamics even at moderate disorder strength). These results are very intriguing, but the methods used in these works are not unbiased: in Refs.~\onlinecite{wahl_signatures_2018,thomson_time_2018,de_tomasi_solving_2019} an approximate unitary diagonalization transformation based on the lowly-entangled nature or localized nature of MBL eigenstates is used, which misses by construction the ETH phase (we note nevertheless that Ref.~\onlinecite{wahl_signatures_2018} presents evidence for the transition in 2d, but without a finite-size or finite-entanglement scaling analysis). The finite-time dynamical results of Ref.~\onlinecite{doggen2020slow} also rely on a method which is entanglement limited. Other exact diagonalization studies in 2d~\cite{geraedts_many_2016,van_nieuwenburg_bloch_2019,wiater_impact_2019} are limited to small systems (lattices with at most 16 lattice sites), such that finite-size scaling is not possible. We also note a recent interesting investigation using exact numerics on a 2d model in the continuum compatible with MBL, albeit with a transition point shifting with the number of particles~\cite{krishna_many_2019}. A quantum Monte Carlo treatment of a Hamiltonian constructed from {\it l-}bits also argues for the existence of MBL in a 2d spin system~\cite{inglis_accessing_2016}. On the experimental side, recent advances in cold-atom experiments present clear dynamical signs of localization (in terms of memory of an initial imbalance of atoms) on the longest accessible time scale in two dimensional systems, either for bosonic atoms in random potential~\cite{choi_exploring_2016} or fermionic gases in quasi-periodic potentials~\cite{bordia_probing_2017}.

In this work, we present a comprehensive study of a two-dimensional strongly interacting disordered model, based on extensive numerical simulations on fairly large systems (square lattices with up to 64 sites). These simulations are exact, unbiased, and allow to study both ETH and MBL regimes and the transition between them. We study an important range of clusters of different sizes and shapes, allowing to follow the evolution of many ETH or MBL indicators with size. This is made possible thanks to two main specificities of our work: we push large-scale parallel numerical methods to their limits (see Sec.~\ref{sec:num}), and we work on a {\it constrained} model (described in Sec.~\ref{sec:model}) that presents specific features appropriate for localization and numerical studies (Sec.~\ref{sec:cons}).

\subsection{Searching for MBL in 2d constrained models}
\label{sec:cons}

\begin{figure*}[t]
\includegraphics[width=0.8\textwidth]{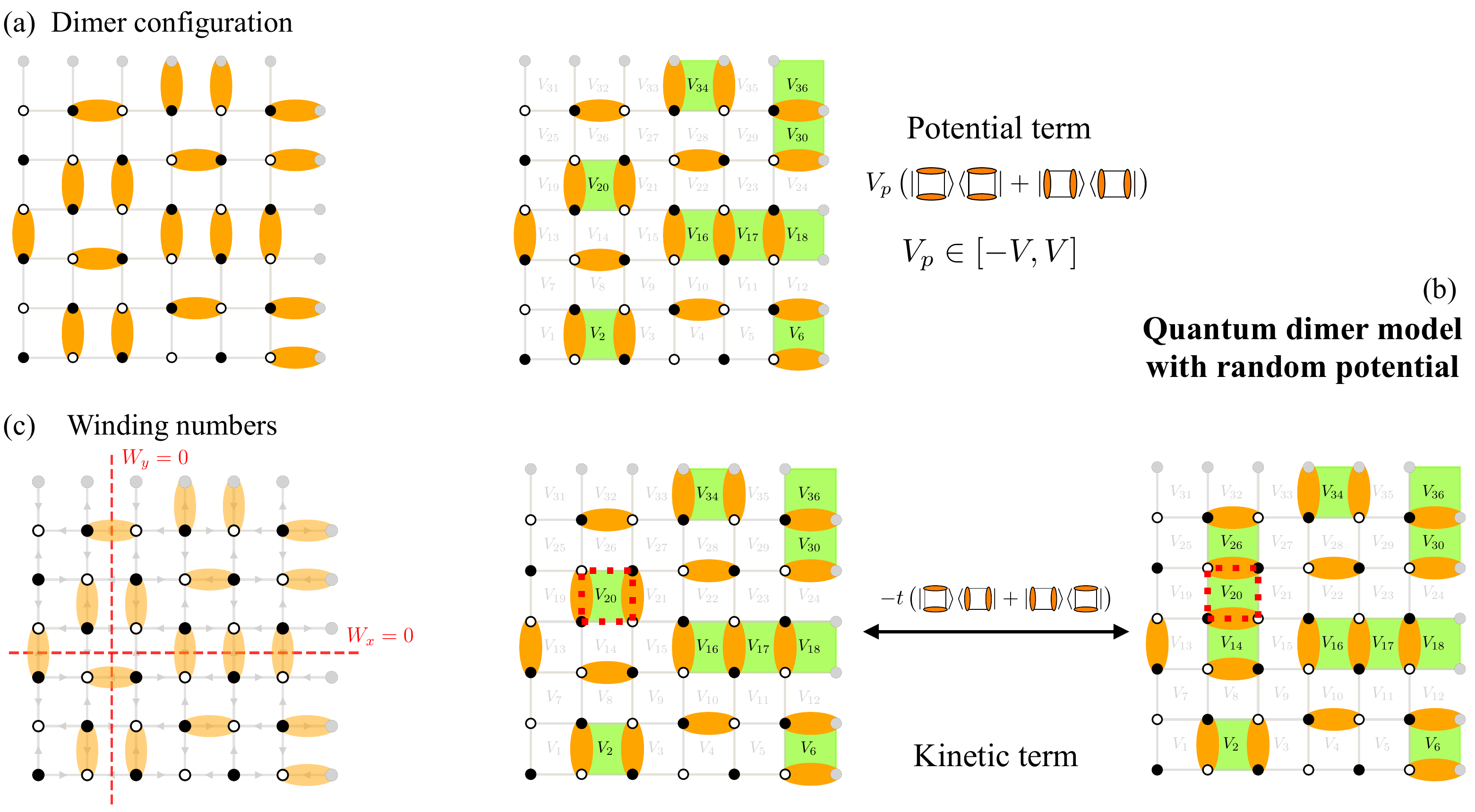}%
\caption{\label{fig:dimers} Quantum dimer models. (a) A configuration of dimers is valid when each lattice site is touched by one and only one dimer. (b) Illustration of the quantum dimer model on the square lattice with a disordered potential Eq.~\ref{eq:H}. Plaquettes $p$ of the lattice where a pair of parallel dimers is present (highlighted in green) contribute a potential $V_p$ (different for each plaquette) to the potential energy of this configuration. The kinetic term flips dimers on such plaquettes such that the orientation of the two dimers on this plaquette is changed. In the bottom right panel, the flip of the plaquette $p=20$ (surrounded by red lines) is represented, resulting in a new configuration with a different potential energy. (c) On the square (bipartite) lattice with periodic boundary conditions, all dimer configurations can be classified by a pair of winding numbers $(W_x,W_y)$ defined as follows. Consider a horizontal (vertical) line --such as represented in the figure-- perpendicular to lattice bonds. $W_x$ ($W_y$) is then equal to the algebraic number ({\it i.e.} $\pm 1$ where the sign is defined by the sublattice on a fixed side of the line) of dimers in this configuration crossing this line. The sign $\pm 1$ is represented by arrows on lattice bonds. Note that the winding numbers do not depend on the position of the line, and they cannot be changed by a local reconfiguration of dimers. The represented configurations all belong to the $(W_x,W_y)=(0,0)$ sector in which all calculations are performed in this work. Periodic boundary conditions are represented with the help of duplicated grey sites on the boundaries. }
\end{figure*}

Constraints exist or emerge in the description of a wide variety of physical systems, ranging from gauge theories, surface physics (adsorption of molecules~\cite{fowler_attempt_1937}), glass physics~\cite{fredrickson_kinetic_1984,ritort_glassy_2003,chandler_dynamics_2010}, quantum or classical magnets~\cite{lacroix_introduction_2011,castelnovo_spin_2012} or atomic physics (such as with Rydberg blockade~\cite{bernien_probing_2017}). Theoretical descriptions are written in terms of models with adapted degrees of freedom, such as dimer, loop or ice models, which exhibit physical features not present in unconstrained counterparts. Consider for instance the dimer model (Fig.~\ref{fig:dimers}) that will be used in this work, where the degrees of freedom are dimers located on bonds of a given lattice (or graph). The constraint is given by the following rule: each site of a lattice must be touched by one and only one dimer. Already at the classical static level, this {\it local} rule enforces long-range, algebraic correlations, between dimers on bipartite lattices~\cite{fisher_statistical_1963,huse_coulomb_2003}. Taking again the dimer example, it is not possible to obtain another dimer configuration by simply moving one dimer: only constrained moves (the smallest of which is represented in Fig.~\ref{fig:dimers}) are possible. This can result in very complex dynamics (see Refs.~\onlinecite{oakes_emergence_2016,das_critical_2005,cepas_multiple_2014} for classical examples).

Recent work highlighted the also unusual {\it quantum} dynamics exhibited by quantum constrained models, either in theory~\cite{chamon_quantum_2005,lan_quantum_2018,turner_weak_2018,feldmeier_dynamical_2019,pancotti_quantum_2020} or experiments~\cite{bernien_probing_2017}, as well as for related lattice gauge theories~\cite{smith_disorder_2017,brenes_many_2017,smith_absence_2017,smith_dynamical_2018,karpov_disorder_2020,Papaefstathiou_disorder_2020,Halimeh_staircase_2020,Halimeh_origin_2020}. On one hand, we would thus expect that constrained models are good natural candidates to exhibit localization due to their enhanced slow dynamics. On the other hand, the local constraint is such that these models are already in a {\it strongly interacting} limit (even without interactions encoded in a Hamiltonian), which would rather favor delocalization and thermalization. This competition provides strong motivation to investigate the possibility of MBL in quantum constrained models. The interplay between constraints, interactions and disorder has been explored in recent works on 1d models~\cite{chen_how_2018,ostmann_localization_2019} (for Ref.~\onlinecite{chen_how_2018}, equivalent to a dimer model on a ladder), which constraints are relevant for the physics of cold-atomic chains in the regime of Rydberg blockade~\cite{bernien_probing_2017}, resulting in rich phase diagrams, with {\it e.g.} different mechanisms to produce MBL~\cite{chen_how_2018}.

Another important aspect (crucial for the numerical approach taken in this work) worth mentioning is the fact that the constraints overall {\it reduce} the size of the phase space for allowed configurations. For instance, the size of the Hilbert space scales as $\sim 1.34^N$ for dimer coverings of square lattices with $N$ sites, instead of $2^N$ for spin 1/2 models (or $4^N$ if no constraints were present). This greatly helps numerics based on exact approaches, as this enlarges the set of available finite samples in 2d. Given that exact diagonalization simulations have been instrumental in the study and understanding of MBL, this is also an important positive feature in the search for MBL in quantum constrained models.

The plan of the paper is as follows. We first describe the model and the 2d lattices used in this work in Sec.~\ref{sec:model}, along with a brief description of the exact numerical methods employed (Sec.~\ref{sec:num}). We then describe in detail physical properties of the eigenstates of this model in Sec.~\ref{sec:eigen}, including spectral and eigenstate statistics, expectation values of local observables as well as half-system entanglement entropy. Sec.~\ref{sec:ml} is devoted to a machine-learning analysis of entanglement spectra. Sec.~\ref{sec:dyn} then presents an analysis of dynamical properties after a quench. Finally, we summarize our results and present perspectives in Sec.~\ref{sec:conc}.

\section{Model and numerical methods}

\subsection{Model}
\label{sec:model}

This work focuses on a quantum dimer model on a square lattice with random potential terms. Quantum dimer models have a long history in condensed matter as low-energy effective models. They have been first introduced by Rokhsar and Kivelson~\cite{rokhsar_superconductivity_1988} in the context of the Resonating Valence Bond theory of high-temperature superconductivity, an interest which has been revived with the use of QDMs to describe experiments on underdoped cuprates~\cite{punk_quantum_2015}. QDMs are also prominent representations of highly frustrated magnets where dimers represent nearest-neighbor spin singlets~\footnote{See R. Moessner, K. S. Raman, Chapter 17 in Ref.~\onlinecite{lacroix_introduction_2011} for a review}. Dimer models are indeed often more tractable than frustrated spin models from which they can be systematically derived~\cite{schwandt_generalized_2010}, while capturing the essential physics. For instance, the existence of 2d quantum spin liquids exhibiting topological order was first demonstrated in a QDM on a non-bipartite lattice~\cite{moessner_resonating_2001}.

Quantum dimer models are defined in a configuration space where only configurations which fulfill the dimer constraint are allowed (dimer coverings, see Fig.~\ref{fig:dimers} (a)). Their Hamiltonian contains a potential term that attributes an energy to every dimer covering and a kinetic term that flips dimers along a loop, allowing to pass from one dimer configuration to another. The specific QDM Hamiltonian that we consider in this work is based on the square lattice and contains a potential term with a random component, taking the following form
\begin{eqnarray}
\label{eq:H}
H  & = & - t \sum_{\square}  \left( |
\begin{tikzpicture}[scale=0.35,baseline={([yshift=-0.6ex]current bounding box)}]
    \draw (0,0) rectangle (1,1);
    \draw[fill=orange] (0.5,1) ellipse (0.56 and 0.12);
    \draw[fill=orange] (0.5,0) ellipse (0.56 and 0.12);
\end{tikzpicture}
\rangle \langle
\begin{tikzpicture}[scale=0.35,baseline={([yshift=-.6ex]current bounding box)}]
    \draw (0,0) rectangle (1,1);
    \draw[fill=orange] (0,0.5) ellipse (0.13 and 0.56);
    \draw[fill=orange] (1,0.5) ellipse (0.13 and 0.56);
\end{tikzpicture} | + |
\begin{tikzpicture}[scale=0.35,baseline={([yshift=-.6ex]current bounding box)}]
    \draw (0,0) rectangle (1,1);
    \draw[fill=orange] (0,0.5) ellipse (0.13 and 0.56);
    \draw[fill=orange] (1,0.5) ellipse (0.13 and 0.56);
\end{tikzpicture}
\rangle \langle
\begin{tikzpicture}[scale=0.35,baseline={([yshift=-0.6ex]current bounding box)}]
    \draw (0,0) rectangle (1,1);
    \draw[fill=orange] (0.5,1) ellipse (0.56 and 0.12);
    \draw[fill=orange] (0.5,0) ellipse (0.56 and 0.12);
\end{tikzpicture} |
\right) \nonumber \\
&  & + \sum_{\square_p}  V_p \left( |
\begin{tikzpicture}[scale=0.35,baseline={([yshift=-0.6ex]current bounding box)}]
    \draw (0,0) rectangle (1,1);
    \draw[fill=orange] (0.5,1) ellipse (0.56 and 0.12);
    \draw[fill=orange] (0.5,0) ellipse (0.56 and 0.12);
\end{tikzpicture}
\rangle \langle
\begin{tikzpicture}[scale=0.35,baseline={([yshift=-0.6ex]current bounding box)}]
    \draw (0,0) rectangle (1,1);
    \draw[fill=orange] (0.5,1) ellipse (0.56 and 0.12);
    \draw[fill=orange] (0.5,0) ellipse (0.56 and 0.12);
\end{tikzpicture}
 | + |
\begin{tikzpicture}[scale=0.35,baseline={([yshift=-.6ex]current bounding box)}]
    \draw (0,0) rectangle (1,1);
    \draw[fill=orange] (0,0.5) ellipse (0.13 and 0.56);
    \draw[fill=orange] (1,0.5) ellipse (0.13 and 0.56);
\end{tikzpicture}
\rangle \langle
\begin{tikzpicture}[scale=0.35,baseline={([yshift=-.6ex]current bounding box)}]
    \draw (0,0) rectangle (1,1);
    \draw[fill=orange] (0,0.5) ellipse (0.13 and 0.56);
    \draw[fill=orange] (1,0.5) ellipse (0.13 and 0.56);
\end{tikzpicture} |
\right).
\end{eqnarray}

The sums run over all plaquettes ${\square_p}$ of the square lattice, $V_p$ is a random potential different for each plaquette which is drawn uniformly from a box distribution $V_p \in [-V,V]$. The kinetic energy scale is set to $t=1$. The various terms are illustrated in Fig.~\ref{fig:dimers} (b). At $V=0$ (no disorder), Ref.~\onlinecite{lan_eigenstate_2017} showed that eigenstates in the middle of the spectrum obey the eigenstate thermalization hypothesis. Quite importantly, there is no limit where Eq.~\ref{eq:H} can be mapped to a model of free particles moving in a random potential (no Anderson localization limit).

Quantum dimer models based on coverings of bipartite lattices (such as the square lattice) have an important property: they display conserved topological numbers, known as winding numbers, when considering systems with periodic boundary conditions. The definition of the two conserved winding numbers (denoted $W_x$ and $W_y$) is given in Fig.~\ref{fig:dimers} (c). It is easy to see that they are conserved by the Hamiltonian Eq.~\ref{eq:H}, and by any local moves ({\it i.e.} that do not go through periodic boundary conditions), hence their topological nature. $W_x$ and $W_y$ both range from $-L/2$ to $L/2$ on a square lattice. The Hamiltonian is block-diagonal, with sectors identified by $(W_x,W_y)$. We note that the resulting abelian $U(1)$ symmetry of the model Eq.~\ref{eq:H} is compatible with many-body localization~\cite{potter_symmetry_2016}. Without loss of generality, we will restrict to the largest sector $(W_x=0,W_y=0)$ in what follows.

\subsection{Lattices and numerical methods}
\label{sec:num}

\begin{table*}
\begin{tabular}{|c|c|c|c|c|c|c|c|c|c|}
  \hline \hline
  Lattice & $4 \times 6$ & $32$ & $4 \times 8$ & $6 \times 6$ & $40$ & $4 \times 10$ & $6 \times 8$ & $52$ & $8 \times 8$ \\
  \hline
  Lattice type & Rectangular & Tilted sq. & Rectangular & Regular sq. & Tilted sq. & Rectangular & Rectangular & Tilted sq. & Regular sq. \\
  \hline
  Hilbert space size $\cal H$  & $1\, 456$ & $13\, 348$ & $17\, 412$ & $44\, 176$ & $139\, 212$ & $216\, 016$ & $1\, 504\, 896$ & $4\, 572\, 468$ &  $153\, 722\, 916$ \\
  \hline \hline
\end{tabular}
  \caption{\label{tab:Hilbert}Two-dimensional lattices used in this work, with their shape (regular or tilted square lattices, or rectangular) and the size of their Hilbert space in the $(W_x=0,W_y=0)$ winding sector.}
\end{table*}

This work uses two-dimensional lattices (represented in Fig.~\ref{fig:Lattices} in the Appendix) with $N$ sites and $2N$ bonds of the following form: regular square lattices with $N=L^2$ sites (with $L=4,6,8$), rectangular lattices with $N=L_x \times L_y$ sites (with $L_{x,y}=4,6,8,10$) as well as tilted square lattices with $N=32, 40$ and $52$ sites (see Appendix for details). Periodic boundary conditions are taken. All the lattices used in this work are presented in Tab.~\ref{tab:Hilbert}, together with the corresponding Hilbert space sizes in the $(W_x=0,W_y=0)$ sector.

We use different numerical methods to treat exactly the Hamiltonian Eq.~\ref{eq:H}. In Sec.~\ref{sec:eigen}, we first use methods which allow to obtain exact eigenstates of $H$. As usually performed in the MBL context, we concentrate on eigenstates with eigenvalue $E$ located right in the middle of the many-body spectrum, that is at $\epsilon=0.5$ with $\epsilon = (E-E_{\rm min})/(E_{\rm max}-E_{\rm min})$ where $E_{\rm min / max}$ are the extremal energies of $H$.
Full numerical diagonalization is possible up to $N=36$ sites. We also use massively parallel shift-invert diagonalization~\cite{pietracaprina_shift-invert_2018} that allows to target eigenstates at $\epsilon=0.5$ on larger samples up to $N=52$. We typically obtain about $100$ eigenstates near $\epsilon=0.5$. Results are averaged over more than 500 (up to 10.000) realizations of disorder. To show the interest of working with constrained models, the Hilbert space of the largest cluster on which we can obtain exact eigenstates ($N=52$) is almost twice as large as the one for $N=24$ spins 1/2 (in the zero magnetization sector), which constitutes the state of the art for the analysis of MBL in spin chains~\cite{pietracaprina_shift-invert_2018,mace_many-body_2019,mace_multifractal_2019}. Obtaining $10$ eigenstates in the middle of the spectrum for one disorder sample at $N=52$ requires about $20$ minutes on 10.000 CPU cores.

In Sec.~\ref{sec:dyn}, we also consider dynamical properties of various observables after a quench. The knowledge of all eigenstates allows to consider properties at arbitrary time up to $N=36$. We also use Krylov expansion time evolution techniques~\cite{nauts_new_1983}, which allow simulations up to time of the order of thousand plaquette flips on large clusters up to $N=64$. Note that simulations on the largest sample $N=64$ are extremely demanding (see sizes of Hilbert space in Tab.~\ref{tab:Hilbert}). In the study of dynamics, we average over between 30 and 1000 realizations of disorder for each simulation set.

\section{Properties of eigenstates}
\label{sec:eigen}

\subsection{Spectral statistics}

\begin{figure}
\includegraphics[width=0.95 \columnwidth]{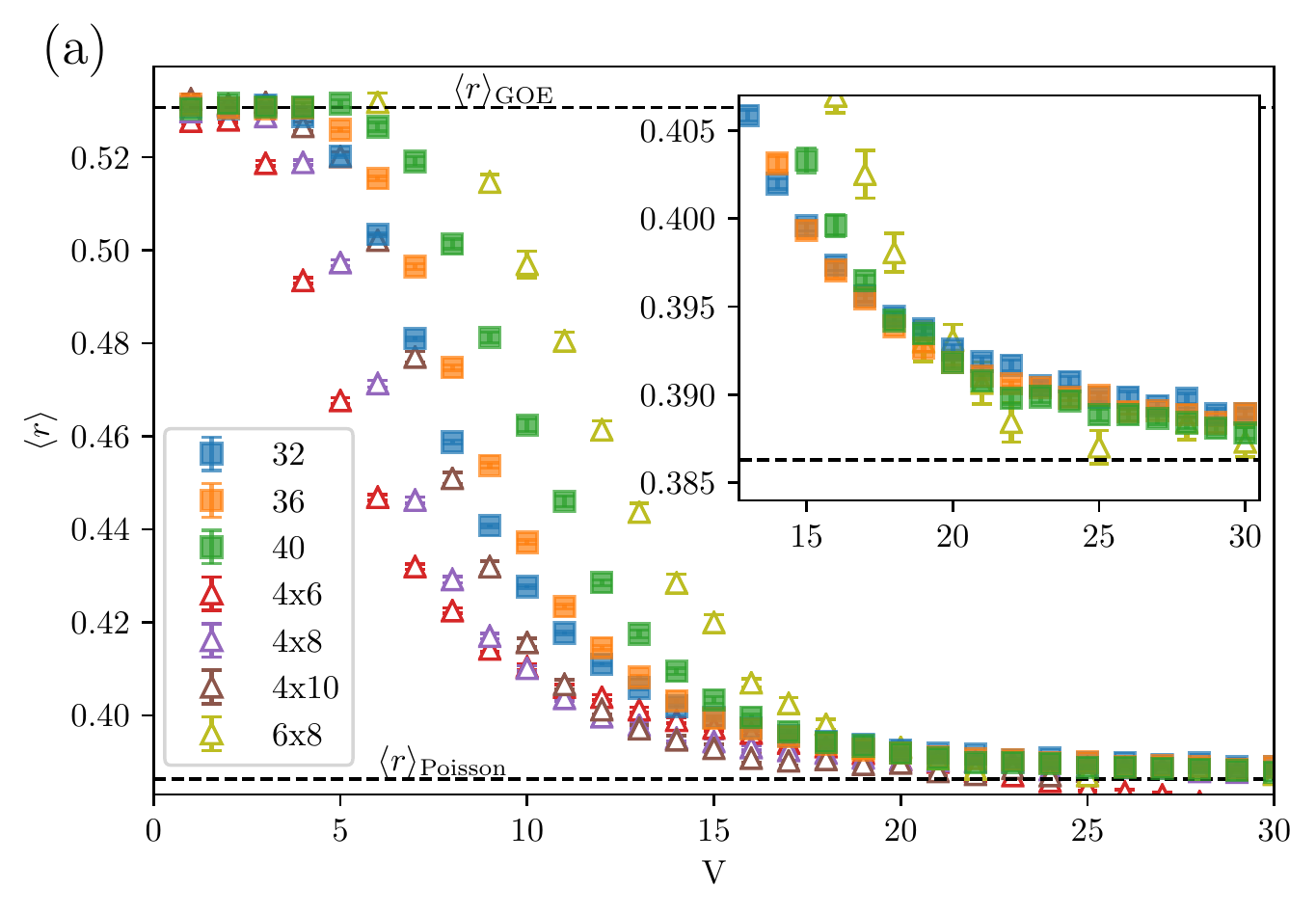}
\includegraphics[width=0.95 \columnwidth]{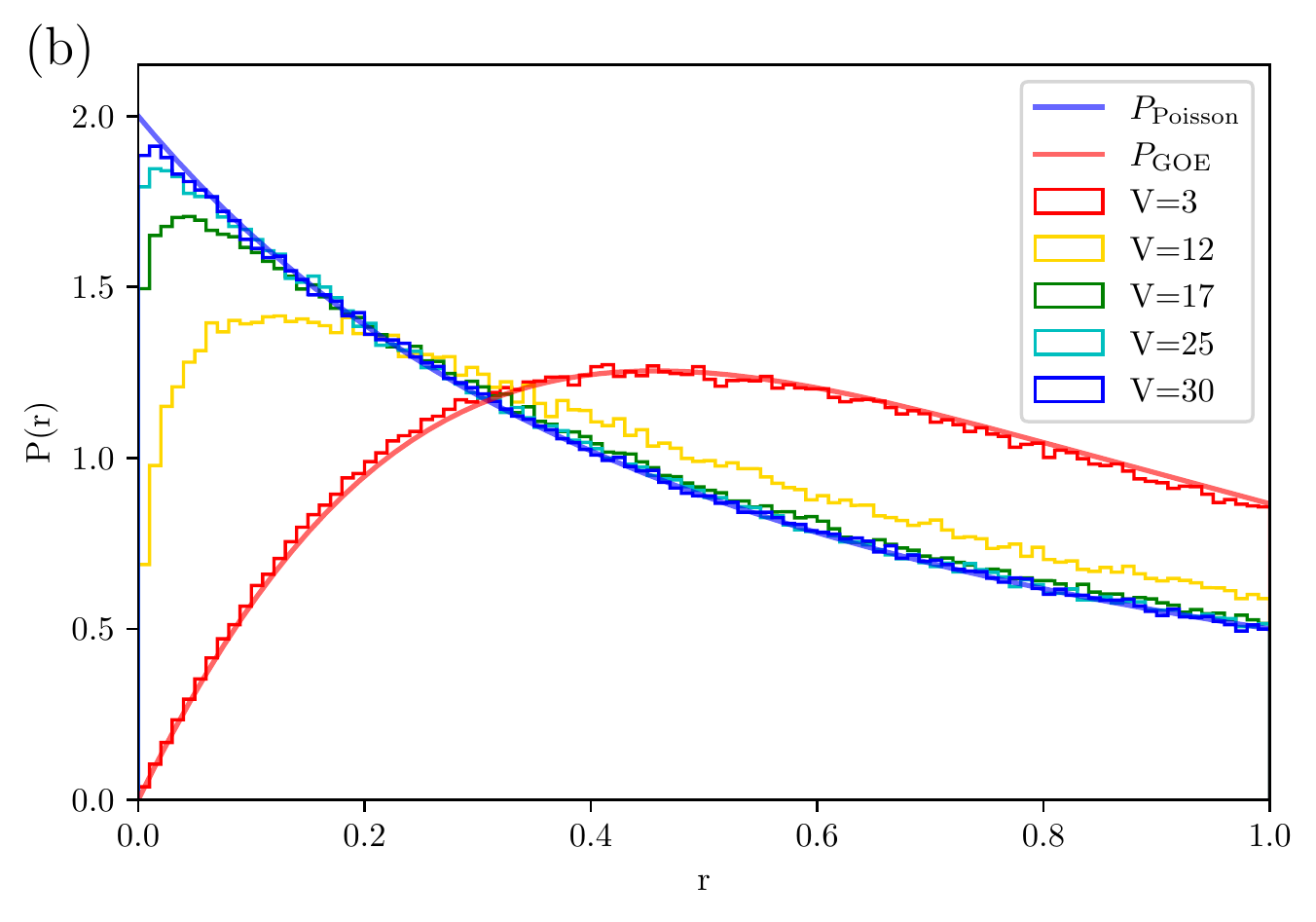}
\caption{\label{fig:gap_ratio} Statistics of the reduced gap ratio $r$ for eigenstates located in the middle of the spectrum $\epsilon=0.5$. (a) Average gap ratio $\langle r \rangle$ for different 2d samples as a function of disorder. Square samples are color highlighted. (b) Probability distribution $P(r)$ for $N=40$ sites tilted square lattice for different values of disorder.}
\end{figure}

We first consider the simplest property of a many-body system, namely its energy spectrum. The nature of the distribution of energy levels can already discriminate between different regimes: ergodic systems display level repulsion while localized systems display a random distribution of level spacing characterized by Poisson statistics. In the many-body context, and in order to counter the fact that the density of states is not uniform, it is useful to consider the distribution of consecutive ratios as introduced in Ref.~\onlinecite{oganesyan_localization_2007}. We define gaps in the many-body spectrum as $s_n = E_n - E_{n-1}$ and consider the consecutive reduced gap ratio $r_n= \frac{\min{(s_n,s_{n+1})}}{\max{(s_n,s_{n+1})}}$, for which $0 \leq r \leq 1$. The distribution $P(r)$ and average value of $\langle r \rangle = \int_0^1 r P(r) dr$, averaged over eigenstates and disorder realizations, have been computed for random matrix ensembles (corresponding to ergodic systems) and for independent energy levels (Poisson statistics)~\cite{atas_distribution_2013}. For real Hamiltonians such as Eq.~\ref{eq:H}, the random matrix ensemble to be considered is the Gaussian Orthogonal Ensemble (GOE). Fig.~\ref{fig:gap_ratio} displays both the distribution $P(r)$ and its average $\langle r \rangle$ for various values of disorder and system sizes. We compare them to predictions for the GOE ensemble~\cite{atas_distribution_2013} $P_{\rm GOE}(r)=\frac{27}{4}\frac{r+r^2}{(1+r+r^2)^{5/2}}$, $\langle r_{\rm GOE} \rangle \simeq 0.5307 $ and Poisson statistics $P_{\rm Poisson}(r)=\frac{2}{(1+r)^2}$, $\langle r_{\mathrm Poisson} \rangle \simeq 0.38629 $. Considering first the distribution (Fig.~\ref{fig:gap_ratio}b), we find that the agreement with $P_{\rm GOE}(r)$ for low values of disorder ($V=3$) and $P_{\rm Poisson}(r)$ for high values ($V=25,30$) is excellent, with no adjustable parameter. Intermediate values of disorder $V=12,17$ present distributions which display a crossover behavior between the two limit distributions for the system size considered $N=40$. The average gap ratio $\langle r \rangle$ also displays an interesting crossover between the two limiting cases, with different system sizes showing an apparent crossing point (see inset of Fig.~\ref{fig:gap_ratio}a for square samples $N=32,36,40$) around $V \simeq 15-20$. We note that rectangular samples also show the same crossover between the GOE and Poisson limits, but do not cross exactly at the same value which can be attributed to their different aspect ratios~\footnote{Considering only the clusters with $L_x=4$ leads to a smaller critical disorder $V \sim 12$ for the one-dimensional geometry of this four-leg tube. Results on other observables are consistent with this estimate.}. The critical value of the gap ratio $\langle r \rangle^* \simeq 0.392$ is smaller than for the 1d standard MBL model~\cite{luitz_many-body_2015}, indicating that the putative transition point looks even closer to the Poissonian localized limit. Note that for the largest disorder $V=30$ considered here, we also checked on the largest system available to full diagonalization ($N=36$) that no band-like structure is present in the energy spectrum, a feature which was found~\cite{papic_many-body_2015} to mimic MBL on too small system sizes.

\subsection{Eigenstate statistics}

\begin{figure}
\includegraphics[width=0.95 \columnwidth]{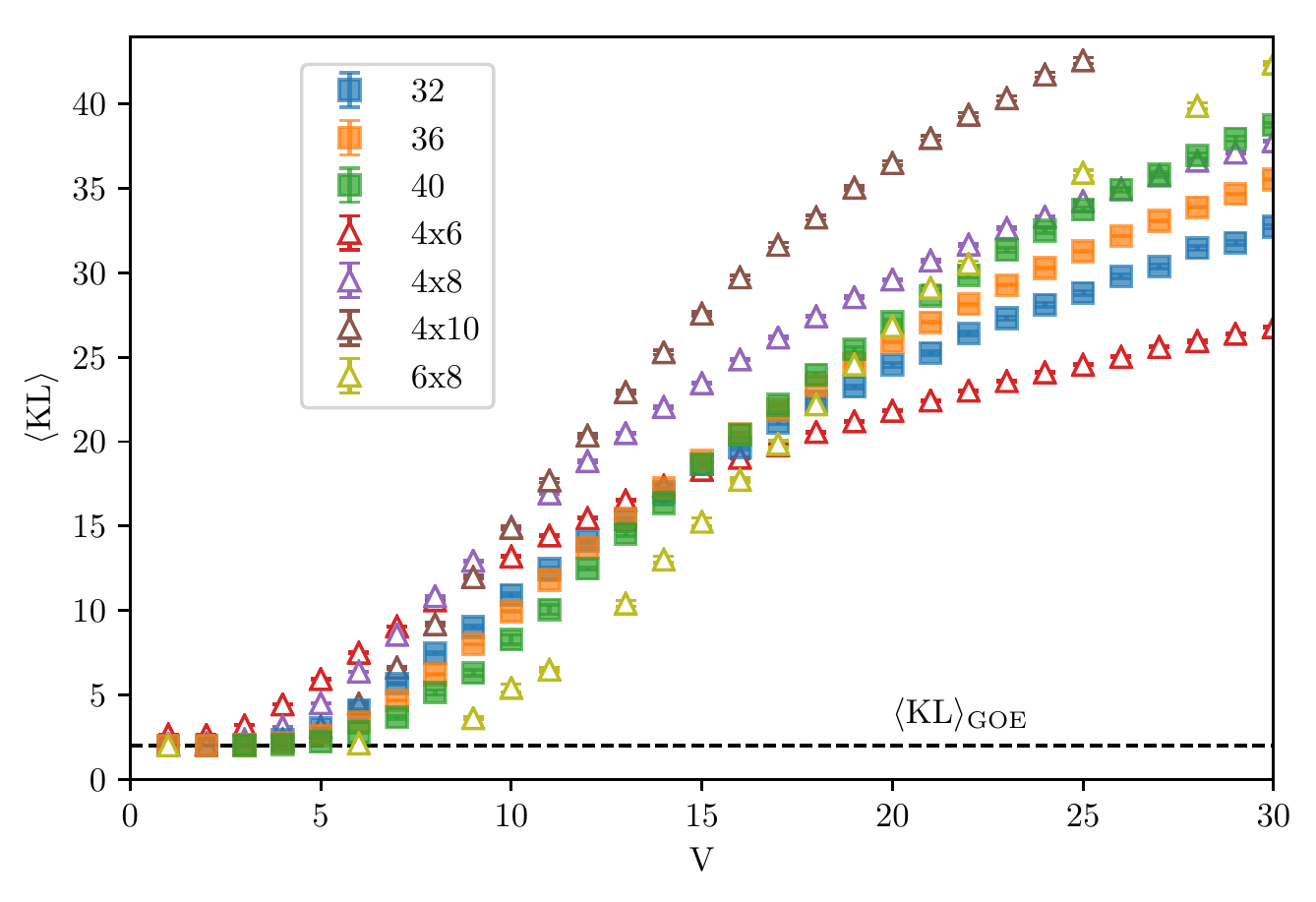}
\caption{\label{fig:KL} Average Kullback-Leibler divergence $\langle \mathrm{KL} \rangle$ as a function of disorder for different system sizes. Square samples are color highlighted.}
\end{figure}

We next consider the similarity between eigenstates at the same energy density. It is well-known that eigenstates in an ergodic phase look 'similar' ({\it e.g.} they present similar expectation values for local observables), while eigenstates in a MBL phase are very different ({\it e.g.} local observables can differ strongly from one eigenstate to the following one in the many-body spectrum). To quantify the degree of similarity, it is useful to consider~\cite{luitz_many-body_2015} the Kullback-Leibler divergence between two eigenstates $| n \rangle $ and $| n' \rangle $ located nearby in the spectrum (here at the same energy density $\epsilon=0.5$). It is defined as $\mathrm{KL}=\sum_i p_i \log (p_i/p'_i)$ where $p_i$ and $p'_i$ are the participation coefficients, {\it i.e.} the square amplitudes $p_i=|\langle  n | i \rangle |^2$ , $p'_i=| \langle  n' | i \rangle |^2$ of eigenstates in a given basis (here the computational basis where dimer occupations are diagonal). The sum runs over the full Hilbert space, and the eigenstates are normalized such that $\sum_i p_i = \sum_i p'_i=1$. Eigenstates at `infinite temperature' ($\epsilon =0.5$) in an ergodic phase should be very close to eigenstates of GOE random matrices, for which one has $\langle \mathrm{KL} \rangle_{\rm GOE}=2$. On the other hand, for eigenstates which are many-body localized, we expect $\langle \mathrm{KL} \rangle$ to diverge with system size as the eigenstates are increasingly different. Fig.~\ref{fig:KL} precisely shows this behavior, with expectation values $\langle \mathrm{KL} \rangle$ converging to their limiting GOE value at low disorder, and diverging for large disorder. We also observe a crossing point at $V \simeq 15$, albeit slightly drifting, for different system sizes. This drifting behavior is also observed for 1d MBL in the random-field Heisenberg model~\cite{luitz_many-body_2015}. The value of $\langle \mathrm{KL} \rangle$ at the transition point is larger than for the 1d spin model~\cite{luitz_many-body_2015}, pointing again to a putative critical point closer to the localized limit.

\subsection{Dimer occupation and columnar imbalance}
\label{sec:occupation}

\begin{figure}
\includegraphics[width=0.95 \columnwidth]{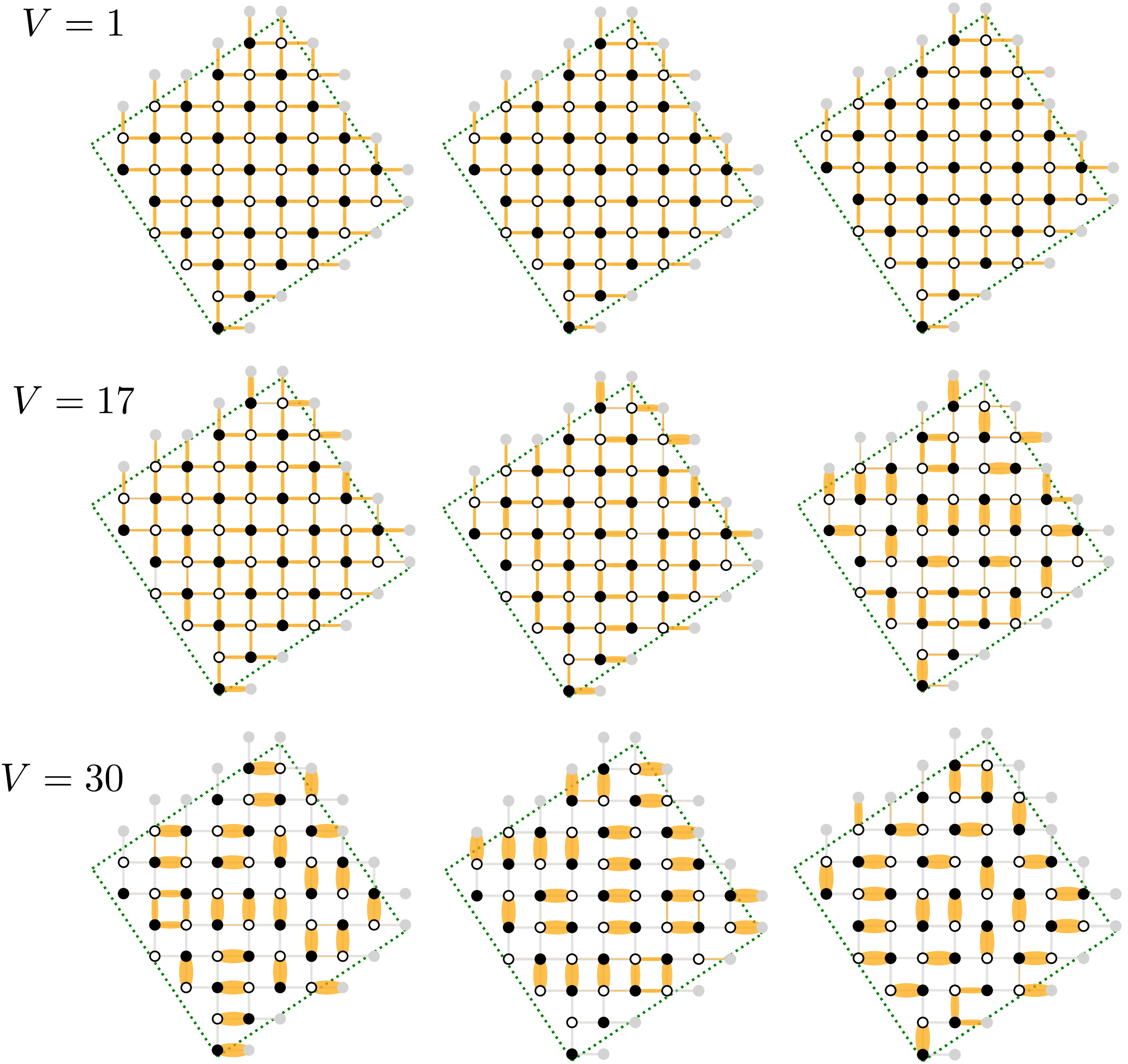}%
\caption{\label{fig:occup} Dimer occupation $\langle n_b \rangle $ for three consecutive eigenstates (from left to right) in the middle of the spectrum, for a single disorder realization for different values of disorder strength (from top to bottom). The width of orange dimers located on each bond is proportional to $\langle n_b \rangle $. Data for the $N=52$ tilted square lattice.}
\end{figure}

{\it Dimer occupation --- } A qualitative understanding of the behavior of eigenstates in the different regimes is provided by considering local observables, and in particular by lattice bond occupation by dimers. For each lattice bond $b$, we compute the dimer occupation $\langle n | n_b | n \rangle$ in the eigenstate $|n\rangle$ with $n_b | i \rangle = 1 $ if the bond $b$ is occupied by a dimer in the basis configuration $|i \rangle$, $0$ otherwise.

Fig.~\ref{fig:occup} represents graphically the dimer occupation $\langle n | n_b | n \rangle$ in three different eigenstates, consecutive in the many-body spectrum, for three values of disorder strength. In the low-disorder phase ($V=1$), all dimer occupations are equally likely: $\langle n_b \rangle \simeq 1/4$ for all bonds and all eigenstates. The dimer occupations at large disorder ($V=30$) mainly overlap with a single dimer configuration, which differs totally from one eigenstate to the next one. Resonances between configurations which differ by one or two plaquette flips can be evidenced by almost equal occupation of parallel dimers around a plaquette in some eigenstates (e.g. left-middle plaquette for the left-most eigenstate at $V=30$). At intermediate disorder ($V=17$), while for the two first left eigenstates dimer occupations look similar (at a superficial level), the right eigenstate configuration is markedly different.

\begin{figure}
\includegraphics[width=0.95 \columnwidth]{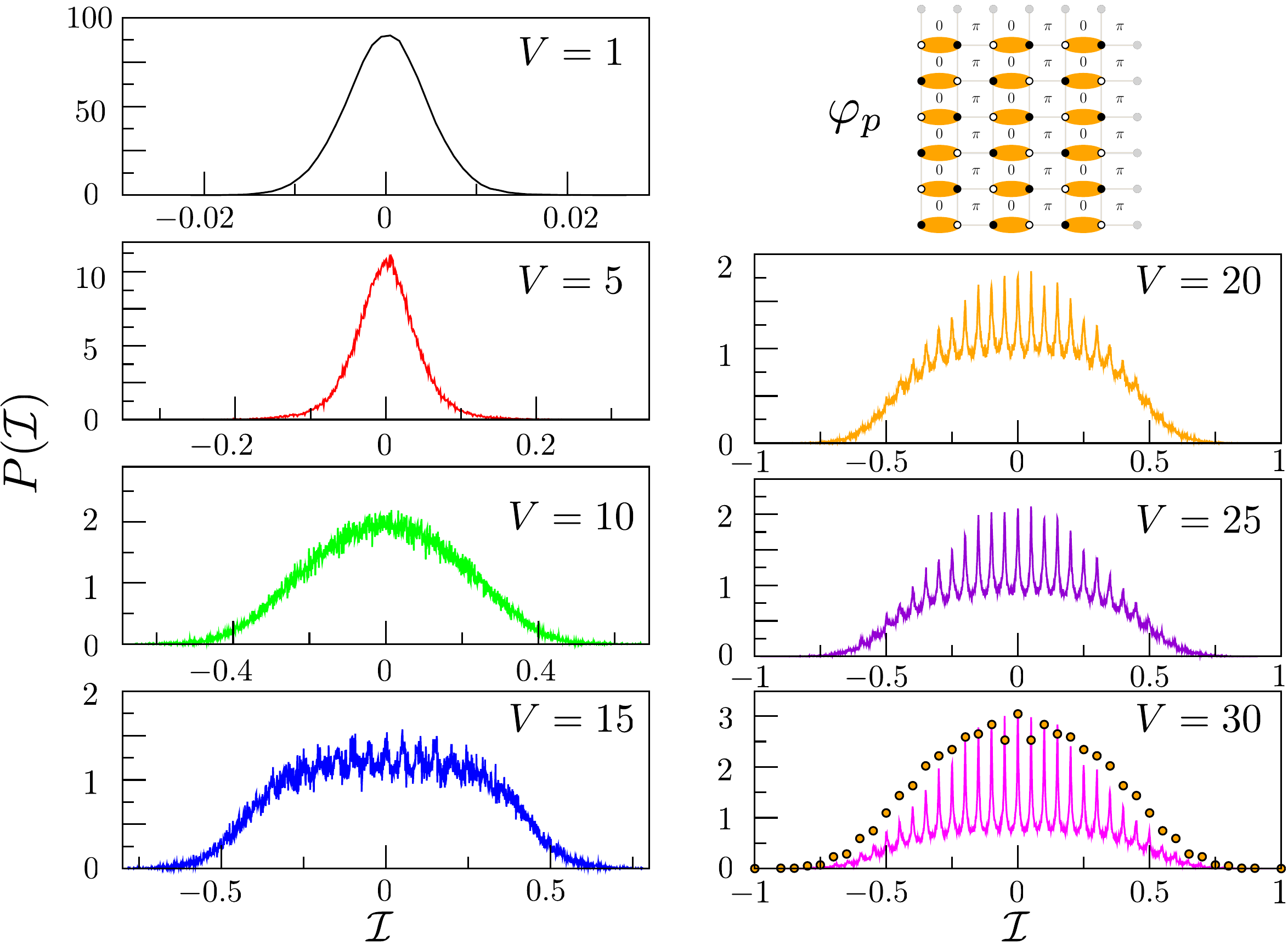}%
\caption{\label{fig:imb} Probability distribution of columnar imbalance $\cal I$ for different values of disorder. On the bottom right panel is also represented (circles) the distribution for pure dimer configurations $P_{\rm conf}({\cal I})$ (normalized such that $P_{\rm conf}({\cal I}=0)=P_{V=30}({\cal I}=0)$. Data for the tilted square lattice with $N=40$ sites, with more than $10.000$ eigenstates for each value of disorder. Inset (top right): definition of the phase $\phi_p$ used in the definition of the columnar imbalance, on top of the columnar configuration for which ${\cal I}=1$.}
\end{figure}

{\it Columnar imbalance --- } We now consider expectation values for another local observable, which will be useful later when discussing dynamics in Sec.~\ref{sec:dyn}. We define the imbalance with respect to some specific state, the columnar configuration represented in the inset of Fig.~\ref{fig:imb}, as:
\begin{equation}
  {\cal I} = \frac{1}{N_p}  \sum_{\square_p} e^{{\rm i} \varphi_p} \left( 2 \left( |
\begin{tikzpicture}[scale=0.35,baseline={([yshift=-0.6ex]current bounding box)}]
    \draw (0,0) rectangle (1,1);
    \draw[fill=orange] (0.5,1) ellipse (0.56 and 0.12);
    \draw[fill=orange] (0.5,0) ellipse (0.56 and 0.12);
\end{tikzpicture}_p
\rangle \langle
\begin{tikzpicture}[scale=0.35,baseline={([yshift=-0.6ex]current bounding box)}]
    \draw (0,0) rectangle (1,1);
    \draw[fill=orange] (0.5,1) ellipse (0.56 and 0.12);
    \draw[fill=orange] (0.5,0) ellipse (0.56 and 0.12);
\end{tikzpicture}_p
 | + |
\begin{tikzpicture}[scale=0.35,baseline={([yshift=-.6ex]current bounding box)}]
    \draw (0,0) rectangle (1,1);
    \draw[fill=orange] (0,0.5) ellipse (0.13 and 0.56);
    \draw[fill=orange] (1,0.5) ellipse (0.13 and 0.56);
\end{tikzpicture}_p
\rangle \langle
\begin{tikzpicture}[scale=0.35,baseline={([yshift=-.6ex]current bounding box)}]
    \draw (0,0) rectangle (1,1);
    \draw[fill=orange] (0,0.5) ellipse (0.13 and 0.56);
    \draw[fill=orange] (1,0.5) ellipse (0.13 and 0.56);
\end{tikzpicture}_p |
\right) - \hat{I} \right).
\label{eq:imb}
\end{equation}

The phase $\varphi_p$ carried out by each plaquette is simply chosen to be $0$ or $\pi$ on a columnar arrangement (see inset of Fig.~\ref{fig:imb}) such that the imbalance of this perfect columnar state is maximal and equal to ${\cal I}=1$.  At thermal equilibrium and infinite temperature, the imbalance vanishes on average.

We display in Fig.~\ref{fig:imb} the probability distribution $P({\cal I})$ of $\langle n | {\cal I} | n \rangle$ taken over eigenstates $|n\rangle$ located at $\epsilon=0.5$ for different values of disorder and computed on the tilted square lattice with $N=40$ sites. For small values of disorder (left column), the probability distribution is gaussian around the mean value $0$, as expected for an ETH phase. The typical width of the distribution (which is very small for the smallest disorder $V=1$) broadens with disorder. Close to $V \simeq 15$, peaks for specific values of the imbalance start to appear in the distribution which become clearly visible for all strong values of disorder $V \geq 20$ (right panels). The distribution of imbalance for pure dimer configurations $P_{\rm conf}({\cal I})$ (obtained by computing ${\cal I}$ for all dimer configurations in the Hilbert space) is represented in the $V=30$ panel for comparison (there are $1/N$ values possible for the imbalance ${\cal I}$). The location of the peaks of $P({\cal I})$ matches precisely those of $P_{\rm conf}({\cal I})$, with however different amplitudes. This indicates that at large $V$, eigenstates are close (but not exactly equal) to pure dimer configurations, in agreement with the qualitative view provided by the snapshots of Fig.~\ref{fig:occup}.

\subsection{Participation entropies}
\label{sec:participation}

\begin{figure}[b]
\includegraphics[width=0.95 \columnwidth]{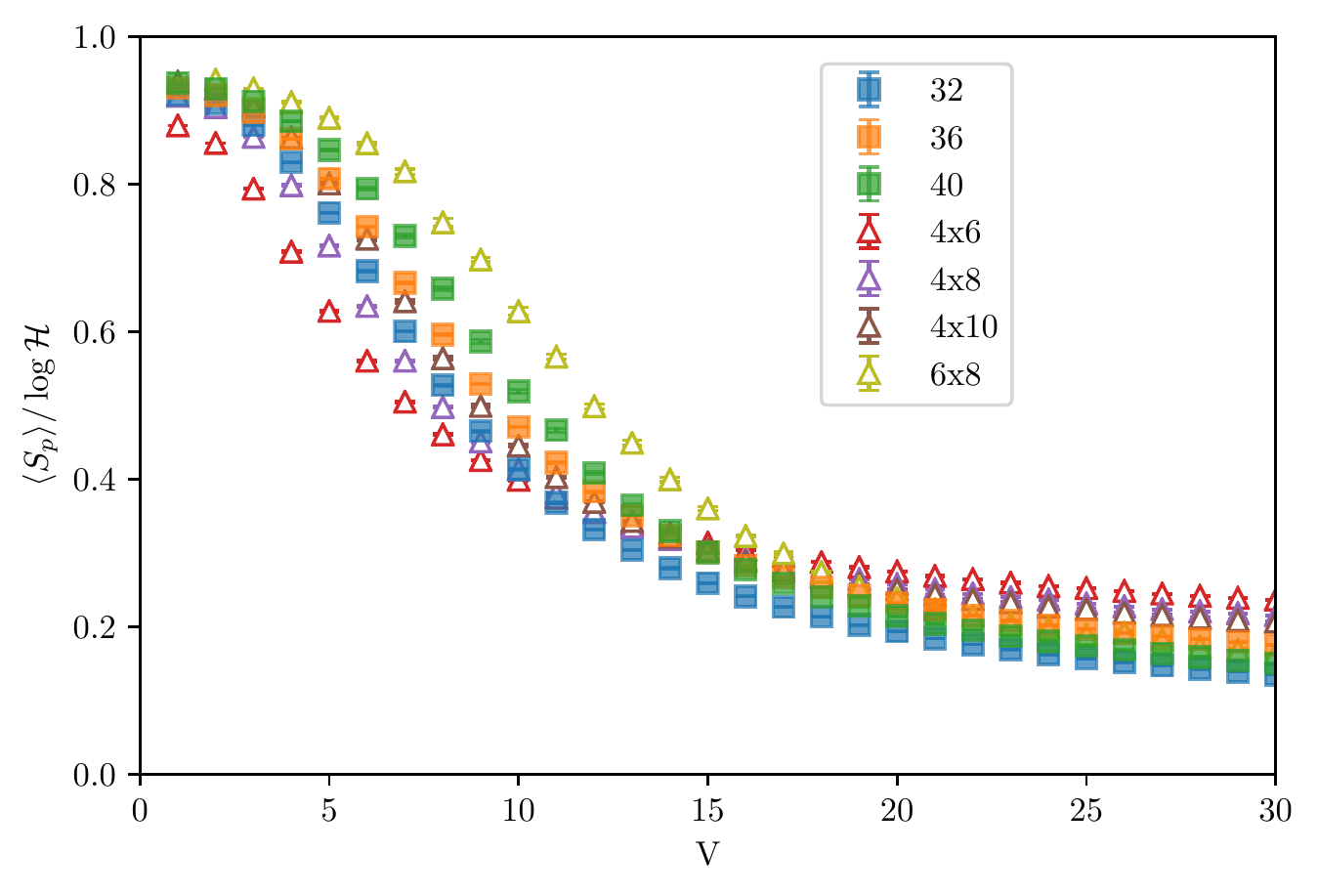}
\caption{\label{fig:part} Average participation entropy divided by the log of the size of Hilbert space as a function of disorder for different system sizes. Square samples are color highlighted.}
\end{figure}

To understand the localization properties of the eigenfunctions in the Hilbert (configuration) space, we next consider the participation entropy
\begin{equation}
S_p (|n\rangle) = - \sum_i p_i \log(p_i)
\end{equation}
where $p_i = |\langle n | i \rangle|^2$ and the sum runs over all basis states $| i \rangle$ of the computational basis. At low disorder, we expect eigenfunctions to be fully delocalized with a participation entropy scaling as $S_p = a \log ({\cal H}) - \dots$ (with $a=1$ for a fully-delocalized state and $\dots$ denoting finite size corrections). In the large disorder regime on the other hand, based on data for the MBL phase in the 1d random field Heisenberg~\cite{mace_multifractal_2019}, we do not expect a true localization in configuration space but instead a volumic, multifractal, behavior where $S_p = a \log (${\cal H}$)$ with $a$ strictly less than 1 (and increasingly small as disorder is increased). Fig.~\ref{fig:part} presents the numerical data for the average participation entropy divided by the log of the size of the Hilbert space $\langle S_p \rangle / \log({\cal H})$. Two regimes can be clearly distinguished: one at low $V$ where data tend to their maximal value ($a=1$) with system size, and another at larger disorder where curves tend to group together towards sensibly smaller values of the volume law coefficient $a$. This is in accordance with the results for the MBL phase in one dimension~\cite{mace_multifractal_2019}. We find that data for $S_p / \log {\cal H}$ for different sizes tend to regroup themselves around $V \simeq 15-20$, consistent with previous estimates of the possible transition point.

\subsection{Entanglement entropy}
\label{sec:entanglement}

Many-body eigenstates in ergodic or localized phases have strikingly different entanglement properties. Consider the von Neumann entanglement entropy of an eigenstate $|n\rangle$ for a system divided in two parts $A$ and $B$:
\begin{equation}
S (|n\rangle)= -{\rm Tr}_A \rho_A \log (\rho_A)
\label{eq:EE}
\end{equation}
with $\rho_A$ the reduced density matrix obtained by tracing out bond degrees of freedom in B $\rho_A=\mathrm{Tr}_B | n \rangle \langle n |$. For our samples, we choose equal parts $A$ and $B$ containing identical number of lattice bonds and sites (see Appendix for a graphical representation of the entanglement cut).

\begin{figure}[b]
\includegraphics[width=0.95 \columnwidth]{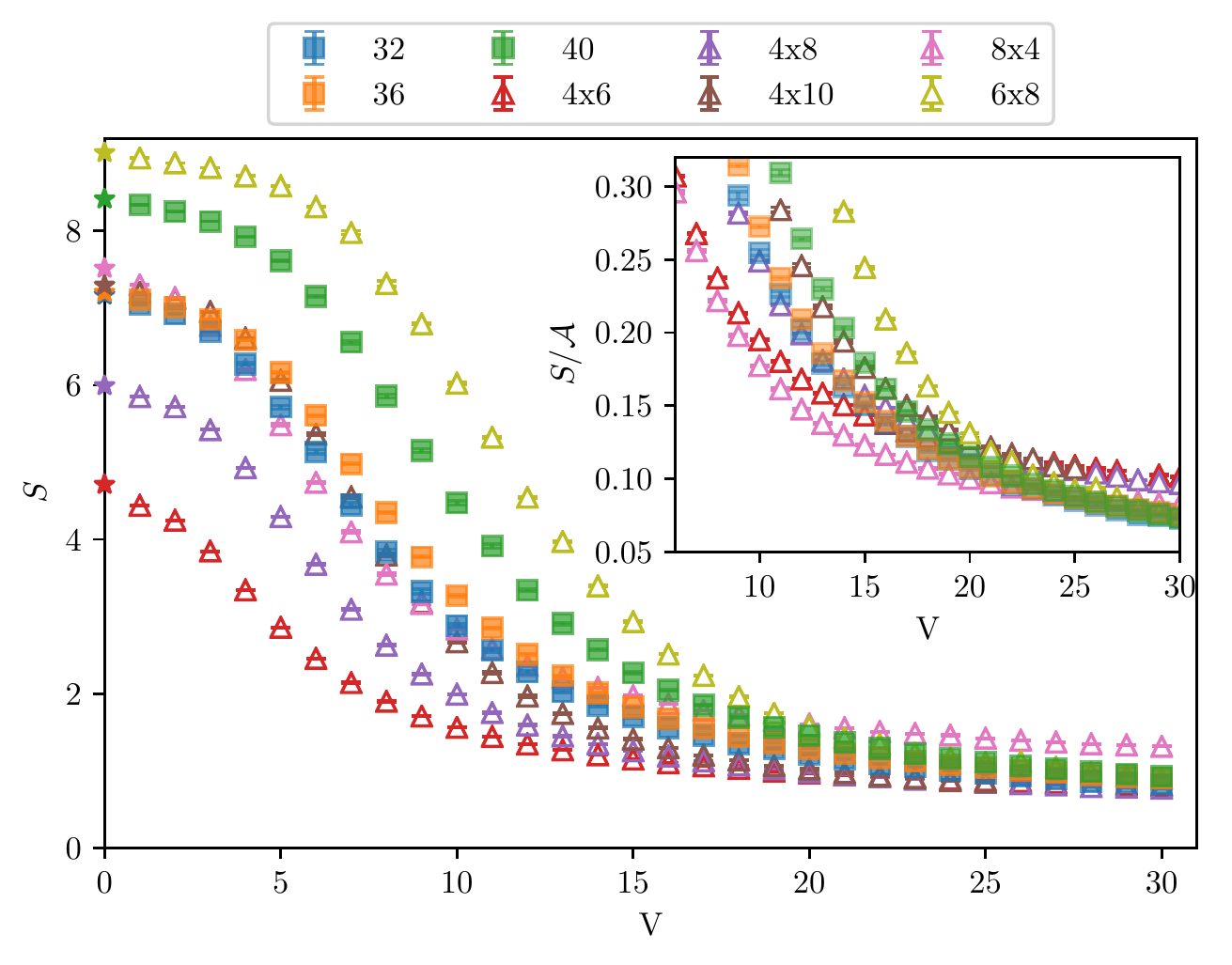}
\caption{\label{fig:S} Average entanglement entropy $S$ of subsystem $A$ as a function of disorder for different system sizes. The star symbol on the $V=0$ axis is the entanglement entropy of a random state $S^{\rm random}$. Inset: Area law scaling of the average entanglement entropy divided by the size of the surface ${\cal A}$ between $A$ and $B$ as a function of disorder (see Appendix for definition of ${\cal A}$ for tilted clusters). Square samples are color highlighted.}
\end{figure}

The entanglement entropy $S$, averaged over eigenstates and realizations of disorder, is represented in Fig.~\ref{fig:S}. In the ETH phase, a {\it volume} law of the entanglement entropy is expected with $S$ scaling with the number of degrees of freedom in $A$. This is what is found in the low-$V$ regime of Fig.~\ref{fig:S}, with the following caveat: due to the constrained nature of the Hilbert space, not all configurations of $A$ are compatible with those of $B$ (see a discussion in Ref.~\onlinecite{morampudi_universal_2020} and in the Appendix) and the number of these `constraint sectors' in $\rho_A$ depends in particular on the {\it area} between $A$ and $B$. This results in different samples with the same volume having different constraints on $\rho_A$ when bipartitioned. This is clearly seen when computing entanglement entropy on {\it random states} in the different samples denoted as $S^{\rm random}$ and represented by the star symbol in Fig.~\ref{fig:S}: for instance, the $8\times 4$ rectangular lattice has a much higher entropy than the $4\times 8$ lattice, even though both have the same volume (number of bonds and sites) for $A$. Fig.~\ref{fig:S} shows that the low-$V$ entanglement entropies of all samples converge to the random-state values, confirming clearly that the low-$V$ phase fulfills the ETH. For larger values of $V$, the entanglement entropy drops down for all samples. The inset of Fig.~\ref{fig:S} shows a reasonable scaling of $S$ with the area ${\cal A}$  at large enough $V$, in agreement with the expected {\it area law} scaling expected in a localized phase. Finite-size effects combined with the constraint effect on $S$ discussed above prevent us from pinpointing a transition/crossover point between the volume and area law scaling with great precision.

\begin{figure}
\includegraphics[width=0.95 \columnwidth]{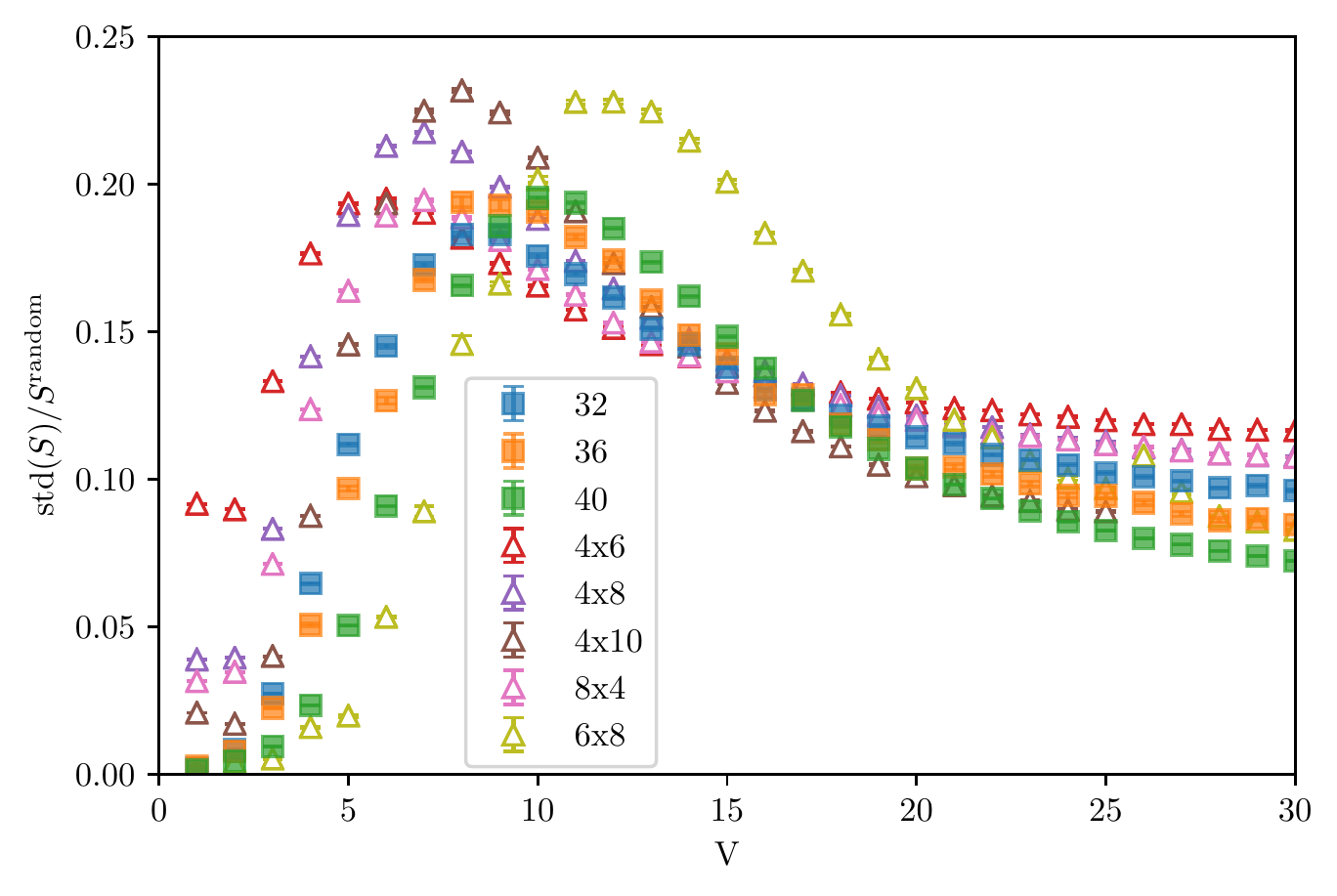}
\caption{\label{fig:Sstd} Standard deviation of entanglement entropy ${\rm std}(S)$ of subsystem $A$ as a function of disorder for different system sizes, scaled by the entanglement entropy of a random state $S^{\rm random}$. Square samples are color highlighted.}
\end{figure}

In previous studies of the behavior of entanglement entropy close to the MBL transition, the {\it variance} of entanglement entropy was shown~\cite{kjall_many-body_2014,khemani_critical_2017} to peak when approaching the transition, which can be rationalized as the result of the crossover between the volume (area) law of entanglement entropy in the ETH (MBL) phase. This is also seen in Fig.~\ref{fig:Sstd} where the standard deviation ${\rm std}(S)$ of the entanglement entropy (computed over all eigenstates and realizations of disorder), scaled by the entanglement entropy of a random state $S^{\rm random}$, is represented. Focusing on the square samples, a right-shift of the peak with system size towards larger values of $V$ is apparent: a similar shift was always found to be present in numerical studies of the MBL transition in 1d~\cite{kjall_many-body_2014,khemani_critical_2017}. For the largest square (rectangular) sample, the peak is located at $V=10$ ($V=12$), providing a lower bound for the possible transition.

\section{Machine learning analysis of entanglement spectra}
\label{sec:ml}

As a complementary approach, we study the quantum dimer model Eq.~\ref{eq:H} using machine learning techniques and follow the supervised scheme introduced in Ref.~\onlinecite{carrasquilla_machine_2017}. As inputs representative of the two phases, we provide entanglement spectra ({\it i.e.} the eigenvalues $\lambda_i$ of the reduced density matrix $\rho_A$) obtained deep in the ETH and the MBL phases and we train a neural network to classify them accordingly. The supervised learning has been used in multiple occasions in the context of delocalization-localization transitions \cite{schindler_probing_2017,li_extracting_2017,venderley_machine_2018,hsu_machine_2018,zhang_interpretable_2019,huembeli_automated_2019} producing results in good qualitative agreement with more conventional analyses. In particular, Refs.~\onlinecite{schindler_probing_2017,venderley_machine_2018,hsu_machine_2018} showed the interest of using labeled entanglement spectra.

\begin{figure}[b]
\includegraphics[width=0.95 \columnwidth]{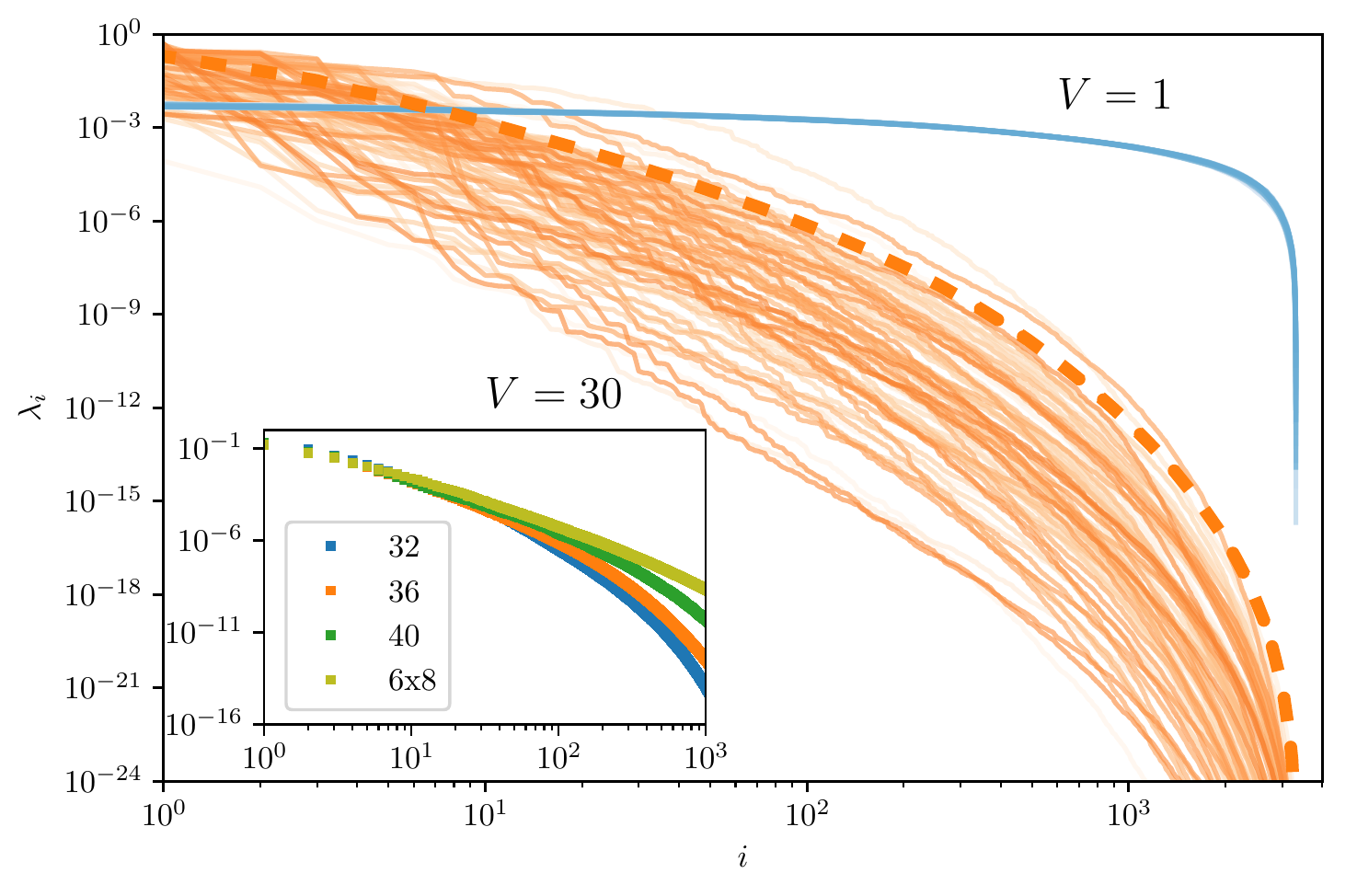}
 \caption{\label{fig:ES} Entanglement spectra (eigenvalues $\lambda_i$ of the reduced density matrix ordered by amplitude) used in the neural network approach to label the ETH ($V=1$, blue colors) and MBL phase ($V=30$, orange colors), for the $N=36$ square sample. For both cases, $~ 100$ spectra obtained from different realizations of disorder are represented. The dashed line represents the average of all spectra for $V=30$. Inset: Higher part of the average entanglement spectra for different sample sizes for $V=30$. The log-log scale highlights a power-law behavior for the larger eigenvalues $\lambda_i$.}
 \end{figure}

For this scheme, we consider the $N=32, 36, 40$ square samples and the largest rectangular sample $N=6\times 8$. For each of them, we provide 10.000 entanglement spectra (2000 for $N=40$, $6\times 8$) including between 100 and 200 disorder realizations per disorder strength at $V=1$ for the ETH-labelled phase and $V=30$ for the MBL-labelled phase. The spectra being rather large (1972 values for $N=32$, up to 21286 for $N=6\times 8$), we found that sorting them allowed for both perfect training and test accuracies for all system sizes. Fig.~\ref{fig:ES} represents the form of the entanglement spectra used in these two limits for the sample $N=36$ for $~100$ different realizations of disorder. The very similar form for various disorder samples in the ETH phase clearly contrasts with the stronger dispersion observed in the MBL phase. In the later, superposing the entanglement spectra for various samples sizes (inset of Fig.~\ref{fig:ES}) highlights that the larger values of the spectrum decay as a power-law, similar to what is found in 1D MBL~\cite{serbyn_power-law_2016}.

 We used a fully-connected neural network consisting of one hidden layer of 32 neurons followed by two softmax output neurons. We follow a cross-validation procedure where we randomly selected half of the dataset to form the training dataset, the rest being assigned to the test set. This process is repeated multiple times, generating new training and test partitions each time. This allowed us to track whether the neural networks were overfitting depending on the training conditions. Namely, we checked that data from $V=1$ and $V=30$ give perfect training and test accuracies for each system size.

\begin{figure}
\includegraphics[width=0.95 \columnwidth]{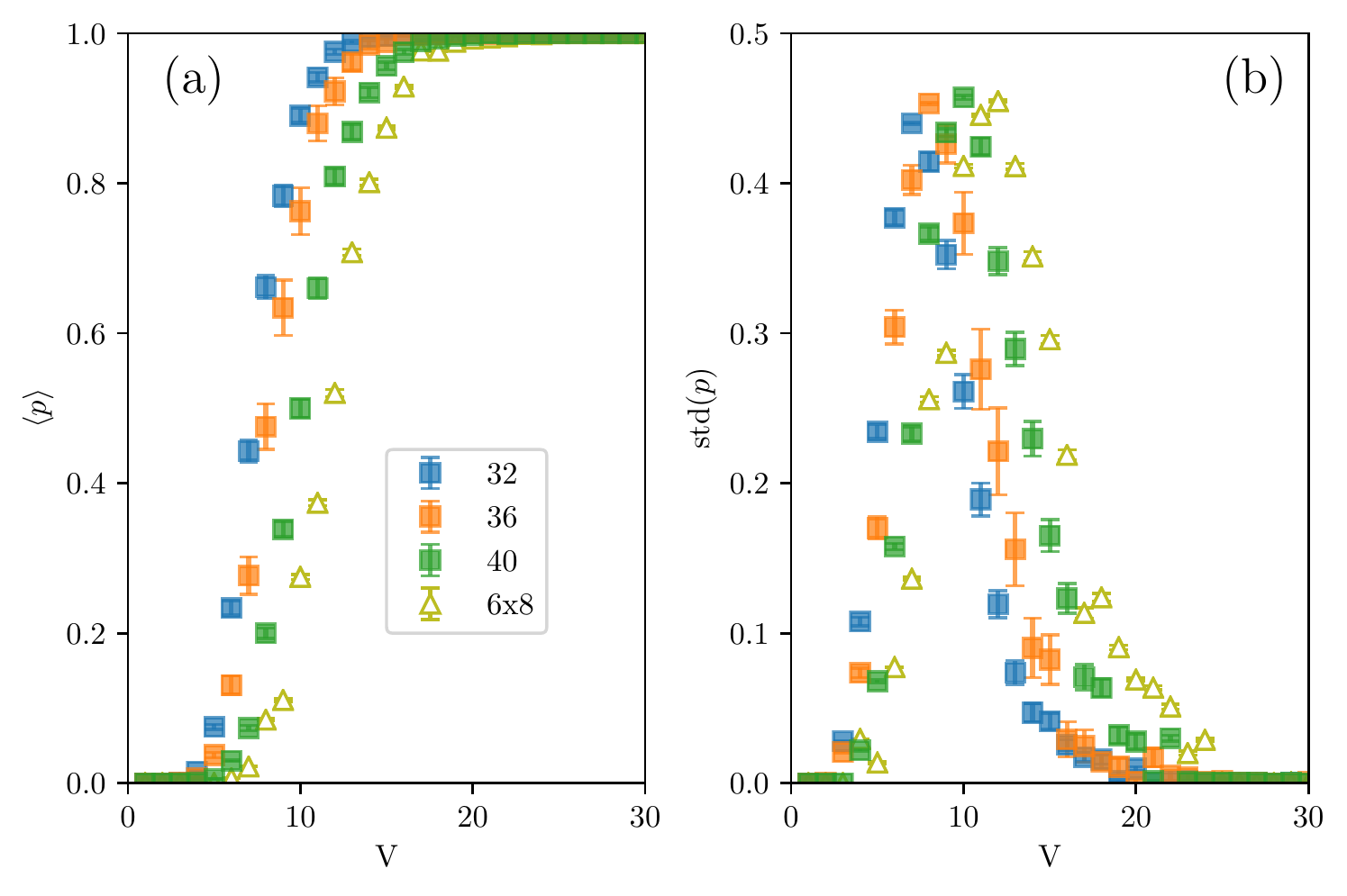}
 \caption{\label{fig:ml}As a function of disorder strength $V$ and for different system sizes: (a) mean and (b) standard deviation (right panel) of the neural network output $p$, defined such that the labels are $p=0$ ($p=1$) in the ETH (MBL) samples at $V=1$ ($V=30$). The neural network predictions involved 5000 entanglement spectra per disorder (including 100 disorder realizations per disorder). Error bars show standard deviation over 10 instances of neural networks from the cross-validation procedure, stopped after 100 epochs. Square samples are color highlighted.}
 \end{figure}

Fig.~\ref{fig:ml} displays features that are consistent with the previous analysis. The left panel displays the average output of the neural network for the different samples as a function of disorder strength. At low $V$, the machine learning analysis validates a fully-ETH phase ({\it i.e. } where all samples are classified ETH) that extends up to a value $V = V_1$, and at large $V$ a fully-MBL phase (with more than $99\%$ accuracy) at large for $V \geq V_2$  ($V_1 \simeq 6$ and $V_2 \gtrsim 20$ for the largest $N=6\times 8$ sample). Notice how these bounding values (in particular $V_1$) shift to higher values of disorder with system size. This reflects that ETH is easier to disrupt on a too small sample, in perfect agreement with the trend in all other observables discussed in Sec.~\ref{sec:eigen}. We find no crossing point with system size in the current data, different from what is observed with the 2d Ising model~\cite{carrasquilla_machine_2017}.

The right panel of Fig.~\ref{fig:ml} shows the standard deviation of the neural network output as a function of disorder. The standard deviation is low in both limits where the phases are well distinguished (at low and large $V$), and peaks at an intermediate value of disorder. The location of the peak (which shifts with system size) is the point where the neural network has most difficulties to classify phases. It can be interpreted as a finite-size estimate of a possible transition point. Notice the similarity between the standard deviation of the neural network output and the standard deviation of the entanglement entropy (Fig.~\ref{fig:Sstd}), in particular the positions of the peaks are almost the same for both quantities for the different sample sizes.

\begin{figure*}[t]
\includegraphics[width=1.5 \columnwidth]{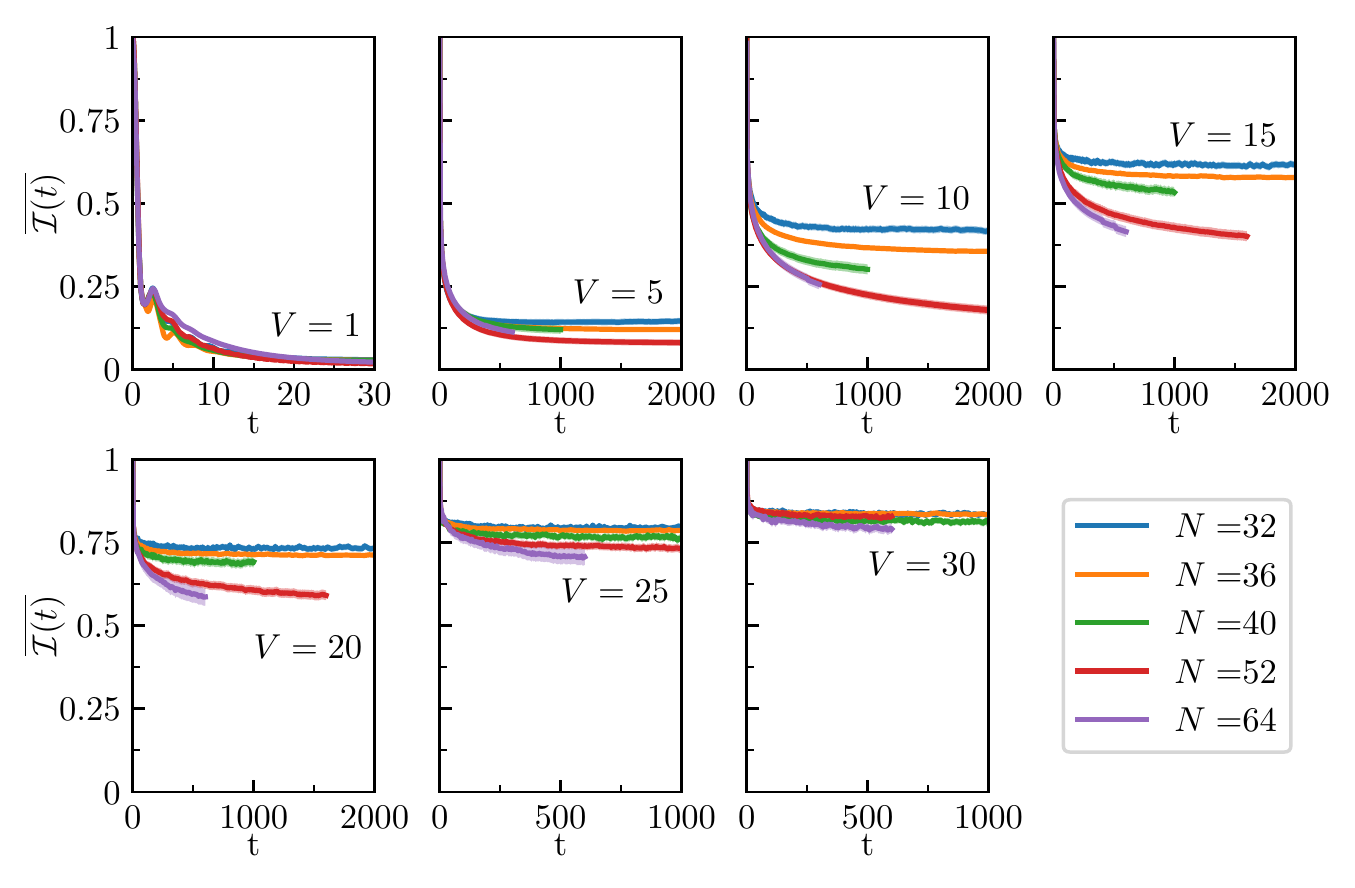}
\caption{\label{fig:imb-time} Decay with time of columnar imbalance $\overline{{\cal I}(t)}$ after a quench from the columnar configuration, for different values of disorder $V$ and different system sizes $N$ of square clusters.}
\end{figure*}

In conclusion of this section, we find that a neural network only fed with entanglement spectra is able to learn how to correctly distinguish the ETH and MBL phases for the quantum dimer model with random potential Eq.~\ref{eq:H} as well as to provide finite-size estimates of the transition point between the two. This automated method gives results in very good agreement with the analysis based on more standard, feature-enginereed, estimators of the phases presented in the previous Sec.~\ref{sec:eigen}. Our results also indicate that the machine learning analysis is also subject to finite-size effects, indicating that the use of numerical data coming from a single size may not be able to provide quantitative results for the study of phase transitions (see a discussion in Ref.~\onlinecite{theveniaut_neural_2019}). One noticeable interest of the neural network analysis (already pinpointed earlier~\cite{schindler_probing_2017,venderley_machine_2018}) is that the required amount of data and overall computational cost is considerably lower than with more traditional observables to obtain approximately similar quality of prediction: for instance good statistics on the gap ratio (Fig.~\ref{fig:gap_ratio}) requires $\simeq 100$ times more realizations of disorder than with the machine learning analysis (Fig.~\ref{fig:ml}).

\begin{figure*}
\includegraphics[width=1.485 \columnwidth]{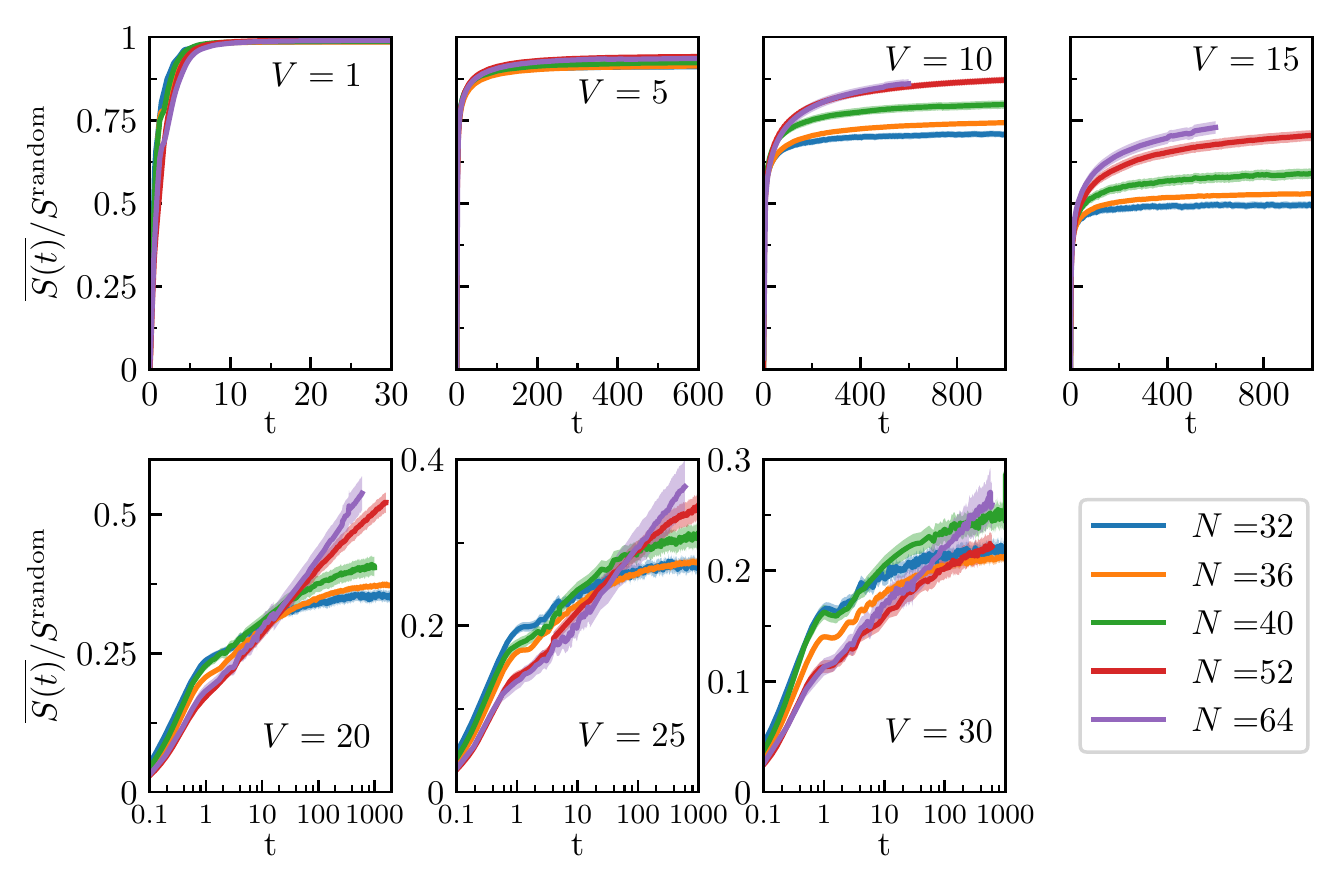}
\caption{\label{fig:Ent-time} Average entanglement entropy $\overline{S(t)}$  after a quench from the columnar configuration, scaled by the (extensive) entanglement entropy of a random state $S^{\rm random}$, for different values of disorder $V$ and different system sizes $N$ of square clusters. The top four panels are represented using the same vertical scale, and a linear scale for the time axis. The bottom three panels use a logarithmic scale for the time axis, highlighting the slow growth of entanglement for the largest values of disorder.}
\end{figure*}

\section{Dynamical properties after a quench}
\label{sec:dyn}

We consider here dynamical properties after a quench of the model Eq.~\ref{eq:H}, in order to probe for the arrested transport dynamics or slow propagation of information expected in a MBL phase. We perform quenches from an initial product state (with no entanglement) and take here the columnar configuration depicted in the inset of Fig.~\ref{fig:imb} (a product state). This columnar configuration is the most flippable state and is well connected to other dimer configurations by the action of the Hamiltonian. We thus intuitively expect this state to rapidly thermalize if it does. The choice of a columnar configuration echoes the experimental protocols for which the system is often initalized in a charge density state~\cite{schreiber_observation_2015,bordia_probing_2017}, or numerical simulations which choose a N\'eel state in spin chains, both states sharing the same properties as the columnar configuration.

\subsection{Dynamics of imbalance from the initial state}
\label{sec:imb_dynamics}

We start with the dynamics of the columnar imbalance defined in Eq.~\ref{eq:imb}. By definition, ${\cal I}(t=0)=1$, whereas ${\cal I}=0$ when averaged over all eigenstates (see Fig.~\ref{fig:imb}). In the long-time limit, ${\cal I}(t)$ is thus expected to vanish in a thermal phase, whereas a non-vanishing long-time value would signal a localized phase. Note that on finite systems, ${\cal I}(t\rightarrow \infty)$ never strictly vanishes. Fig.~\ref{fig:imb-time} displays the columnar imbalance as a function of time, with panels corresponding to different disorder strengths $V$. In the low-disorder limit, the imbalance quickly decays to a vanishing value on short time-scales (this is most readily seen at $V=1$). When disorder is increased, decays become slower and for intermediate disorder values $V=10,15,20$, the saturation value is not reached for the largest clusters within the time scale studied  (for $N=32,36$ data not shown at longer time show that the saturation is reached within less than a percent for $t=1000$ for all disorders considered). We can nevertheless observe that the imbalance clearly decreases with system size for the largest time available, in a manner compatible with a vanishing value in the thermodynamic limit for $V\lesssim 15$. For larger values of disorder, one clearly observes that the saturation value is much larger and, within available time and cluster sizes, the imbalance does not tend to vanish: this is most speaking for $V=30$ where more than $80\%$ of the initial imbalance is retained even on the largest clusters. Pinpointing exactly a transition point with dynamics is difficult due to the time scale limitations for the largest clusters, but the present data appear to be compatible with a transition/crossover for values of $V$ between $15$ and $20$, similar to what is found with the analysis of eigenstate data in Sec.~\ref{sec:eigen}.

\begin{figure*}
\includegraphics[width=1.485 \columnwidth]{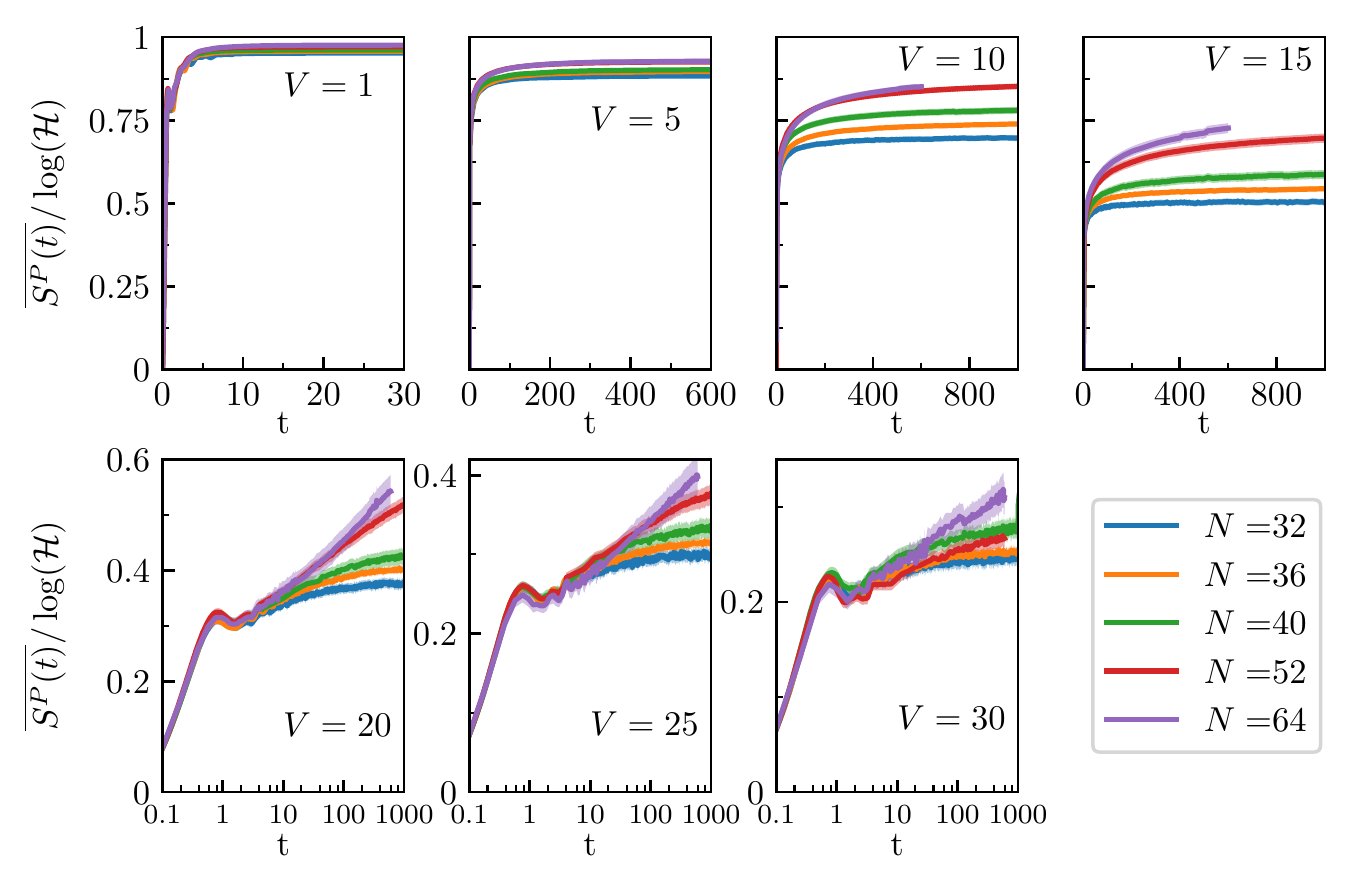}
\caption{\label{fig:Part-time} Growth of the average participation entropy $\overline{S_p(t)}$, normalized by its maximal value $\log({\cal H})$ after a quench from the columnar configuration for different values of disorder $V$ and different system sizes $N$. As in Fig.~\ref{fig:Ent-time}, the top four panels use the same vertical scale, and a linear scale for the time axis whereas the bottom three panels use a log scale for the time axis.}
\end{figure*}

\subsection{Entanglement and participation entropies dynamics}
\label{sec:ent_dynamics}

The nature of the growth of entanglement entropy $S(t)$ after a quench is a hallmark feature to distinguish MBL in one-dimension, where it takes a logarithmic form with time~\cite{znidaric_many-body_2008,bardarson_unbounded_2012},  from an ETH regime (with ballistic growth  $S(t)$~\cite{kim_ballistic_2013}) or an Anderson localized regime (where it saturates to a constant $O(1)$). In both  MBL and ETH regimes, the saturation value at long time is expected to fulfill a {\it volume} law, with a prefactor equal to (smaller than) the corresponding thermal entropy for the ETH (MBL) case~\cite{serbyn_universal_2013,nanduri_entanglement_2014}.

We represent in Fig.~\ref{fig:Ent-time} the dynamics after a quench from the columnar configuration  of the average entanglement entropy $\overline{S(t)}$ normalized by the entanglement entropy of a random state $S^\mathrm{random}$ (which is the thermal entropy up to $O(1)$ corrections) for different disorder strenghts. The ETH regime with a rapid growth of entanglement is readily identified at low disorder $V=1$, where $\overline{S(t)}$ reaches its saturation value (close to the maximum value reached by a random state) in less than $~ 10$ plaquette flips. In the top panels where the same scale is used for the vertical axis, the entanglement entropy is observed to grow slower towards its saturation value, with visible finite-size effects. Again the time and size limitations do not strictly allow to draw definitive conclusions, but it appears very likely that for large enough clusters, the entanglement entropy will reach its thermal value at long time for $V \lesssim 15$. On the bottom panels (disorders $V=20,25,30$) on the other hand, the entanglement entropy develops at an even much slower pace. The logarithmic horizontal scale chosen for the bottom panels highlights the striking feature that the growth of entanglement entropy appears to behave as $\log(t)$ for the largest clusters which are not hampered by finite-size saturation effects.  Note that for the largest data at $V=15$, a logarithmic growth is also compatible with the data (not shown). Naively exporting the argument (based on the exponentially-decaying interactions between local integral of motions) that explains the logarithmic growth of entanglement in 1d MBL, we would also expect a logarithmic growth in 2d for the entanglement cut geometry chosen here, as entanglement can grow only in one direction.

We finally present the evolution after a quench of the participation entropy $\overline{S_p(t)}$ in Fig.~\ref{fig:Part-time}. The dynamics of the participation entropy is almost identical to the one of the entanglement entropy discussed above: at small disorder, rapid saturation to the maximum attainable value $\log({\cal H})$, slower growth at intermediate values and logarithmic growth for the largest samples for strong disorder (note again the logarithmic scale on the bottom panels of Fig.~\ref{fig:Part-time}). The only noticeable difference in the dynamics of both entropies comes from different finite-size effects at short time, with participation data showing an improved nesting of curves as system size is increased which is not observed for entanglement. From participation entropy, we conclude again on slow dynamics starting from $V \gtrsim 15-20$ consistent with the existence of a localized phase.

\section{Discussion and perspectives}
\label{sec:conc}

We presented a comprehensive and extensive large-scale exact numerical study of the localization properties of a two-dimensional constrained quantum many-body system with disorder. All data, both for eigenstate or dynamical properties, can be interpreted as consistent with the existence of two distinct phases: an ETH phase at low disorder and a many-body localized phase at strong disorder, which are separated by a transition located around disorder strength $V \simeq 15-20$.

Our evidence for a MBL transition in the 2d quantum dimer model comes from numerical simulations on finite lattices. Of course, finite-size simulations can always be argued to artificially detect a MBL phase even when only a ETH phase occurs in the thermodynamic limit. We would like to emphasize that the level of numerical evidence for a MBL transition in the 2d quantum dimer model is in our opinion and experience similar to the one obtained for the standard model of MBL in one dimension~\cite{pal_many-body_2010,luitz_many-body_2015}, with similar or larger Hilbert space sizes and time scales probed in the current work.
This is of course not a definite proof that a MBL transition occurs in the thermodynamic limit.

What is perhaps more important with respect to potential experiments is the fact that even if the behavior at large values of disorder (say $V\geq 25$) is ultimately ergodic, the time scales and / or system sizes needed to probe ergodicity would be extremely large. An experimental realization of Eq.~\ref{eq:H} or of a similar constrained model in 2d would see localization for all practical purposes on the time scales available in the lab. We deliberately avoided to attempt to perform a data collapse analysis to extract {\it e.g.} critical exponents due to the two following limiting points: the moderate {\it linear} system sizes that we can reach as well as the non-uniform aspect ratio between available samples would likely provide biased estimates for critical exponents. For one-dimensional MBL, the larger linear length scales that can be reached have been argued (see e.g. Ref.~\onlinecite{khemani_two_2017}) not to be large enough to provide correct estimates of asymptotic critical behavior. The situation is likely to be the same here. With these numerical limitations in mind, we can nevertheless observe that critical values of the gap ratio or of the Kullback-Leibler divergence of eigenstates are closer to their Poisson than their ETH limits, indicating that the putative transition point is even less ergodic than for the one-dimensional MBL transition in the random-field Heisenberg spin chain~\cite{luitz_many-body_2015}.

Going back to the {\it l-}bits arguments which predict the asymptotic unstability of MBL in two dimensions~\cite{de_roeck_stability_2017,de_roeck_many-body_2017,potirniche_exploration_2019}, we remark that we do not know how to simply construct {\it l}-bits for the quantum dimer model. The crucial point is that there is no tensor product structure for the Hilbert space of quantum dimer models, thus constructing a basis of commuting operators with local support appears ill-defined. As the bubble argument is explicitly based on {\it l-}bits, it cannot strictly speaking apply here.

Regarding experiments, the relative strong disorder needed to induce the MBL phase does not provide immediate relevance of our results in frustrated magnets, where QDMs naturally emerge. However, the artificial realizations of quantum dimer models (and other related 2d constrained models) have been extensively studied in cold-atomic and mesoscopic systems. Precise experimental protocols have been proposed using Ryberg atoms~\cite{glaetzle_quantum_2014,celi_emerging_2020}, trapped ions~\cite{nath_hexagonal_2015}, large-spin ultracold atoms~\cite{sundar_quantum_2019}, Josephson Junction arrays~\cite{ioffe_topologically_2002} or superconducting quantum circuits~\cite{marcos_two-dimensional_2014}. Further explorations are needed to understand which experimental platform is the most suited to implement random potential or interactions. Constrained models in 1d have already been implemented (with no randomness) exploiting Rydberg blockade in atomic chains~\cite{bernien_probing_2017}, already exhibiting rich unusual quantum dynamics.

There are several perspectives opened up by our work. First, the roadmap to two-dimensional MBL can be exploited using the QDM on other two-dimensional lattices, allowing to test for universality and to search for other features.  A recent investigation~\cite{pietracaprina_probing_2020} considered the honeycomb lattice, which has an effective smaller local Hilbert space allowing to reach larger samples, and obtained results similar to ours. The QDM on the kagome lattice is also an interesting case worth pursuing at it possesses conserved $Z_2$ quantum numbers, allowing the exciting possibility of 2d topological order in MBL states~\cite{moessner_resonating_2001}. The possibility of MBL in other quantum constrained models such as quantum ice or loop models also provide an interesting follow-up.

The large-scale numerical techniques that we use in the present work would be useful to investigate other aspects of the model Eq.~\ref{eq:H} not addressed here, in particular to make connection with 1d models where MBL is best understood. The current investigation focused on the 'infinite temperature limit', that is excited states in the middle of the spectrum ($\epsilon=0.5$). It would be interesting to construct a full energy-disorder phase diagram to see if a many-body mobility edge exists as in 1d spin chain models~\cite{luitz_many-body_2015}. Also the detailed dynamics occurring in the ETH phase remains to be investigated. The relatively slow dynamics exhibited around $V \sim 10,15$ in Figures~\ref{fig:imb-time}, \ref{fig:Ent-time}, \ref{fig:Part-time} could be signs of a precursor subdiffusive regime, similar to what is observed in 1d systems~\cite{agarwal_anomalous_2015,luitz_extended_2016,znidaric_diffusive_2016}. Finally, it would be intriguing to see if the machine learning techniques used in Sec.~\ref{sec:ml} would be able to distinguish 1d from 2d MBL, and if not, to use neural networks trained on 1d spin chain models to probe 2d MBL (transfer learning).

On the analytical side, the absence of an obvious scheme for constructing {\it l-}bits points towards the interest of building an alternative effective description for the many-body localized phase in such 2d constrained models. Recall as well that there is no free-particle limit to start with in the model Eq.~\ref{eq:H}. The height description~\cite{blote_roughening_1982} of dimer coverings (and of constrained models in general) appears as a good candidate to develop such a framework. Finally from a purely classical perspective, we note that the strong disorder limit $V \rightarrow \infty$ constitutes an interesting classical statistical mechanics problem (with for instance non-trivial degeneracies), unsolved to the best of our knowledge and distinct from other interacting dimer models~\cite{alet_interacting_2005}.

\begin{acknowledgments}
We gratefully thank Juan P. Garrahan and Stephen Powell for useful discussions at an initial stage of this project. We also thank Nicolas Laflorencie for useful comments on the manuscript and Nicolas Mac\'e for an insightful remark on the scaling of entanglement growth.
H.T. is supported by a grant from the Fondation CFM pour la Recherche. Z.L. acknowledges support from EPSRC Grant No. EP/M019691/1.
This work benefited from the support of the project THERMOLOC ANR-16-CE30-0023-02 of the French National Research Agency (ANR) and by the French Programme Investissements d'Avenir under the program ANR-11-IDEX-0002-02, reference ANR-10-LABX-0037-NEXT. We acknowledge PRACE for awarding access to HLRS's Hazel Hen computer based in Stuttgart, Germany under grant number 2016153659, as well as the use of HPC resources from CALMIP (grants 2017-P0677 and 2018-P0677) and GENCI (Grant 2018-A0030500225). The shift-invert~\cite{pietracaprina_shift-invert_2018} and Krylov time evolution numerical calculations are based on the linear algebra libraries PETSc~\cite{petsc-efficient,petsc-user-ref}, SLEPc~\cite{slepc-toms}, MUMPS~\cite{amestoy_fully_2001,amestoy_hybrid_2006} and STRUMPACK~\cite{ghysels_efficient_2016,ghysels_robust_2017}. The neural network analysis used the library TensorFlow~\cite{tensorflow2015-whitepaper}.
\end{acknowledgments}

\appendix*
\label{sec:appendix}
\section{Lattices}

\subsection{Lattice geometries}

\begin{figure}
\includegraphics[width=\columnwidth]{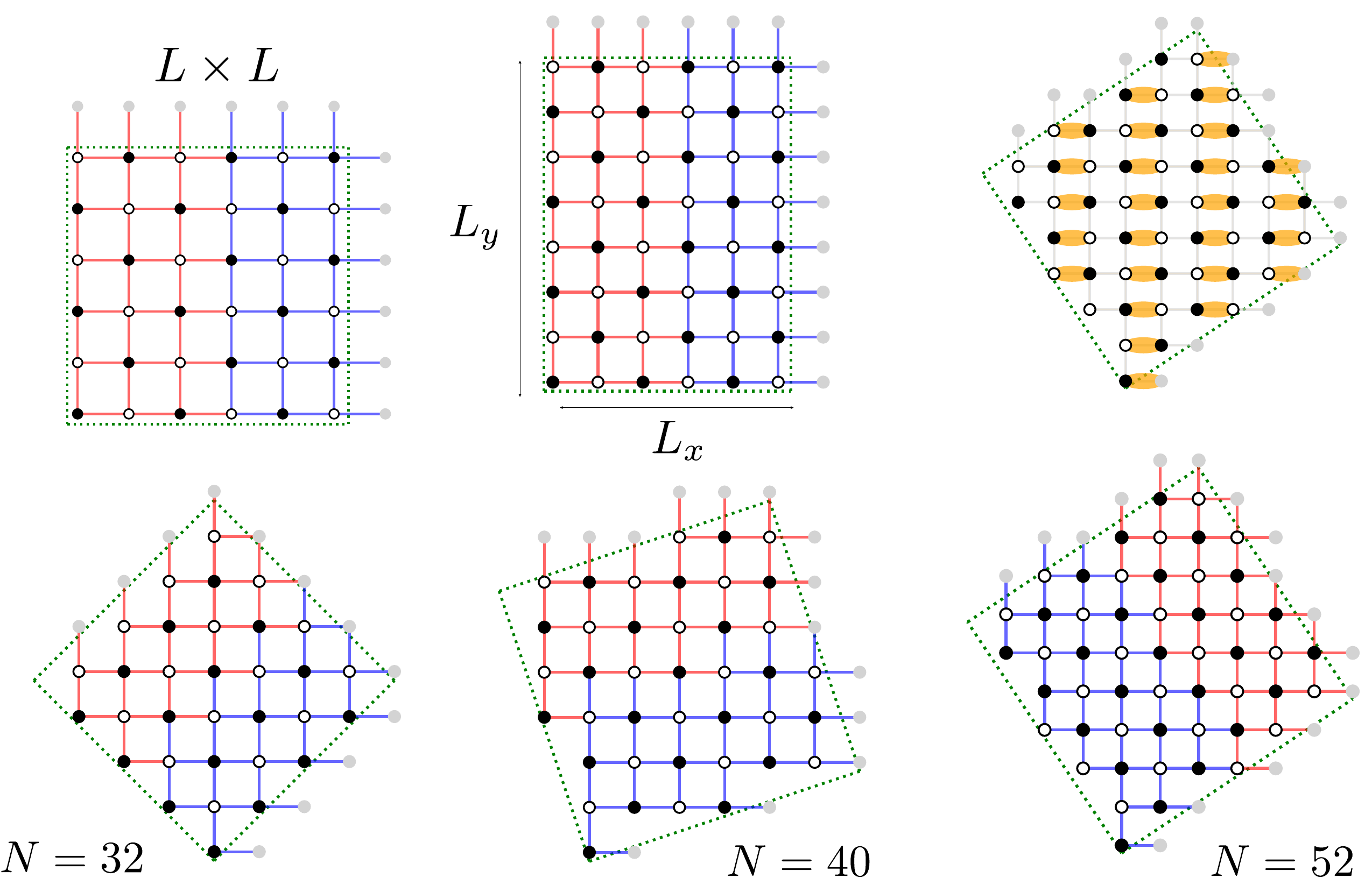}%
\caption{\label{fig:Lattices} Lattices used in this work: regular square lattices with $N=L^2$ sites (top left panel with $L=6$), rectangular lattices with $N=L_x \times L_y$ (top middle panel with $L_x=6$ and $L_y=8$), and tilted square lattices (bottom row with $N=32, 40, 52$ sites). The red / blue bond decomposition is used for the entanglement partition.  Top right: columnar state for the tilted $N=52$ square lattice.}
\end{figure}

We use regular and tilted square lattices as well as rectangular lattices with $N$ sites and $2N$ bonds.
Regular square lattices have $N=L^2$ sites, and we used samples with $L=4,6,8$. Rectangular lattices, which have $N=L_x \times L_y$ sites (with $L_{x,y}=4,6,8,10$), have a non-unity aspect ratio, which can affect the analysis of finite-size effects. Finally, tilted square lattices host $N=p^2+q^2$ sites, and we use the combination $p=4,q=4$ (with $N=32$ sites), $p=6,q=2$ ($N=40$), and $p=6,q=4$ ($N=52$). Tilted square lattices have a unity aspect ratio, but they do not have all symmetries of the full regular square: this is however irrelevant as the Hamiltonian Eq.~\ref{eq:H} does not allow any lattice symmetries due to disorder. All these lattices, which are represented in Fig.~\ref{fig:Lattices}, have in common that they host a $(W_x,W_y)=(0,0)$ winding sector: this is true when $L,L_x,L_y,p$ and $q$ are even. One can also easily define columnar state on these samples (see example for $N=52$ in Fig.~\ref{fig:Lattices}).

\subsection{Entanglement cut}

When computing bipartite entanglement properties, there is some freedom in choosing where is located the entanglement cut, {\it i.e.} which degrees of freedom belong to $A$ or $B$ in Eq.~\ref{eq:EE}. We choose the cuts displayed in Fig.~\ref{fig:Lattices}, where all lattices are divided in two parts with equal number of bonds. With this cut, the area ${\cal A}$ between $A$ and $B$ is expected to scale with $L$ for regular square lattices, $L_y$ for rectangular lattices, and $\sqrt{p^2+q^2}$ for tilted square lattices.

This choice of cuts for the regular square, rectangular and $N=32$ tilted square lattice also induces the following property: one of the two windings numbers (the one associated to a line parallel to the cut, {\it e.g.} $W_y$ for the rectangular lattices) is conserved by the cut, but not the other one. The reduced density matrix on $A$ is block-diagonal with blocks labelled by winding numbers in $A$ ($W_y^A$ which is forced to be $0$, and $W_x^A$), which we take advantage of in numerical computations. The existence of blocks in the reduced density matrices and its effect on entanglement has been discussed in other models with conservation laws~\cite{vidmar_entanglement_2017,garrison_does_2018,zhou_operator_2017,morampudi_universal_2020}.

We note also a supplementary specificity of constrained models for entanglement properties (see discussion in Ref.~\onlinecite{morampudi_universal_2020}). Due to the constraint and even within the conserved-number blocks (e.g. configurations of $A$ with $W_x^A$ and configurations of B with $W_x^B$ such that $W_x^A+W_x^B=W_x=0$), not all configurations of $A$ are compatible with those of $B$. This property has an important consequence, for instance when computing volume-law entanglement, as different samples with the same volume can have different constraints on configurations in $A$ and $B$ when bipartitioned. This is illustrated in the main text with the example of the $8 \times 4$ and $4 \times 8$ rectangular samples. This property can also cause difficulties in comparing scaling of entanglement with system size for different samples.

\bibliography{Dimers_MBL}

\begin{thebibliography}{137}%
\makeatletter
\providecommand \@ifxundefined [1]{%
 \@ifx{#1\undefined}
}%
\providecommand \@ifnum [1]{%
 \ifnum #1\expandafter \@firstoftwo
 \else \expandafter \@secondoftwo
 \fi
}%
\providecommand \@ifx [1]{%
 \ifx #1\expandafter \@firstoftwo
 \else \expandafter \@secondoftwo
 \fi
}%
\providecommand \natexlab [1]{#1}%
\providecommand \enquote  [1]{``#1''}%
\providecommand \bibnamefont  [1]{#1}%
\providecommand \bibfnamefont [1]{#1}%
\providecommand \citenamefont [1]{#1}%
\providecommand \href@noop [0]{\@secondoftwo}%
\providecommand \href [0]{\begingroup \@sanitize@url \@href}%
\providecommand \@href[1]{\@@startlink{#1}\@@href}%
\providecommand \@@href[1]{\endgroup#1\@@endlink}%
\providecommand \@sanitize@url [0]{\catcode `\\12\catcode `\$12\catcode
  `\&12\catcode `\#12\catcode `\^12\catcode `\_12\catcode `\%12\relax}%
\providecommand \@@startlink[1]{}%
\providecommand \@@endlink[0]{}%
\providecommand \url  [0]{\begingroup\@sanitize@url \@url }%
\providecommand \@url [1]{\endgroup\@href {#1}{\urlprefix }}%
\providecommand \urlprefix  [0]{URL }%
\providecommand \Eprint [0]{\href }%
\providecommand \doibase [0]{http://dx.doi.org/}%
\providecommand \selectlanguage [0]{\@gobble}%
\providecommand \bibinfo  [0]{\@secondoftwo}%
\providecommand \bibfield  [0]{\@secondoftwo}%
\providecommand \translation [1]{[#1]}%
\providecommand \BibitemOpen [0]{}%
\providecommand \bibitemStop [0]{}%
\providecommand \bibitemNoStop [0]{.\EOS\space}%
\providecommand \EOS [0]{\spacefactor3000\relax}%
\providecommand \BibitemShut  [1]{\csname bibitem#1\endcsname}%
\let\auto@bib@innerbib\@empty
\bibitem [{\citenamefont {Neumann}(1929)}]{neumann_beweis_1929}%
  \BibitemOpen
  \bibfield  {author} {\bibinfo {author} {\bibfnamefont {J.~v.}\ \bibnamefont
  {Neumann}},\ }\href {\doibase 10.1007/BF01339852} {\bibfield  {journal}
  {\bibinfo  {journal} {Z. Phys.}\ }\textbf {\bibinfo {volume} {57}},\ \bibinfo
  {pages} {30} (\bibinfo {year} {1929})}\BibitemShut {NoStop}%
\bibitem [{\citenamefont {Goldstein}\ \emph {et~al.}(2010)\citenamefont
  {Goldstein}, \citenamefont {Lebowitz}, \citenamefont {Tumulka},\ and\
  \citenamefont {Zanghì}}]{goldstein_long-time_2010}%
  \BibitemOpen
  \bibfield  {author} {\bibinfo {author} {\bibfnamefont {S.}~\bibnamefont
  {Goldstein}}, \bibinfo {author} {\bibfnamefont {J.~L.}\ \bibnamefont
  {Lebowitz}}, \bibinfo {author} {\bibfnamefont {R.}~\bibnamefont {Tumulka}}, \
  and\ \bibinfo {author} {\bibfnamefont {N.}~\bibnamefont {Zanghì}},\ }\href
  {\doibase 10.1140/epjh/e2010-00007-7} {\bibfield  {journal} {\bibinfo
  {journal} {Eur. Phys. J. H}\ }\textbf {\bibinfo {volume} {35}},\ \bibinfo
  {pages} {173} (\bibinfo {year} {2010})}\BibitemShut {NoStop}%
\bibitem [{\citenamefont {Gogolin}\ and\ \citenamefont
  {Eisert}(2016)}]{gogolin_equilibration_2016}%
  \BibitemOpen
  \bibfield  {author} {\bibinfo {author} {\bibfnamefont {C.}~\bibnamefont
  {Gogolin}}\ and\ \bibinfo {author} {\bibfnamefont {J.}~\bibnamefont
  {Eisert}},\ }\href {\doibase 10.1088/0034-4885/79/5/056001} {\bibfield
  {journal} {\bibinfo  {journal} {Rep. Prog. Phys.}\ }\textbf {\bibinfo
  {volume} {79}},\ \bibinfo {pages} {056001} (\bibinfo {year}
  {2016})}\BibitemShut {NoStop}%
\bibitem [{\citenamefont {Bloch}\ \emph {et~al.}(2008)\citenamefont {Bloch},
  \citenamefont {Dalibard},\ and\ \citenamefont
  {Zwerger}}]{bloch_many-body_2008}%
  \BibitemOpen
  \bibfield  {author} {\bibinfo {author} {\bibfnamefont {I.}~\bibnamefont
  {Bloch}}, \bibinfo {author} {\bibfnamefont {J.}~\bibnamefont {Dalibard}}, \
  and\ \bibinfo {author} {\bibfnamefont {W.}~\bibnamefont {Zwerger}},\ }\href
  {\doibase 10.1103/RevModPhys.80.885} {\bibfield  {journal} {\bibinfo
  {journal} {Rev. Mod. Phys.}\ }\textbf {\bibinfo {volume} {80}},\ \bibinfo
  {pages} {885} (\bibinfo {year} {2008})}\BibitemShut {NoStop}%
\bibitem [{\citenamefont {Gross}\ and\ \citenamefont
  {Bloch}(2017)}]{gross_quantum_2017}%
  \BibitemOpen
  \bibfield  {author} {\bibinfo {author} {\bibfnamefont {C.}~\bibnamefont
  {Gross}}\ and\ \bibinfo {author} {\bibfnamefont {I.}~\bibnamefont {Bloch}},\
  }\href {\doibase 10.1126/science.aal3837} {\bibfield  {journal} {\bibinfo
  {journal} {Science}\ }\textbf {\bibinfo {volume} {357}},\ \bibinfo {pages}
  {995} (\bibinfo {year} {2017})}\BibitemShut {NoStop}%
\bibitem [{\citenamefont {Deutsch}(1991)}]{deutsch_quantum_1991}%
  \BibitemOpen
  \bibfield  {author} {\bibinfo {author} {\bibfnamefont {J.~M.}\ \bibnamefont
  {Deutsch}},\ }\href {\doibase 10.1103/PhysRevA.43.2046} {\bibfield  {journal}
  {\bibinfo  {journal} {Phys. Rev. A}\ }\textbf {\bibinfo {volume} {43}},\
  \bibinfo {pages} {2046} (\bibinfo {year} {1991})}\BibitemShut {NoStop}%
\bibitem [{\citenamefont {Srednicki}(1994)}]{srednicki_chaos_1994}%
  \BibitemOpen
  \bibfield  {author} {\bibinfo {author} {\bibfnamefont {M.}~\bibnamefont
  {Srednicki}},\ }\href {\doibase 10.1103/PhysRevE.50.888} {\bibfield
  {journal} {\bibinfo  {journal} {Phys. Rev. E}\ }\textbf {\bibinfo {volume}
  {50}},\ \bibinfo {pages} {888} (\bibinfo {year} {1994})}\BibitemShut
  {NoStop}%
\bibitem [{\citenamefont {Rigol}\ \emph {et~al.}(2008)\citenamefont {Rigol},
  \citenamefont {Dunjko},\ and\ \citenamefont
  {Olshanii}}]{rigol_thermalization_2008}%
  \BibitemOpen
  \bibfield  {author} {\bibinfo {author} {\bibfnamefont {M.}~\bibnamefont
  {Rigol}}, \bibinfo {author} {\bibfnamefont {V.}~\bibnamefont {Dunjko}}, \
  and\ \bibinfo {author} {\bibfnamefont {M.}~\bibnamefont {Olshanii}},\ }\href
  {\doibase 10.1038/nature06838} {\bibfield  {journal} {\bibinfo  {journal}
  {Nature}\ }\textbf {\bibinfo {volume} {452}},\ \bibinfo {pages} {854}
  (\bibinfo {year} {2008})}\BibitemShut {NoStop}%
\bibitem [{\citenamefont {Deutsch}(2018)}]{deutsch_eigenstate_2018}%
  \BibitemOpen
  \bibfield  {author} {\bibinfo {author} {\bibfnamefont {J.~M.}\ \bibnamefont
  {Deutsch}},\ }\href {\doibase 10.1088/1361-6633/aac9f1} {\bibfield  {journal}
  {\bibinfo  {journal} {Rep. Prog. Phys.}\ }\textbf {\bibinfo {volume} {81}},\
  \bibinfo {pages} {082001} (\bibinfo {year} {2018})}\BibitemShut {NoStop}%
\bibitem [{\citenamefont {Basko}\ \emph {et~al.}(2006)\citenamefont {Basko},
  \citenamefont {Aleiner},\ and\ \citenamefont
  {Altshuler}}]{basko_metalinsulator_2006}%
  \BibitemOpen
  \bibfield  {author} {\bibinfo {author} {\bibfnamefont {D.~M.}\ \bibnamefont
  {Basko}}, \bibinfo {author} {\bibfnamefont {I.~L.}\ \bibnamefont {Aleiner}},
  \ and\ \bibinfo {author} {\bibfnamefont {B.~L.}\ \bibnamefont {Altshuler}},\
  }\href {\doibase 10.1016/j.aop.2005.11.014} {\bibfield  {journal} {\bibinfo
  {journal} {Ann. Phys. (N.Y.)}\ }\textbf {\bibinfo {volume} {321}},\ \bibinfo
  {pages} {1126} (\bibinfo {year} {2006})}\BibitemShut {NoStop}%
\bibitem [{\citenamefont {Gornyi}\ \emph {et~al.}(2005)\citenamefont {Gornyi},
  \citenamefont {Mirlin},\ and\ \citenamefont
  {Polyakov}}]{gornyi_interacting_2005}%
  \BibitemOpen
  \bibfield  {author} {\bibinfo {author} {\bibfnamefont {I.~V.}\ \bibnamefont
  {Gornyi}}, \bibinfo {author} {\bibfnamefont {A.~D.}\ \bibnamefont {Mirlin}},
  \ and\ \bibinfo {author} {\bibfnamefont {D.~G.}\ \bibnamefont {Polyakov}},\
  }\href {\doibase 10.1103/PhysRevLett.95.206603} {\bibfield  {journal}
  {\bibinfo  {journal} {Phys. Rev. Lett.}\ }\textbf {\bibinfo {volume} {95}},\
  \bibinfo {pages} {206603} (\bibinfo {year} {2005})}\BibitemShut {NoStop}%
\bibitem [{\citenamefont {Bauer}\ and\ \citenamefont
  {Nayak}(2013)}]{bauer_area_2013}%
  \BibitemOpen
  \bibfield  {author} {\bibinfo {author} {\bibfnamefont {B.}~\bibnamefont
  {Bauer}}\ and\ \bibinfo {author} {\bibfnamefont {C.}~\bibnamefont {Nayak}},\
  }\href {\doibase 10.1088/1742-5468/2013/09/P09005} {\bibfield  {journal}
  {\bibinfo  {journal} {J. Stat. Mech.}\ }\textbf {\bibinfo {volume} {2013}},\
  \bibinfo {pages} {P09005} (\bibinfo {year} {2013})}\BibitemShut {NoStop}%
\bibitem [{\citenamefont {Huse}\ \emph {et~al.}(2013)\citenamefont {Huse},
  \citenamefont {Nandkishore}, \citenamefont {Oganesyan}, \citenamefont {Pal},\
  and\ \citenamefont {Sondhi}}]{huse_localization-protected_2013}%
  \BibitemOpen
  \bibfield  {author} {\bibinfo {author} {\bibfnamefont {D.~A.}\ \bibnamefont
  {Huse}}, \bibinfo {author} {\bibfnamefont {R.}~\bibnamefont {Nandkishore}},
  \bibinfo {author} {\bibfnamefont {V.}~\bibnamefont {Oganesyan}}, \bibinfo
  {author} {\bibfnamefont {A.}~\bibnamefont {Pal}}, \ and\ \bibinfo {author}
  {\bibfnamefont {S.~L.}\ \bibnamefont {Sondhi}},\ }\href {\doibase
  10.1103/PhysRevB.88.014206} {\bibfield  {journal} {\bibinfo  {journal} {Phys.
  Rev. B}\ }\textbf {\bibinfo {volume} {88}},\ \bibinfo {pages} {014206}
  (\bibinfo {year} {2013})}\BibitemShut {NoStop}%
\bibitem [{\citenamefont {Bahri}\ \emph {et~al.}(2015)\citenamefont {Bahri},
  \citenamefont {Vosk}, \citenamefont {Altman},\ and\ \citenamefont
  {Vishwanath}}]{bahri_localization_2015}%
  \BibitemOpen
  \bibfield  {author} {\bibinfo {author} {\bibfnamefont {Y.}~\bibnamefont
  {Bahri}}, \bibinfo {author} {\bibfnamefont {R.}~\bibnamefont {Vosk}},
  \bibinfo {author} {\bibfnamefont {E.}~\bibnamefont {Altman}}, \ and\ \bibinfo
  {author} {\bibfnamefont {A.}~\bibnamefont {Vishwanath}},\ }\href {\doibase
  10.1038/ncomms8341} {\bibfield  {journal} {\bibinfo  {journal} {Nat. Comm.}\
  }\textbf {\bibinfo {volume} {6}},\ \bibinfo {pages} {7341} (\bibinfo {year}
  {2015})}\BibitemShut {NoStop}%
\bibitem [{\citenamefont {Pekker}\ \emph {et~al.}(2014)\citenamefont {Pekker},
  \citenamefont {Refael}, \citenamefont {Altman}, \citenamefont {Demler},\ and\
  \citenamefont {Oganesyan}}]{pekker_hilbert-glass_2014}%
  \BibitemOpen
  \bibfield  {author} {\bibinfo {author} {\bibfnamefont {D.}~\bibnamefont
  {Pekker}}, \bibinfo {author} {\bibfnamefont {G.}~\bibnamefont {Refael}},
  \bibinfo {author} {\bibfnamefont {E.}~\bibnamefont {Altman}}, \bibinfo
  {author} {\bibfnamefont {E.}~\bibnamefont {Demler}}, \ and\ \bibinfo {author}
  {\bibfnamefont {V.}~\bibnamefont {Oganesyan}},\ }\href
  {https://link.aps.org/doi/10.1103/PhysRevX.4.011052} {\bibfield  {journal}
  {\bibinfo  {journal} {Phys. Rev. X}\ }\textbf {\bibinfo {volume} {4}},\
  \bibinfo {pages} {011052} (\bibinfo {year} {2014})}\BibitemShut {NoStop}%
\bibitem [{\citenamefont {Kj\"all}\ \emph {et~al.}(2014)\citenamefont
  {Kj\"all}, \citenamefont {Bardarson},\ and\ \citenamefont
  {Pollmann}}]{kjall_many-body_2014}%
  \BibitemOpen
  \bibfield  {author} {\bibinfo {author} {\bibfnamefont {J.~A.}\ \bibnamefont
  {Kj\"all}}, \bibinfo {author} {\bibfnamefont {J.~H.}\ \bibnamefont
  {Bardarson}}, \ and\ \bibinfo {author} {\bibfnamefont {F.}~\bibnamefont
  {Pollmann}},\ }\href
  {https://link.aps.org/doi/10.1103/PhysRevLett.113.107204} {\bibfield
  {journal} {\bibinfo  {journal} {Phys. Rev. Lett.}\ }\textbf {\bibinfo
  {volume} {113}},\ \bibinfo {pages} {107204} (\bibinfo {year}
  {2014})}\BibitemShut {NoStop}%
\bibitem [{\citenamefont {Chandran}\ \emph {et~al.}(2014)\citenamefont
  {Chandran}, \citenamefont {Khemani}, \citenamefont {Laumann},\ and\
  \citenamefont {Sondhi}}]{chandran_many-body_2014}%
  \BibitemOpen
  \bibfield  {author} {\bibinfo {author} {\bibfnamefont {A.}~\bibnamefont
  {Chandran}}, \bibinfo {author} {\bibfnamefont {V.}~\bibnamefont {Khemani}},
  \bibinfo {author} {\bibfnamefont {C.~R.}\ \bibnamefont {Laumann}}, \ and\
  \bibinfo {author} {\bibfnamefont {S.~L.}\ \bibnamefont {Sondhi}},\ }\href
  {https://journals.aps.org/prb/abstract/10.1103/PhysRevB.89.144201} {\bibfield
   {journal} {\bibinfo  {journal} {Phys. Rev. B}\ }\textbf {\bibinfo {volume}
  {89}},\ \bibinfo {pages} {144201} (\bibinfo {year} {2014})}\BibitemShut
  {NoStop}%
\bibitem [{\citenamefont {{\v Z}nidari{\v c}}\ \emph
  {et~al.}(2008)\citenamefont {{\v Z}nidari{\v c}}, \citenamefont {Prosen},\
  and\ \citenamefont {Prelov{\v s}ek}}]{znidaric_many-body_2008}%
  \BibitemOpen
  \bibfield  {author} {\bibinfo {author} {\bibfnamefont {M.}~\bibnamefont {{\v
  Z}nidari{\v c}}}, \bibinfo {author} {\bibfnamefont {T.}~\bibnamefont
  {Prosen}}, \ and\ \bibinfo {author} {\bibfnamefont {P.}~\bibnamefont
  {Prelov{\v s}ek}},\ }\href
  {http://link.aps.org/doi/10.1103/PhysRevB.77.064426} {\bibfield  {journal}
  {\bibinfo  {journal} {Phys. Rev. B}\ }\textbf {\bibinfo {volume} {77}},\
  \bibinfo {pages} {064426} (\bibinfo {year} {2008})}\BibitemShut {NoStop}%
\bibitem [{\citenamefont {Bardarson}\ \emph {et~al.}(2012)\citenamefont
  {Bardarson}, \citenamefont {Pollmann},\ and\ \citenamefont
  {Moore}}]{bardarson_unbounded_2012}%
  \BibitemOpen
  \bibfield  {author} {\bibinfo {author} {\bibfnamefont {J.~H.}\ \bibnamefont
  {Bardarson}}, \bibinfo {author} {\bibfnamefont {F.}~\bibnamefont {Pollmann}},
  \ and\ \bibinfo {author} {\bibfnamefont {J.~E.}\ \bibnamefont {Moore}},\
  }\href {\doibase 10.1103/PhysRevLett.109.017202} {\bibfield  {journal}
  {\bibinfo  {journal} {Phys. Rev. Lett.}\ }\textbf {\bibinfo {volume} {109}},\
  \bibinfo {pages} {017202} (\bibinfo {year} {2012})}\BibitemShut {NoStop}%
\bibitem [{\citenamefont {Anderson}(1958)}]{anderson_absence_1958}%
  \BibitemOpen
  \bibfield  {author} {\bibinfo {author} {\bibfnamefont {P.~W.}\ \bibnamefont
  {Anderson}},\ }\href {\doibase 10.1103/PhysRev.109.1492} {\bibfield
  {journal} {\bibinfo  {journal} {Phys. Rev.}\ }\textbf {\bibinfo {volume}
  {109}},\ \bibinfo {pages} {1492} (\bibinfo {year} {1958})}\BibitemShut
  {NoStop}%
\bibitem [{\citenamefont {Pal}\ and\ \citenamefont
  {Huse}(2010)}]{pal_many-body_2010}%
  \BibitemOpen
  \bibfield  {author} {\bibinfo {author} {\bibfnamefont {A.}~\bibnamefont
  {Pal}}\ and\ \bibinfo {author} {\bibfnamefont {D.~A.}\ \bibnamefont {Huse}},\
  }\href {\doibase 10.1103/PhysRevB.82.174411} {\bibfield  {journal} {\bibinfo
  {journal} {Phys. Rev. B}\ }\textbf {\bibinfo {volume} {82}},\ \bibinfo
  {pages} {174411} (\bibinfo {year} {2010})}\BibitemShut {NoStop}%
\bibitem [{\citenamefont {Luitz}\ \emph {et~al.}(2015)\citenamefont {Luitz},
  \citenamefont {Laflorencie},\ and\ \citenamefont
  {Alet}}]{luitz_many-body_2015}%
  \BibitemOpen
  \bibfield  {author} {\bibinfo {author} {\bibfnamefont {D.~J.}\ \bibnamefont
  {Luitz}}, \bibinfo {author} {\bibfnamefont {N.}~\bibnamefont {Laflorencie}},
  \ and\ \bibinfo {author} {\bibfnamefont {F.}~\bibnamefont {Alet}},\ }\href
  {\doibase 10.1103/PhysRevB.91.081103} {\bibfield  {journal} {\bibinfo
  {journal} {Phys. Rev. B}\ }\textbf {\bibinfo {volume} {91}},\ \bibinfo
  {pages} {081103} (\bibinfo {year} {2015})}\BibitemShut {NoStop}%
\bibitem [{\citenamefont {Wahl}\ \emph {et~al.}(2017)\citenamefont {Wahl},
  \citenamefont {Pal},\ and\ \citenamefont {Simon}}]{wahl_efficient_2017}%
  \BibitemOpen
  \bibfield  {author} {\bibinfo {author} {\bibfnamefont {T.~B.}\ \bibnamefont
  {Wahl}}, \bibinfo {author} {\bibfnamefont {A.}~\bibnamefont {Pal}}, \ and\
  \bibinfo {author} {\bibfnamefont {S.~H.}\ \bibnamefont {Simon}},\ }\href
  {\doibase 10.1103/PhysRevX.7.021018} {\bibfield  {journal} {\bibinfo
  {journal} {Phys. Rev. X}\ }\textbf {\bibinfo {volume} {7}},\ \bibinfo {pages}
  {021018} (\bibinfo {year} {2017})}\BibitemShut {NoStop}%
\bibitem [{\citenamefont {Vosk}\ \emph {et~al.}(2015)\citenamefont {Vosk},
  \citenamefont {Huse},\ and\ \citenamefont {Altman}}]{vosk_theory_2015}%
  \BibitemOpen
  \bibfield  {author} {\bibinfo {author} {\bibfnamefont {R.}~\bibnamefont
  {Vosk}}, \bibinfo {author} {\bibfnamefont {D.~A.}\ \bibnamefont {Huse}}, \
  and\ \bibinfo {author} {\bibfnamefont {E.}~\bibnamefont {Altman}},\ }\href
  {\doibase 10.1103/PhysRevX.5.031032} {\bibfield  {journal} {\bibinfo
  {journal} {Phys. Rev. X}\ }\textbf {\bibinfo {volume} {5}},\ \bibinfo {pages}
  {031032} (\bibinfo {year} {2015})}\BibitemShut {NoStop}%
\bibitem [{\citenamefont {Potter}\ \emph {et~al.}(2015)\citenamefont {Potter},
  \citenamefont {Vasseur},\ and\ \citenamefont
  {Parameswaran}}]{potter_universal_2015}%
  \BibitemOpen
  \bibfield  {author} {\bibinfo {author} {\bibfnamefont {A.~C.}\ \bibnamefont
  {Potter}}, \bibinfo {author} {\bibfnamefont {R.}~\bibnamefont {Vasseur}}, \
  and\ \bibinfo {author} {\bibfnamefont {S.}~\bibnamefont {Parameswaran}},\
  }\href {\doibase 10.1103/PhysRevX.5.031033} {\bibfield  {journal} {\bibinfo
  {journal} {Phys. Rev. X}\ }\textbf {\bibinfo {volume} {5}},\ \bibinfo {pages}
  {031033} (\bibinfo {year} {2015})}\BibitemShut {NoStop}%
\bibitem [{\citenamefont {Dumitrescu}\ \emph {et~al.}(2017)\citenamefont
  {Dumitrescu}, \citenamefont {Vasseur},\ and\ \citenamefont
  {Potter}}]{dumitrescu_scaling_2017}%
  \BibitemOpen
  \bibfield  {author} {\bibinfo {author} {\bibfnamefont {P.~T.}\ \bibnamefont
  {Dumitrescu}}, \bibinfo {author} {\bibfnamefont {R.}~\bibnamefont {Vasseur}},
  \ and\ \bibinfo {author} {\bibfnamefont {A.~C.}\ \bibnamefont {Potter}},\
  }\href {\doibase 10.1103/PhysRevLett.119.110604} {\bibfield  {journal}
  {\bibinfo  {journal} {Phys. Rev. Lett.}\ }\textbf {\bibinfo {volume} {119}},\
  \bibinfo {pages} {110604} (\bibinfo {year} {2017})}\BibitemShut {NoStop}%
\bibitem [{\citenamefont {Thiery}\ \emph {et~al.}(2018)\citenamefont {Thiery},
  \citenamefont {Huveneers}, \citenamefont {M\"uller},\ and\ \citenamefont
  {De~Roeck}}]{thiery_many-body_2018}%
  \BibitemOpen
  \bibfield  {author} {\bibinfo {author} {\bibfnamefont {T.}~\bibnamefont
  {Thiery}}, \bibinfo {author} {\bibfnamefont {F.}~\bibnamefont {Huveneers}},
  \bibinfo {author} {\bibfnamefont {M.}~\bibnamefont {M\"uller}}, \ and\
  \bibinfo {author} {\bibfnamefont {W.}~\bibnamefont {De~Roeck}},\ }\href
  {\doibase 10.1103/PhysRevLett.121.140601} {\bibfield  {journal} {\bibinfo
  {journal} {Phys. Rev. Lett.}\ }\textbf {\bibinfo {volume} {121}},\ \bibinfo
  {pages} {140601} (\bibinfo {year} {2018})}\BibitemShut {NoStop}%
\bibitem [{\citenamefont {Zhang}\ and\ \citenamefont
  {Yao}(2018)}]{zhang_universal_2018}%
  \BibitemOpen
  \bibfield  {author} {\bibinfo {author} {\bibfnamefont {S.-X.}\ \bibnamefont
  {Zhang}}\ and\ \bibinfo {author} {\bibfnamefont {H.}~\bibnamefont {Yao}},\
  }\href {\doibase 10.1103/PhysRevLett.121.206601} {\bibfield  {journal}
  {\bibinfo  {journal} {Phys. Rev. Lett.}\ }\textbf {\bibinfo {volume} {121}},\
  \bibinfo {pages} {206601} (\bibinfo {year} {2018})}\BibitemShut {NoStop}%
\bibitem [{\citenamefont
  {Imbrie}(2016{\natexlab{a}})}]{imbrie_diagonalization_2016}%
  \BibitemOpen
  \bibfield  {author} {\bibinfo {author} {\bibfnamefont {J.~Z.}\ \bibnamefont
  {Imbrie}},\ }\href {https://link.aps.org/doi/10.1103/PhysRevLett.117.027201}
  {\bibfield  {journal} {\bibinfo  {journal} {Phys. Rev. Lett.}\ }\textbf
  {\bibinfo {volume} {117}},\ \bibinfo {pages} {027201} (\bibinfo {year}
  {2016}{\natexlab{a}})}\BibitemShut {NoStop}%
\bibitem [{\citenamefont {Imbrie}(2016{\natexlab{b}})}]{imbrie_many-body_2016}%
  \BibitemOpen
  \bibfield  {author} {\bibinfo {author} {\bibfnamefont {J.~Z.}\ \bibnamefont
  {Imbrie}},\ }\href {\doibase 10.1007/s10955-016-1508-x} {\bibfield  {journal}
  {\bibinfo  {journal} {J. Stat. Phys.}\ }\textbf {\bibinfo {volume} {163}},\
  \bibinfo {pages} {998} (\bibinfo {year} {2016}{\natexlab{b}})}\BibitemShut
  {NoStop}%
\bibitem [{\citenamefont {Schreiber}\ \emph {et~al.}(2015)\citenamefont
  {Schreiber}, \citenamefont {Hodgman}, \citenamefont {Bordia}, \citenamefont
  {Lüschen}, \citenamefont {Fischer}, \citenamefont {Vosk}, \citenamefont
  {Altman}, \citenamefont {Schneider},\ and\ \citenamefont
  {Bloch}}]{schreiber_observation_2015}%
  \BibitemOpen
  \bibfield  {author} {\bibinfo {author} {\bibfnamefont {M.}~\bibnamefont
  {Schreiber}}, \bibinfo {author} {\bibfnamefont {S.~S.}\ \bibnamefont
  {Hodgman}}, \bibinfo {author} {\bibfnamefont {P.}~\bibnamefont {Bordia}},
  \bibinfo {author} {\bibfnamefont {H.~P.}\ \bibnamefont {Lüschen}}, \bibinfo
  {author} {\bibfnamefont {M.~H.}\ \bibnamefont {Fischer}}, \bibinfo {author}
  {\bibfnamefont {R.}~\bibnamefont {Vosk}}, \bibinfo {author} {\bibfnamefont
  {E.}~\bibnamefont {Altman}}, \bibinfo {author} {\bibfnamefont
  {U.}~\bibnamefont {Schneider}}, \ and\ \bibinfo {author} {\bibfnamefont
  {I.}~\bibnamefont {Bloch}},\ }\href {\doibase 10.1126/science.aaa7432}
  {\bibfield  {journal} {\bibinfo  {journal} {Science}\ }\textbf {\bibinfo
  {volume} {349}},\ \bibinfo {pages} {842} (\bibinfo {year}
  {2015})}\BibitemShut {NoStop}%
\bibitem [{\citenamefont {Smith}\ \emph {et~al.}(2016)\citenamefont {Smith},
  \citenamefont {Lee}, \citenamefont {Richerme}, \citenamefont {Neyenhuis},
  \citenamefont {Hess}, \citenamefont {Hauke}, \citenamefont {Heyl},
  \citenamefont {Huse},\ and\ \citenamefont {Monroe}}]{smith_many-body_2016}%
  \BibitemOpen
  \bibfield  {author} {\bibinfo {author} {\bibfnamefont {J.}~\bibnamefont
  {Smith}}, \bibinfo {author} {\bibfnamefont {A.}~\bibnamefont {Lee}}, \bibinfo
  {author} {\bibfnamefont {P.}~\bibnamefont {Richerme}}, \bibinfo {author}
  {\bibfnamefont {B.}~\bibnamefont {Neyenhuis}}, \bibinfo {author}
  {\bibfnamefont {P.~W.}\ \bibnamefont {Hess}}, \bibinfo {author}
  {\bibfnamefont {P.}~\bibnamefont {Hauke}}, \bibinfo {author} {\bibfnamefont
  {M.}~\bibnamefont {Heyl}}, \bibinfo {author} {\bibfnamefont {D.~A.}\
  \bibnamefont {Huse}}, \ and\ \bibinfo {author} {\bibfnamefont
  {C.}~\bibnamefont {Monroe}},\ }\href {\doibase 10.1038/nphys3783} {\bibfield
  {journal} {\bibinfo  {journal} {Nat. Phys.}\ }\textbf {\bibinfo {volume}
  {12}},\ \bibinfo {pages} {907} (\bibinfo {year} {2016})}\BibitemShut
  {NoStop}%
\bibitem [{\citenamefont {Lukin}\ \emph {et~al.}(2018)\citenamefont {Lukin},
  \citenamefont {Rispoli}, \citenamefont {Schittko}, \citenamefont {Tai},
  \citenamefont {Kaufman}, \citenamefont {Choi}, \citenamefont {Khemani},
  \citenamefont {Léonard},\ and\ \citenamefont
  {Greiner}}]{lukin_probing_2018}%
  \BibitemOpen
  \bibfield  {author} {\bibinfo {author} {\bibfnamefont {A.}~\bibnamefont
  {Lukin}}, \bibinfo {author} {\bibfnamefont {M.}~\bibnamefont {Rispoli}},
  \bibinfo {author} {\bibfnamefont {R.}~\bibnamefont {Schittko}}, \bibinfo
  {author} {\bibfnamefont {M.~E.}\ \bibnamefont {Tai}}, \bibinfo {author}
  {\bibfnamefont {A.~M.}\ \bibnamefont {Kaufman}}, \bibinfo {author}
  {\bibfnamefont {S.}~\bibnamefont {Choi}}, \bibinfo {author} {\bibfnamefont
  {V.}~\bibnamefont {Khemani}}, \bibinfo {author} {\bibfnamefont
  {J.}~\bibnamefont {Léonard}}, \ and\ \bibinfo {author} {\bibfnamefont
  {M.}~\bibnamefont {Greiner}},\ }\href {http://arxiv.org/abs/1805.09819}
  {\bibfield  {journal} {\bibinfo  {journal} {arXiv:1805.09819}\ } (\bibinfo
  {year} {2018})}\BibitemShut {NoStop}%
\bibitem [{\citenamefont {Huse}\ \emph {et~al.}(2014)\citenamefont {Huse},
  \citenamefont {Nandkishore},\ and\ \citenamefont
  {Oganesyan}}]{huse_phenomenology_2014}%
  \BibitemOpen
  \bibfield  {author} {\bibinfo {author} {\bibfnamefont {D.~A.}\ \bibnamefont
  {Huse}}, \bibinfo {author} {\bibfnamefont {R.}~\bibnamefont {Nandkishore}}, \
  and\ \bibinfo {author} {\bibfnamefont {V.}~\bibnamefont {Oganesyan}},\ }\href
  {\doibase 10.1103/PhysRevB.90.174202} {\bibfield  {journal} {\bibinfo
  {journal} {Phys. Rev. B}\ }\textbf {\bibinfo {volume} {90}},\ \bibinfo
  {pages} {174202} (\bibinfo {year} {2014})}\BibitemShut {NoStop}%
\bibitem [{\citenamefont {Serbyn}\ \emph
  {et~al.}(2013{\natexlab{a}})\citenamefont {Serbyn}, \citenamefont {Papi\'c},\
  and\ \citenamefont {Abanin}}]{serbyn_local_2013}%
  \BibitemOpen
  \bibfield  {author} {\bibinfo {author} {\bibfnamefont {M.}~\bibnamefont
  {Serbyn}}, \bibinfo {author} {\bibfnamefont {Z.}~\bibnamefont {Papi\'c}}, \
  and\ \bibinfo {author} {\bibfnamefont {D.~A.}\ \bibnamefont {Abanin}},\
  }\href {\doibase 10.1103/PhysRevLett.111.127201} {\bibfield  {journal}
  {\bibinfo  {journal} {Phys. Rev. Lett.}\ }\textbf {\bibinfo {volume} {111}},\
  \bibinfo {pages} {127201} (\bibinfo {year} {2013}{\natexlab{a}})}\BibitemShut
  {NoStop}%
\bibitem [{\citenamefont {Imbrie}\ \emph {et~al.}(2017)\citenamefont {Imbrie},
  \citenamefont {Ros},\ and\ \citenamefont {Scardicchio}}]{imbrie_local_2017}%
  \BibitemOpen
  \bibfield  {author} {\bibinfo {author} {\bibfnamefont {J.~Z.}\ \bibnamefont
  {Imbrie}}, \bibinfo {author} {\bibfnamefont {V.}~\bibnamefont {Ros}}, \ and\
  \bibinfo {author} {\bibfnamefont {A.}~\bibnamefont {Scardicchio}},\ }\href
  {\doibase 10.1002/andp.201600278} {\bibfield  {journal} {\bibinfo  {journal}
  {Ann. Phys. (Berl.)}\ }\textbf {\bibinfo {volume} {529}},\ \bibinfo {pages}
  {1600278} (\bibinfo {year} {2017})}\BibitemShut {NoStop}%
\bibitem [{\citenamefont {Nandkishore}\ and\ \citenamefont
  {Huse}(2015)}]{nandkishore_many-body_2015}%
  \BibitemOpen
  \bibfield  {author} {\bibinfo {author} {\bibfnamefont {R.}~\bibnamefont
  {Nandkishore}}\ and\ \bibinfo {author} {\bibfnamefont {D.~A.}\ \bibnamefont
  {Huse}},\ }\href {\doibase 10.1146/annurev-conmatphys-031214-014726}
  {\bibfield  {journal} {\bibinfo  {journal} {Ann. Rev. Cond. Matt.}\ }\textbf
  {\bibinfo {volume} {6}},\ \bibinfo {pages} {15} (\bibinfo {year}
  {2015})}\BibitemShut {NoStop}%
\bibitem [{\citenamefont {Altman}\ and\ \citenamefont
  {Vosk}(2015)}]{altman_universal_2015}%
  \BibitemOpen
  \bibfield  {author} {\bibinfo {author} {\bibfnamefont {E.}~\bibnamefont
  {Altman}}\ and\ \bibinfo {author} {\bibfnamefont {R.}~\bibnamefont {Vosk}},\
  }\href {\doibase 10.1146/annurev-conmatphys-031214-014701} {\bibfield
  {journal} {\bibinfo  {journal} {Ann. Rev. Cond. Matt.}\ }\textbf {\bibinfo
  {volume} {6}},\ \bibinfo {pages} {383} (\bibinfo {year} {2015})}\BibitemShut
  {NoStop}%
\bibitem [{\citenamefont {Abanin}\ and\ \citenamefont
  {Papi\'c}(2017)}]{abanin_recent_2017}%
  \BibitemOpen
  \bibfield  {author} {\bibinfo {author} {\bibfnamefont {D.~A.}\ \bibnamefont
  {Abanin}}\ and\ \bibinfo {author} {\bibfnamefont {Z.}~\bibnamefont
  {Papi\'c}},\ }\href {\doibase 10.1002/andp.201700169} {\bibfield  {journal}
  {\bibinfo  {journal} {Ann. Phys. (Berl.)}\ }\textbf {\bibinfo {volume}
  {529}},\ \bibinfo {pages} {1700169} (\bibinfo {year} {2017})}\BibitemShut
  {NoStop}%
\bibitem [{\citenamefont {Parameswaran}\ \emph {et~al.}(2017)\citenamefont
  {Parameswaran}, \citenamefont {Potter},\ and\ \citenamefont
  {Vasseur}}]{parameswaran_eigenstate_2017}%
  \BibitemOpen
  \bibfield  {author} {\bibinfo {author} {\bibfnamefont {S.~A.}\ \bibnamefont
  {Parameswaran}}, \bibinfo {author} {\bibfnamefont {A.~C.}\ \bibnamefont
  {Potter}}, \ and\ \bibinfo {author} {\bibfnamefont {R.}~\bibnamefont
  {Vasseur}},\ }\href
  {http://onlinelibrary.wiley.com/doi/10.1002/andp.201600302/abstract}
  {\bibfield  {journal} {\bibinfo  {journal} {Ann. Phys. (Berl.)}\ ,\ \bibinfo
  {pages} {1600302}} (\bibinfo {year} {2017})}\BibitemShut {NoStop}%
\bibitem [{\citenamefont {Alet}\ and\ \citenamefont
  {Laflorencie}(2018)}]{alet_many-body_2018}%
  \BibitemOpen
  \bibfield  {author} {\bibinfo {author} {\bibfnamefont {F.}~\bibnamefont
  {Alet}}\ and\ \bibinfo {author} {\bibfnamefont {N.}~\bibnamefont
  {Laflorencie}},\ }\href {\doibase 10.1016/j.crhy.2018.03.003} {\bibfield
  {journal} {\bibinfo  {journal} {C. R. Phys.}\ }\textbf {\bibinfo {volume}
  {19}},\ \bibinfo {pages} {498} (\bibinfo {year} {2018})}\BibitemShut
  {NoStop}%
\bibitem [{\citenamefont {Abanin}\ \emph {et~al.}(2019)\citenamefont {Abanin},
  \citenamefont {Altman}, \citenamefont {Bloch},\ and\ \citenamefont
  {Serbyn}}]{abanin_ergodicity_2019}%
  \BibitemOpen
  \bibfield  {author} {\bibinfo {author} {\bibfnamefont {D.~A.}\ \bibnamefont
  {Abanin}}, \bibinfo {author} {\bibfnamefont {E.}~\bibnamefont {Altman}},
  \bibinfo {author} {\bibfnamefont {I.}~\bibnamefont {Bloch}}, \ and\ \bibinfo
  {author} {\bibfnamefont {M.}~\bibnamefont {Serbyn}},\ }\href {\doibase
  10.1103/RevModPhys.91.021001} {\bibfield  {journal} {\bibinfo  {journal}
  {Rev. Mod. Phys.}\ }\textbf {\bibinfo {volume} {91}},\ \bibinfo {pages}
  {021001} (\bibinfo {year} {2019})}\BibitemShut {NoStop}%
\bibitem [{\citenamefont {Evers}\ and\ \citenamefont
  {Mirlin}(2008)}]{evers_anderson_2008}%
  \BibitemOpen
  \bibfield  {author} {\bibinfo {author} {\bibfnamefont {F.}~\bibnamefont
  {Evers}}\ and\ \bibinfo {author} {\bibfnamefont {A.~D.}\ \bibnamefont
  {Mirlin}},\ }\href {\doibase 10.1103/RevModPhys.80.1355} {\bibfield
  {journal} {\bibinfo  {journal} {Rev. Mod. Phys.}\ }\textbf {\bibinfo {volume}
  {80}},\ \bibinfo {pages} {1355} (\bibinfo {year} {2008})}\BibitemShut
  {NoStop}%
\bibitem [{\citenamefont {De~Roeck}\ and\ \citenamefont
  {Huveneers}(2017)}]{de_roeck_stability_2017}%
  \BibitemOpen
  \bibfield  {author} {\bibinfo {author} {\bibfnamefont {W.}~\bibnamefont
  {De~Roeck}}\ and\ \bibinfo {author} {\bibfnamefont {F.}~\bibnamefont
  {Huveneers}},\ }\href {\doibase 10.1103/PhysRevB.95.155129} {\bibfield
  {journal} {\bibinfo  {journal} {Phys. Rev. B}\ }\textbf {\bibinfo {volume}
  {95}},\ \bibinfo {pages} {155129} (\bibinfo {year} {2017})}\BibitemShut
  {NoStop}%
\bibitem [{\citenamefont {De~Roeck}\ and\ \citenamefont
  {Imbrie}(2017)}]{de_roeck_many-body_2017}%
  \BibitemOpen
  \bibfield  {author} {\bibinfo {author} {\bibfnamefont {W.}~\bibnamefont
  {De~Roeck}}\ and\ \bibinfo {author} {\bibfnamefont {J.~Z.}\ \bibnamefont
  {Imbrie}},\ }\href
  {http://rsta.royalsocietypublishing.org/content/375/2108/20160422} {\bibfield
   {journal} {\bibinfo  {journal} {Phil. Trans. R. Soc. A}\ }\textbf {\bibinfo
  {volume} {375}} (\bibinfo {year} {2017})}\BibitemShut {NoStop}%
\bibitem [{\citenamefont {Potirniche}\ \emph {et~al.}(2019)\citenamefont
  {Potirniche}, \citenamefont {Banerjee},\ and\ \citenamefont
  {Altman}}]{potirniche_exploration_2019}%
  \BibitemOpen
  \bibfield  {author} {\bibinfo {author} {\bibfnamefont {I.-D.}\ \bibnamefont
  {Potirniche}}, \bibinfo {author} {\bibfnamefont {S.}~\bibnamefont
  {Banerjee}}, \ and\ \bibinfo {author} {\bibfnamefont {E.}~\bibnamefont
  {Altman}},\ }\href {\doibase 10.1103/PhysRevB.99.205149} {\bibfield
  {journal} {\bibinfo  {journal} {Phys. Rev. B}\ }\textbf {\bibinfo {volume}
  {99}},\ \bibinfo {pages} {205149} (\bibinfo {year} {2019})}\BibitemShut
  {NoStop}%
\bibitem [{\citenamefont {Wahl}\ \emph {et~al.}(2018)\citenamefont {Wahl},
  \citenamefont {Pal},\ and\ \citenamefont {Simon}}]{wahl_signatures_2018}%
  \BibitemOpen
  \bibfield  {author} {\bibinfo {author} {\bibfnamefont {T.~B.}\ \bibnamefont
  {Wahl}}, \bibinfo {author} {\bibfnamefont {A.}~\bibnamefont {Pal}}, \ and\
  \bibinfo {author} {\bibfnamefont {S.~H.}\ \bibnamefont {Simon}},\ }\href
  {\doibase 10.1038/s41567-018-0339-x} {\bibfield  {journal} {\bibinfo
  {journal} {Nat. Phys.}\ ,\ \bibinfo {pages} {1}} (\bibinfo {year}
  {2018})}\BibitemShut {NoStop}%
\bibitem [{\citenamefont {Thomson}\ and\ \citenamefont
  {Schir\'o}(2018)}]{thomson_time_2018}%
  \BibitemOpen
  \bibfield  {author} {\bibinfo {author} {\bibfnamefont {S.~J.}\ \bibnamefont
  {Thomson}}\ and\ \bibinfo {author} {\bibfnamefont {M.}~\bibnamefont
  {Schir\'o}},\ }\href {\doibase 10.1103/PhysRevB.97.060201} {\bibfield
  {journal} {\bibinfo  {journal} {Phys. Rev. B}\ }\textbf {\bibinfo {volume}
  {97}},\ \bibinfo {pages} {060201} (\bibinfo {year} {2018})}\BibitemShut
  {NoStop}%
\bibitem [{\citenamefont {De~Tomasi}\ \emph {et~al.}(2019)\citenamefont
  {De~Tomasi}, \citenamefont {Pollmann},\ and\ \citenamefont
  {Heyl}}]{de_tomasi_solving_2019}%
  \BibitemOpen
  \bibfield  {author} {\bibinfo {author} {\bibfnamefont {G.}~\bibnamefont
  {De~Tomasi}}, \bibinfo {author} {\bibfnamefont {F.}~\bibnamefont {Pollmann}},
  \ and\ \bibinfo {author} {\bibfnamefont {M.}~\bibnamefont {Heyl}},\ }\href
  {\doibase 10.1103/PhysRevB.99.241114} {\bibfield  {journal} {\bibinfo
  {journal} {Phys. Rev. B}\ }\textbf {\bibinfo {volume} {99}},\ \bibinfo
  {pages} {241114} (\bibinfo {year} {2019})}\BibitemShut {NoStop}%
\bibitem [{\citenamefont {Doggen}\ \emph {et~al.}(2020)\citenamefont {Doggen},
  \citenamefont {Gornyi}, \citenamefont {Mirlin},\ and\ \citenamefont
  {Polyakov}}]{doggen2020slow}%
  \BibitemOpen
  \bibfield  {author} {\bibinfo {author} {\bibfnamefont {E.~V.~H.}\
  \bibnamefont {Doggen}}, \bibinfo {author} {\bibfnamefont {I.~V.}\
  \bibnamefont {Gornyi}}, \bibinfo {author} {\bibfnamefont {A.~D.}\
  \bibnamefont {Mirlin}}, \ and\ \bibinfo {author} {\bibfnamefont {D.~G.}\
  \bibnamefont {Polyakov}},\ }\href {https://arxiv.org/abs/2002.07635}
  {\bibfield  {journal} {\bibinfo  {journal} {arXiv:2002.07635}\ } (\bibinfo
  {year} {2020})}\BibitemShut {NoStop}%
\bibitem [{\citenamefont {Geraedts}\ \emph {et~al.}(2016)\citenamefont
  {Geraedts}, \citenamefont {Nandkishore},\ and\ \citenamefont
  {Regnault}}]{geraedts_many_2016}%
  \BibitemOpen
  \bibfield  {author} {\bibinfo {author} {\bibfnamefont {S.~D.}\ \bibnamefont
  {Geraedts}}, \bibinfo {author} {\bibfnamefont {R.}~\bibnamefont
  {Nandkishore}}, \ and\ \bibinfo {author} {\bibfnamefont {N.}~\bibnamefont
  {Regnault}},\ }\href {\doibase 10.1103/PhysRevB.93.174202} {\bibfield
  {journal} {\bibinfo  {journal} {Phys. Rev. B}\ }\textbf {\bibinfo {volume}
  {93}},\ \bibinfo {pages} {174202} (\bibinfo {year} {2016})}\BibitemShut
  {NoStop}%
\bibitem [{\citenamefont {van Nieuwenburg}\ \emph {et~al.}(2019)\citenamefont
  {van Nieuwenburg}, \citenamefont {Baum},\ and\ \citenamefont
  {Refael}}]{van_nieuwenburg_bloch_2019}%
  \BibitemOpen
  \bibfield  {author} {\bibinfo {author} {\bibfnamefont {E.}~\bibnamefont {van
  Nieuwenburg}}, \bibinfo {author} {\bibfnamefont {Y.}~\bibnamefont {Baum}}, \
  and\ \bibinfo {author} {\bibfnamefont {G.}~\bibnamefont {Refael}},\ }\href
  {\doibase 10.1073/pnas.1819316116} {\bibfield  {journal} {\bibinfo  {journal}
  {Proceedings of the National Academy of Sciences}\ }\textbf {\bibinfo
  {volume} {116}},\ \bibinfo {pages} {9269} (\bibinfo {year}
  {2019})}\BibitemShut {NoStop}%
\bibitem [{\citenamefont {Wiater}\ and\ \citenamefont
  {Zakrzewski}(2018)}]{wiater_impact_2019}%
  \BibitemOpen
  \bibfield  {author} {\bibinfo {author} {\bibfnamefont {D.}~\bibnamefont
  {Wiater}}\ and\ \bibinfo {author} {\bibfnamefont {J.}~\bibnamefont
  {Zakrzewski}},\ }\href {\doibase 10.1103/PhysRevB.98.094202} {\bibfield
  {journal} {\bibinfo  {journal} {Phys. Rev. B}\ }\textbf {\bibinfo {volume}
  {98}},\ \bibinfo {pages} {094202} (\bibinfo {year} {2018})}\BibitemShut
  {NoStop}%
\bibitem [{\citenamefont {Krishna}\ \emph {et~al.}(2019)\citenamefont
  {Krishna}, \citenamefont {Ippoliti},\ and\ \citenamefont
  {Bhatt}}]{krishna_many_2019}%
  \BibitemOpen
  \bibfield  {author} {\bibinfo {author} {\bibfnamefont {A.}~\bibnamefont
  {Krishna}}, \bibinfo {author} {\bibfnamefont {M.}~\bibnamefont {Ippoliti}}, \
  and\ \bibinfo {author} {\bibfnamefont {R.~N.}\ \bibnamefont {Bhatt}},\ }\href
  {\doibase 10.1103/PhysRevB.99.041111} {\bibfield  {journal} {\bibinfo
  {journal} {Phys. Rev. B}\ }\textbf {\bibinfo {volume} {99}},\ \bibinfo
  {pages} {041111} (\bibinfo {year} {2019})}\BibitemShut {NoStop}%
\bibitem [{\citenamefont {Inglis}\ and\ \citenamefont
  {Pollet}(2016)}]{inglis_accessing_2016}%
  \BibitemOpen
  \bibfield  {author} {\bibinfo {author} {\bibfnamefont {S.}~\bibnamefont
  {Inglis}}\ and\ \bibinfo {author} {\bibfnamefont {L.}~\bibnamefont
  {Pollet}},\ }\href {\doibase 10.1103/PhysRevLett.117.120402} {\bibfield
  {journal} {\bibinfo  {journal} {Phys. Rev. Lett.}\ }\textbf {\bibinfo
  {volume} {117}},\ \bibinfo {pages} {120402} (\bibinfo {year}
  {2016})}\BibitemShut {NoStop}%
\bibitem [{\citenamefont {Choi}\ \emph {et~al.}(2016)\citenamefont {Choi},
  \citenamefont {Hild}, \citenamefont {Zeiher}, \citenamefont {Schauß},
  \citenamefont {Rubio-Abadal}, \citenamefont {Yefsah}, \citenamefont
  {Khemani}, \citenamefont {Huse}, \citenamefont {Bloch},\ and\ \citenamefont
  {Gross}}]{choi_exploring_2016}%
  \BibitemOpen
  \bibfield  {author} {\bibinfo {author} {\bibfnamefont {J.-y.}\ \bibnamefont
  {Choi}}, \bibinfo {author} {\bibfnamefont {S.}~\bibnamefont {Hild}}, \bibinfo
  {author} {\bibfnamefont {J.}~\bibnamefont {Zeiher}}, \bibinfo {author}
  {\bibfnamefont {P.}~\bibnamefont {Schauß}}, \bibinfo {author} {\bibfnamefont
  {A.}~\bibnamefont {Rubio-Abadal}}, \bibinfo {author} {\bibfnamefont
  {T.}~\bibnamefont {Yefsah}}, \bibinfo {author} {\bibfnamefont
  {V.}~\bibnamefont {Khemani}}, \bibinfo {author} {\bibfnamefont {D.~A.}\
  \bibnamefont {Huse}}, \bibinfo {author} {\bibfnamefont {I.}~\bibnamefont
  {Bloch}}, \ and\ \bibinfo {author} {\bibfnamefont {C.}~\bibnamefont
  {Gross}},\ }\href {\doibase 10.1126/science.aaf8834} {\bibfield  {journal}
  {\bibinfo  {journal} {Science}\ }\textbf {\bibinfo {volume} {352}},\ \bibinfo
  {pages} {1547} (\bibinfo {year} {2016})}\BibitemShut {NoStop}%
\bibitem [{\citenamefont {Bordia}\ \emph {et~al.}(2017)\citenamefont {Bordia},
  \citenamefont {L\"uschen}, \citenamefont {Scherg}, \citenamefont
  {Gopalakrishnan}, \citenamefont {Knap}, \citenamefont {Schneider},\ and\
  \citenamefont {Bloch}}]{bordia_probing_2017}%
  \BibitemOpen
  \bibfield  {author} {\bibinfo {author} {\bibfnamefont {P.}~\bibnamefont
  {Bordia}}, \bibinfo {author} {\bibfnamefont {H.}~\bibnamefont {L\"uschen}},
  \bibinfo {author} {\bibfnamefont {S.}~\bibnamefont {Scherg}}, \bibinfo
  {author} {\bibfnamefont {S.}~\bibnamefont {Gopalakrishnan}}, \bibinfo
  {author} {\bibfnamefont {M.}~\bibnamefont {Knap}}, \bibinfo {author}
  {\bibfnamefont {U.}~\bibnamefont {Schneider}}, \ and\ \bibinfo {author}
  {\bibfnamefont {I.}~\bibnamefont {Bloch}},\ }\href {\doibase
  10.1103/PhysRevX.7.041047} {\bibfield  {journal} {\bibinfo  {journal} {Phys.
  Rev. X}\ }\textbf {\bibinfo {volume} {7}},\ \bibinfo {pages} {041047}
  (\bibinfo {year} {2017})}\BibitemShut {NoStop}%
\bibitem [{\citenamefont {Fowler}\ and\ \citenamefont
  {Rushbrooke}(1937)}]{fowler_attempt_1937}%
  \BibitemOpen
  \bibfield  {author} {\bibinfo {author} {\bibfnamefont {R.~H.}\ \bibnamefont
  {Fowler}}\ and\ \bibinfo {author} {\bibfnamefont {G.~S.}\ \bibnamefont
  {Rushbrooke}},\ }\href {\doibase 10.1039/TF9373301272} {\bibfield  {journal}
  {\bibinfo  {journal} {Trans. Faraday Soc.}\ }\textbf {\bibinfo {volume}
  {33}},\ \bibinfo {pages} {1272} (\bibinfo {year} {1937})}\BibitemShut
  {NoStop}%
\bibitem [{\citenamefont {Fredrickson}\ and\ \citenamefont
  {Andersen}(1984)}]{fredrickson_kinetic_1984}%
  \BibitemOpen
  \bibfield  {author} {\bibinfo {author} {\bibfnamefont {G.~H.}\ \bibnamefont
  {Fredrickson}}\ and\ \bibinfo {author} {\bibfnamefont {H.~C.}\ \bibnamefont
  {Andersen}},\ }\href {\doibase 10.1103/PhysRevLett.53.1244} {\bibfield
  {journal} {\bibinfo  {journal} {Phys. Rev. Lett.}\ }\textbf {\bibinfo
  {volume} {53}},\ \bibinfo {pages} {1244} (\bibinfo {year}
  {1984})}\BibitemShut {NoStop}%
\bibitem [{\citenamefont {Ritort}\ and\ \citenamefont
  {Sollich}(2003)}]{ritort_glassy_2003}%
  \BibitemOpen
  \bibfield  {author} {\bibinfo {author} {\bibfnamefont {F.}~\bibnamefont
  {Ritort}}\ and\ \bibinfo {author} {\bibfnamefont {P.}~\bibnamefont
  {Sollich}},\ }\href {\doibase 10.1080/0001873031000093582} {\bibfield
  {journal} {\bibinfo  {journal} {Adv. Phys.}\ }\textbf {\bibinfo {volume}
  {52}},\ \bibinfo {pages} {219} (\bibinfo {year} {2003})}\BibitemShut
  {NoStop}%
\bibitem [{\citenamefont {Chandler}\ and\ \citenamefont
  {Garrahan}(2010)}]{chandler_dynamics_2010}%
  \BibitemOpen
  \bibfield  {author} {\bibinfo {author} {\bibfnamefont {D.}~\bibnamefont
  {Chandler}}\ and\ \bibinfo {author} {\bibfnamefont {J.~P.}\ \bibnamefont
  {Garrahan}},\ }\href {\doibase 10.1146/annurev.physchem.040808.090405}
  {\bibfield  {journal} {\bibinfo  {journal} {Annu. Rev. Phys. Chem.}\ }\textbf
  {\bibinfo {volume} {61}},\ \bibinfo {pages} {191} (\bibinfo {year}
  {2010})}\BibitemShut {NoStop}%
\bibitem [{\citenamefont {Lacroix}\ \emph {et~al.}(2011)\citenamefont
  {Lacroix}, \citenamefont {Mendels},\ and\ \citenamefont
  {Mila}}]{lacroix_introduction_2011}%
  \BibitemOpen
  \bibinfo {editor} {\bibfnamefont {C.}~\bibnamefont {Lacroix}}, \bibinfo
  {editor} {\bibfnamefont {P.}~\bibnamefont {Mendels}}, \ and\ \bibinfo
  {editor} {\bibfnamefont {F.}~\bibnamefont {Mila}},\ eds.,\ \href
  {www.springer.com/la/book/9783642105883} {\emph {\bibinfo {title}
  {Introduction to Frustrated Magnetism: Materials, Experiments, Theory}}},\
  Springer Series in Solid-State Sciences\ (\bibinfo  {publisher}
  {Springer-Verlag},\ \bibinfo {address} {Berlin Heidelberg},\ \bibinfo {year}
  {2011})\BibitemShut {NoStop}%
\bibitem [{\citenamefont {Castelnovo}\ \emph {et~al.}(2012)\citenamefont
  {Castelnovo}, \citenamefont {Moessner},\ and\ \citenamefont
  {Sondhi}}]{castelnovo_spin_2012}%
  \BibitemOpen
  \bibfield  {author} {\bibinfo {author} {\bibfnamefont {C.}~\bibnamefont
  {Castelnovo}}, \bibinfo {author} {\bibfnamefont {R.}~\bibnamefont
  {Moessner}}, \ and\ \bibinfo {author} {\bibfnamefont {S.}~\bibnamefont
  {Sondhi}},\ }\href {\doibase 10.1146/annurev-conmatphys-020911-125058}
  {\bibfield  {journal} {\bibinfo  {journal} {Annu. Rev. Cond. Matt. Phys.}\
  }\textbf {\bibinfo {volume} {3}},\ \bibinfo {pages} {35} (\bibinfo {year}
  {2012})}\BibitemShut {NoStop}%
\bibitem [{\citenamefont {Bernien}\ \emph {et~al.}(2017)\citenamefont
  {Bernien}, \citenamefont {Schwartz}, \citenamefont {Keesling}, \citenamefont
  {Levine}, \citenamefont {Omran}, \citenamefont {Pichler}, \citenamefont
  {Choi}, \citenamefont {Zibrov}, \citenamefont {Endres}, \citenamefont
  {Greiner}, \citenamefont {Vuletić},\ and\ \citenamefont
  {Lukin}}]{bernien_probing_2017}%
  \BibitemOpen
  \bibfield  {author} {\bibinfo {author} {\bibfnamefont {H.}~\bibnamefont
  {Bernien}}, \bibinfo {author} {\bibfnamefont {S.}~\bibnamefont {Schwartz}},
  \bibinfo {author} {\bibfnamefont {A.}~\bibnamefont {Keesling}}, \bibinfo
  {author} {\bibfnamefont {H.}~\bibnamefont {Levine}}, \bibinfo {author}
  {\bibfnamefont {A.}~\bibnamefont {Omran}}, \bibinfo {author} {\bibfnamefont
  {H.}~\bibnamefont {Pichler}}, \bibinfo {author} {\bibfnamefont
  {S.}~\bibnamefont {Choi}}, \bibinfo {author} {\bibfnamefont {A.~S.}\
  \bibnamefont {Zibrov}}, \bibinfo {author} {\bibfnamefont {M.}~\bibnamefont
  {Endres}}, \bibinfo {author} {\bibfnamefont {M.}~\bibnamefont {Greiner}},
  \bibinfo {author} {\bibfnamefont {V.}~\bibnamefont {Vuletić}}, \ and\
  \bibinfo {author} {\bibfnamefont {M.~D.}\ \bibnamefont {Lukin}},\ }\href
  {\doibase 10.1038/nature24622} {\bibfield  {journal} {\bibinfo  {journal}
  {Nature}\ }\textbf {\bibinfo {volume} {551}},\ \bibinfo {pages} {579}
  (\bibinfo {year} {2017})}\BibitemShut {NoStop}%
\bibitem [{\citenamefont {Fisher}\ and\ \citenamefont
  {Stephenson}(1963)}]{fisher_statistical_1963}%
  \BibitemOpen
  \bibfield  {author} {\bibinfo {author} {\bibfnamefont {M.~E.}\ \bibnamefont
  {Fisher}}\ and\ \bibinfo {author} {\bibfnamefont {J.}~\bibnamefont
  {Stephenson}},\ }\href {\doibase 10.1103/PhysRev.132.1411} {\bibfield
  {journal} {\bibinfo  {journal} {Phys. Rev.}\ }\textbf {\bibinfo {volume}
  {132}},\ \bibinfo {pages} {1411} (\bibinfo {year} {1963})}\BibitemShut
  {NoStop}%
\bibitem [{\citenamefont {Huse}\ \emph {et~al.}(2003)\citenamefont {Huse},
  \citenamefont {Krauth}, \citenamefont {Moessner},\ and\ \citenamefont
  {Sondhi}}]{huse_coulomb_2003}%
  \BibitemOpen
  \bibfield  {author} {\bibinfo {author} {\bibfnamefont {D.~A.}\ \bibnamefont
  {Huse}}, \bibinfo {author} {\bibfnamefont {W.}~\bibnamefont {Krauth}},
  \bibinfo {author} {\bibfnamefont {R.}~\bibnamefont {Moessner}}, \ and\
  \bibinfo {author} {\bibfnamefont {S.~L.}\ \bibnamefont {Sondhi}},\ }\href
  {\doibase 10.1103/PhysRevLett.91.167004} {\bibfield  {journal} {\bibinfo
  {journal} {Phys. Rev. Lett.}\ }\textbf {\bibinfo {volume} {91}},\ \bibinfo
  {pages} {167004} (\bibinfo {year} {2003})}\BibitemShut {NoStop}%
\bibitem [{\citenamefont {Oakes}\ \emph {et~al.}(2016)\citenamefont {Oakes},
  \citenamefont {Garrahan},\ and\ \citenamefont
  {Powell}}]{oakes_emergence_2016}%
  \BibitemOpen
  \bibfield  {author} {\bibinfo {author} {\bibfnamefont {T.}~\bibnamefont
  {Oakes}}, \bibinfo {author} {\bibfnamefont {J.~P.}\ \bibnamefont {Garrahan}},
  \ and\ \bibinfo {author} {\bibfnamefont {S.}~\bibnamefont {Powell}},\ }\href
  {\doibase 10.1103/PhysRevE.93.032129} {\bibfield  {journal} {\bibinfo
  {journal} {Phys. Rev. E}\ }\textbf {\bibinfo {volume} {93}},\ \bibinfo
  {pages} {032129} (\bibinfo {year} {2016})}\BibitemShut {NoStop}%
\bibitem [{\citenamefont {Das}\ \emph {et~al.}(2005)\citenamefont {Das},
  \citenamefont {Farrell}, \citenamefont {Kondev},\ and\ \citenamefont
  {Chakraborty}}]{das_critical_2005}%
  \BibitemOpen
  \bibfield  {author} {\bibinfo {author} {\bibfnamefont {D.}~\bibnamefont
  {Das}}, \bibinfo {author} {\bibfnamefont {G.}~\bibnamefont {Farrell}},
  \bibinfo {author} {\bibfnamefont {J.}~\bibnamefont {Kondev}}, \ and\ \bibinfo
  {author} {\bibfnamefont {B.}~\bibnamefont {Chakraborty}},\ }\href {\doibase
  10.1021/jp051636l} {\bibfield  {journal} {\bibinfo  {journal} {J. Phys. Chem.
  B}\ }\textbf {\bibinfo {volume} {109}},\ \bibinfo {pages} {21413} (\bibinfo
  {year} {2005})}\BibitemShut {NoStop}%
\bibitem [{\citenamefont {C\'epas}(2014)}]{cepas_multiple_2014}%
  \BibitemOpen
  \bibfield  {author} {\bibinfo {author} {\bibfnamefont {O.}~\bibnamefont
  {C\'epas}},\ }\href {\doibase 10.1103/PhysRevB.90.064404} {\bibfield
  {journal} {\bibinfo  {journal} {Phys. Rev. B}\ }\textbf {\bibinfo {volume}
  {90}},\ \bibinfo {pages} {064404} (\bibinfo {year} {2014})}\BibitemShut
  {NoStop}%
\bibitem [{\citenamefont {Chamon}(2005)}]{chamon_quantum_2005}%
  \BibitemOpen
  \bibfield  {author} {\bibinfo {author} {\bibfnamefont {C.}~\bibnamefont
  {Chamon}},\ }\href {\doibase 10.1103/PhysRevLett.94.040402} {\bibfield
  {journal} {\bibinfo  {journal} {Phys. Rev. Lett.}\ }\textbf {\bibinfo
  {volume} {94}},\ \bibinfo {pages} {040402} (\bibinfo {year}
  {2005})}\BibitemShut {NoStop}%
\bibitem [{\citenamefont {Lan}\ \emph {et~al.}(2018)\citenamefont {Lan},
  \citenamefont {van Horssen}, \citenamefont {Powell},\ and\ \citenamefont
  {Garrahan}}]{lan_quantum_2018}%
  \BibitemOpen
  \bibfield  {author} {\bibinfo {author} {\bibfnamefont {Z.}~\bibnamefont
  {Lan}}, \bibinfo {author} {\bibfnamefont {M.}~\bibnamefont {van Horssen}},
  \bibinfo {author} {\bibfnamefont {S.}~\bibnamefont {Powell}}, \ and\ \bibinfo
  {author} {\bibfnamefont {J.~P.}\ \bibnamefont {Garrahan}},\ }\href {\doibase
  10.1103/PhysRevLett.121.040603} {\bibfield  {journal} {\bibinfo  {journal}
  {Phys. Rev. Lett.}\ }\textbf {\bibinfo {volume} {121}},\ \bibinfo {pages}
  {040603} (\bibinfo {year} {2018})}\BibitemShut {NoStop}%
\bibitem [{\citenamefont {Turner}\ \emph {et~al.}(2018)\citenamefont {Turner},
  \citenamefont {Michailidis}, \citenamefont {Abanin}, \citenamefont {Serbyn},\
  and\ \citenamefont {Papić}}]{turner_weak_2018}%
  \BibitemOpen
  \bibfield  {author} {\bibinfo {author} {\bibfnamefont {C.~J.}\ \bibnamefont
  {Turner}}, \bibinfo {author} {\bibfnamefont {A.~A.}\ \bibnamefont
  {Michailidis}}, \bibinfo {author} {\bibfnamefont {D.~A.}\ \bibnamefont
  {Abanin}}, \bibinfo {author} {\bibfnamefont {M.}~\bibnamefont {Serbyn}}, \
  and\ \bibinfo {author} {\bibfnamefont {Z.}~\bibnamefont {Papić}},\ }\href
  {\doibase 10.1038/s41567-018-0137-5} {\bibfield  {journal} {\bibinfo
  {journal} {Nature Physics}\ }\textbf {\bibinfo {volume} {14}},\ \bibinfo
  {pages} {745} (\bibinfo {year} {2018})}\BibitemShut {NoStop}%
\bibitem [{\citenamefont {Feldmeier}\ \emph {et~al.}(2019)\citenamefont
  {Feldmeier}, \citenamefont {Pollmann},\ and\ \citenamefont
  {Knap}}]{feldmeier_dynamical_2019}%
  \BibitemOpen
  \bibfield  {author} {\bibinfo {author} {\bibfnamefont {J.}~\bibnamefont
  {Feldmeier}}, \bibinfo {author} {\bibfnamefont {F.}~\bibnamefont {Pollmann}},
  \ and\ \bibinfo {author} {\bibfnamefont {M.}~\bibnamefont {Knap}},\ }\href
  {\doibase 10.1103/PhysRevLett.123.040601} {\bibfield  {journal} {\bibinfo
  {journal} {Phys. Rev. Lett.}\ }\textbf {\bibinfo {volume} {123}},\ \bibinfo
  {pages} {040601} (\bibinfo {year} {2019})}\BibitemShut {NoStop}%
\bibitem [{\citenamefont {Pancotti}\ \emph {et~al.}(2020)\citenamefont
  {Pancotti}, \citenamefont {Giudice}, \citenamefont {Cirac}, \citenamefont
  {Garrahan},\ and\ \citenamefont {Ba\~nuls}}]{pancotti_quantum_2020}%
  \BibitemOpen
  \bibfield  {author} {\bibinfo {author} {\bibfnamefont {N.}~\bibnamefont
  {Pancotti}}, \bibinfo {author} {\bibfnamefont {G.}~\bibnamefont {Giudice}},
  \bibinfo {author} {\bibfnamefont {J.~I.}\ \bibnamefont {Cirac}}, \bibinfo
  {author} {\bibfnamefont {J.~P.}\ \bibnamefont {Garrahan}}, \ and\ \bibinfo
  {author} {\bibfnamefont {M.~C.}\ \bibnamefont {Ba\~nuls}},\ }\href {\doibase
  10.1103/PhysRevX.10.021051} {\bibfield  {journal} {\bibinfo  {journal} {Phys.
  Rev. X}\ }\textbf {\bibinfo {volume} {10}},\ \bibinfo {pages} {021051}
  (\bibinfo {year} {2020})}\BibitemShut {NoStop}%
\bibitem [{\citenamefont {Smith}\ \emph
  {et~al.}(2017{\natexlab{a}})\citenamefont {Smith}, \citenamefont {Knolle},
  \citenamefont {Kovrizhin},\ and\ \citenamefont
  {Moessner}}]{smith_disorder_2017}%
  \BibitemOpen
  \bibfield  {author} {\bibinfo {author} {\bibfnamefont {A.}~\bibnamefont
  {Smith}}, \bibinfo {author} {\bibfnamefont {J.}~\bibnamefont {Knolle}},
  \bibinfo {author} {\bibfnamefont {D.~L.}\ \bibnamefont {Kovrizhin}}, \ and\
  \bibinfo {author} {\bibfnamefont {R.}~\bibnamefont {Moessner}},\ }\href
  {\doibase 10.1103/PhysRevLett.118.266601} {\bibfield  {journal} {\bibinfo
  {journal} {Phys. Rev. Lett.}\ }\textbf {\bibinfo {volume} {118}},\ \bibinfo
  {pages} {266601} (\bibinfo {year} {2017}{\natexlab{a}})}\BibitemShut
  {NoStop}%
\bibitem [{\citenamefont {Brenes}\ \emph {et~al.}(2018)\citenamefont {Brenes},
  \citenamefont {Dalmonte}, \citenamefont {Heyl},\ and\ \citenamefont
  {Scardicchio}}]{brenes_many_2017}%
  \BibitemOpen
  \bibfield  {author} {\bibinfo {author} {\bibfnamefont {M.}~\bibnamefont
  {Brenes}}, \bibinfo {author} {\bibfnamefont {M.}~\bibnamefont {Dalmonte}},
  \bibinfo {author} {\bibfnamefont {M.}~\bibnamefont {Heyl}}, \ and\ \bibinfo
  {author} {\bibfnamefont {A.}~\bibnamefont {Scardicchio}},\ }\href {\doibase
  10.1103/PhysRevLett.120.030601} {\bibfield  {journal} {\bibinfo  {journal}
  {Phys. Rev. Lett.}\ }\textbf {\bibinfo {volume} {120}},\ \bibinfo {pages}
  {030601} (\bibinfo {year} {2018})}\BibitemShut {NoStop}%
\bibitem [{\citenamefont {Smith}\ \emph
  {et~al.}(2017{\natexlab{b}})\citenamefont {Smith}, \citenamefont {Knolle},
  \citenamefont {Moessner},\ and\ \citenamefont
  {Kovrizhin}}]{smith_absence_2017}%
  \BibitemOpen
  \bibfield  {author} {\bibinfo {author} {\bibfnamefont {A.}~\bibnamefont
  {Smith}}, \bibinfo {author} {\bibfnamefont {J.}~\bibnamefont {Knolle}},
  \bibinfo {author} {\bibfnamefont {R.}~\bibnamefont {Moessner}}, \ and\
  \bibinfo {author} {\bibfnamefont {D.~L.}\ \bibnamefont {Kovrizhin}},\ }\href
  {\doibase 10.1103/PhysRevLett.119.176601} {\bibfield  {journal} {\bibinfo
  {journal} {Phys. Rev. Lett.}\ }\textbf {\bibinfo {volume} {119}},\ \bibinfo
  {pages} {176601} (\bibinfo {year} {2017}{\natexlab{b}})}\BibitemShut
  {NoStop}%
\bibitem [{\citenamefont {Smith}\ \emph {et~al.}(2018)\citenamefont {Smith},
  \citenamefont {Knolle}, \citenamefont {Moessner},\ and\ \citenamefont
  {Kovrizhin}}]{smith_dynamical_2018}%
  \BibitemOpen
  \bibfield  {author} {\bibinfo {author} {\bibfnamefont {A.}~\bibnamefont
  {Smith}}, \bibinfo {author} {\bibfnamefont {J.}~\bibnamefont {Knolle}},
  \bibinfo {author} {\bibfnamefont {R.}~\bibnamefont {Moessner}}, \ and\
  \bibinfo {author} {\bibfnamefont {D.~L.}\ \bibnamefont {Kovrizhin}},\ }\href
  {\doibase 10.1103/PhysRevB.97.245137} {\bibfield  {journal} {\bibinfo
  {journal} {Phys. Rev. B}\ }\textbf {\bibinfo {volume} {97}},\ \bibinfo
  {pages} {245137} (\bibinfo {year} {2018})}\BibitemShut {NoStop}%
\bibitem [{\citenamefont {{Karpov}}\ \emph {et~al.}(2020)\citenamefont
  {{Karpov}}, \citenamefont {{Verdel}}, \citenamefont {{Huang}}, \citenamefont
  {{Schmitt}},\ and\ \citenamefont {{Heyl}}}]{karpov_disorder_2020}%
  \BibitemOpen
  \bibfield  {author} {\bibinfo {author} {\bibfnamefont {P.}~\bibnamefont
  {{Karpov}}}, \bibinfo {author} {\bibfnamefont {R.}~\bibnamefont {{Verdel}}},
  \bibinfo {author} {\bibfnamefont {Y.~P.}\ \bibnamefont {{Huang}}}, \bibinfo
  {author} {\bibfnamefont {M.}~\bibnamefont {{Schmitt}}}, \ and\ \bibinfo
  {author} {\bibfnamefont {M.}~\bibnamefont {{Heyl}}},\ }\href
  {https://arxiv.org/abs/2003.04901} {\bibfield  {journal} {\bibinfo  {journal}
  {arXiv:2003.04901}\ } (\bibinfo {year} {2020})}\BibitemShut {NoStop}%
\bibitem [{\citenamefont {{Papaefstathiou}}\ \emph {et~al.}(2020)\citenamefont
  {{Papaefstathiou}}, \citenamefont {{Smith}},\ and\ \citenamefont
  {{Knolle}}}]{Papaefstathiou_disorder_2020}%
  \BibitemOpen
  \bibfield  {author} {\bibinfo {author} {\bibfnamefont {I.}~\bibnamefont
  {{Papaefstathiou}}}, \bibinfo {author} {\bibfnamefont {A.}~\bibnamefont
  {{Smith}}}, \ and\ \bibinfo {author} {\bibfnamefont {J.}~\bibnamefont
  {{Knolle}}},\ }\href {https://arxiv.org/abs/2003.12497} {\bibfield  {journal}
  {\bibinfo  {journal} {arXiv:2003.12497}\ } (\bibinfo {year}
  {2020})}\BibitemShut {NoStop}%
\bibitem [{\citenamefont {{Halimeh}}\ and\ \citenamefont
  {{Hauke}}(2020{\natexlab{a}})}]{Halimeh_staircase_2020}%
  \BibitemOpen
  \bibfield  {author} {\bibinfo {author} {\bibfnamefont {J.~C.}\ \bibnamefont
  {{Halimeh}}}\ and\ \bibinfo {author} {\bibfnamefont {P.}~\bibnamefont
  {{Hauke}}},\ }\href {https://arxiv.org/abs/2004.07248} {\bibfield  {journal}
  {\bibinfo  {journal} {arXiv:2004.07248}\ } (\bibinfo {year}
  {2020}{\natexlab{a}})}\BibitemShut {NoStop}%
\bibitem [{\citenamefont {{Halimeh}}\ and\ \citenamefont
  {{Hauke}}(2020{\natexlab{b}})}]{Halimeh_origin_2020}%
  \BibitemOpen
  \bibfield  {author} {\bibinfo {author} {\bibfnamefont {J.~C.}\ \bibnamefont
  {{Halimeh}}}\ and\ \bibinfo {author} {\bibfnamefont {P.}~\bibnamefont
  {{Hauke}}},\ }\href {https://arxiv.org/abs/2004.07254} {\bibfield  {journal}
  {\bibinfo  {journal} {arXiv:2004.07254}\ } (\bibinfo {year}
  {2020}{\natexlab{b}})}\BibitemShut {NoStop}%
\bibitem [{\citenamefont {Chen}\ \emph {et~al.}(2018)\citenamefont {Chen},
  \citenamefont {Burnell},\ and\ \citenamefont {Chandran}}]{chen_how_2018}%
  \BibitemOpen
  \bibfield  {author} {\bibinfo {author} {\bibfnamefont {C.}~\bibnamefont
  {Chen}}, \bibinfo {author} {\bibfnamefont {F.}~\bibnamefont {Burnell}}, \
  and\ \bibinfo {author} {\bibfnamefont {A.}~\bibnamefont {Chandran}},\ }\href
  {\doibase 10.1103/PhysRevLett.121.085701} {\bibfield  {journal} {\bibinfo
  {journal} {Phys. Rev. Lett.}\ }\textbf {\bibinfo {volume} {121}},\ \bibinfo
  {pages} {085701} (\bibinfo {year} {2018})}\BibitemShut {NoStop}%
\bibitem [{\citenamefont {Ostmann}\ \emph {et~al.}(2019)\citenamefont
  {Ostmann}, \citenamefont {Marcuzzi}, \citenamefont {Garrahan},\ and\
  \citenamefont {Lesanovsky}}]{ostmann_localization_2019}%
  \BibitemOpen
  \bibfield  {author} {\bibinfo {author} {\bibfnamefont {M.}~\bibnamefont
  {Ostmann}}, \bibinfo {author} {\bibfnamefont {M.}~\bibnamefont {Marcuzzi}},
  \bibinfo {author} {\bibfnamefont {J.~P.}\ \bibnamefont {Garrahan}}, \ and\
  \bibinfo {author} {\bibfnamefont {I.}~\bibnamefont {Lesanovsky}},\ }\href
  {\doibase 10.1103/PhysRevA.99.060101} {\bibfield  {journal} {\bibinfo
  {journal} {Phys. Rev. A}\ }\textbf {\bibinfo {volume} {99}},\ \bibinfo
  {pages} {060101} (\bibinfo {year} {2019})}\BibitemShut {NoStop}%
\bibitem [{\citenamefont {Rokhsar}\ and\ \citenamefont
  {Kivelson}(1988)}]{rokhsar_superconductivity_1988}%
  \BibitemOpen
  \bibfield  {author} {\bibinfo {author} {\bibfnamefont {D.~S.}\ \bibnamefont
  {Rokhsar}}\ and\ \bibinfo {author} {\bibfnamefont {S.~A.}\ \bibnamefont
  {Kivelson}},\ }\href {\doibase 10.1103/PhysRevLett.61.2376} {\bibfield
  {journal} {\bibinfo  {journal} {Phys. Rev. Lett.}\ }\textbf {\bibinfo
  {volume} {61}},\ \bibinfo {pages} {2376} (\bibinfo {year}
  {1988})}\BibitemShut {NoStop}%
\bibitem [{\citenamefont {Punk}\ \emph {et~al.}(2015)\citenamefont {Punk},
  \citenamefont {Allais},\ and\ \citenamefont {Sachdev}}]{punk_quantum_2015}%
  \BibitemOpen
  \bibfield  {author} {\bibinfo {author} {\bibfnamefont {M.}~\bibnamefont
  {Punk}}, \bibinfo {author} {\bibfnamefont {A.}~\bibnamefont {Allais}}, \ and\
  \bibinfo {author} {\bibfnamefont {S.}~\bibnamefont {Sachdev}},\ }\href
  {\doibase 10.1073/pnas.1512206112} {\bibfield  {journal} {\bibinfo  {journal}
  {Proc. Natl. Acad. Sci. USA}\ }\textbf {\bibinfo {volume} {112}},\ \bibinfo
  {pages} {9552} (\bibinfo {year} {2015})}\BibitemShut {NoStop}%
\bibitem [{Note1()}]{Note1}%
  \BibitemOpen
  \bibinfo {note} {See R. Moessner, K. S. Raman, Chapter 17 in Ref.~\protect
  \rev@citealpnum {lacroix_introduction_2011} for a review}\BibitemShut
  {NoStop}%
\bibitem [{\citenamefont {Schwandt}\ \emph {et~al.}(2010)\citenamefont
  {Schwandt}, \citenamefont {Mambrini},\ and\ \citenamefont
  {Poilblanc}}]{schwandt_generalized_2010}%
  \BibitemOpen
  \bibfield  {author} {\bibinfo {author} {\bibfnamefont {D.}~\bibnamefont
  {Schwandt}}, \bibinfo {author} {\bibfnamefont {M.}~\bibnamefont {Mambrini}},
  \ and\ \bibinfo {author} {\bibfnamefont {D.}~\bibnamefont {Poilblanc}},\
  }\href {\doibase 10.1103/PhysRevB.81.214413} {\bibfield  {journal} {\bibinfo
  {journal} {Phys. Rev. B}\ }\textbf {\bibinfo {volume} {81}},\ \bibinfo
  {pages} {214413} (\bibinfo {year} {2010})}\BibitemShut {NoStop}%
\bibitem [{\citenamefont {Moessner}\ and\ \citenamefont
  {Sondhi}(2001)}]{moessner_resonating_2001}%
  \BibitemOpen
  \bibfield  {author} {\bibinfo {author} {\bibfnamefont {R.}~\bibnamefont
  {Moessner}}\ and\ \bibinfo {author} {\bibfnamefont {S.~L.}\ \bibnamefont
  {Sondhi}},\ }\href {\doibase 10.1103/PhysRevLett.86.1881} {\bibfield
  {journal} {\bibinfo  {journal} {Phys. Rev. Lett.}\ }\textbf {\bibinfo
  {volume} {86}},\ \bibinfo {pages} {1881} (\bibinfo {year}
  {2001})}\BibitemShut {NoStop}%
\bibitem [{\citenamefont {Lan}\ and\ \citenamefont
  {Powell}(2017)}]{lan_eigenstate_2017}%
  \BibitemOpen
  \bibfield  {author} {\bibinfo {author} {\bibfnamefont {Z.}~\bibnamefont
  {Lan}}\ and\ \bibinfo {author} {\bibfnamefont {S.}~\bibnamefont {Powell}},\
  }\href {\doibase 10.1103/PhysRevB.96.115140} {\bibfield  {journal} {\bibinfo
  {journal} {Phys. Rev. B}\ }\textbf {\bibinfo {volume} {96}},\ \bibinfo
  {pages} {115140} (\bibinfo {year} {2017})}\BibitemShut {NoStop}%
\bibitem [{\citenamefont {Potter}\ and\ \citenamefont
  {Vasseur}(2016)}]{potter_symmetry_2016}%
  \BibitemOpen
  \bibfield  {author} {\bibinfo {author} {\bibfnamefont {A.~C.}\ \bibnamefont
  {Potter}}\ and\ \bibinfo {author} {\bibfnamefont {R.}~\bibnamefont
  {Vasseur}},\ }\href {\doibase 10.1103/PhysRevB.94.224206} {\bibfield
  {journal} {\bibinfo  {journal} {Phys. Rev. B}\ }\textbf {\bibinfo {volume}
  {94}},\ \bibinfo {pages} {224206} (\bibinfo {year} {2016})}\BibitemShut
  {NoStop}%
\bibitem [{\citenamefont {Pietracaprina}\ \emph {et~al.}(2018)\citenamefont
  {Pietracaprina}, \citenamefont {Mac\'e}, \citenamefont {Luitz},\ and\
  \citenamefont {Alet}}]{pietracaprina_shift-invert_2018}%
  \BibitemOpen
  \bibfield  {author} {\bibinfo {author} {\bibfnamefont {F.}~\bibnamefont
  {Pietracaprina}}, \bibinfo {author} {\bibfnamefont {N.}~\bibnamefont
  {Mac\'e}}, \bibinfo {author} {\bibfnamefont {D.~J.}\ \bibnamefont {Luitz}}, \
  and\ \bibinfo {author} {\bibfnamefont {F.}~\bibnamefont {Alet}},\ }\href
  {\doibase 10.21468/SciPostPhys.5.5.045} {\bibfield  {journal} {\bibinfo
  {journal} {SciPost Physics}\ }\textbf {\bibinfo {volume} {5}},\ \bibinfo
  {pages} {045} (\bibinfo {year} {2018})}\BibitemShut {NoStop}%
\bibitem [{\citenamefont {Macé}\ \emph {et~al.}(2019)\citenamefont {Macé},
  \citenamefont {Laflorencie},\ and\ \citenamefont
  {Alet}}]{mace_many-body_2019}%
  \BibitemOpen
  \bibfield  {author} {\bibinfo {author} {\bibfnamefont {N.}~\bibnamefont
  {Macé}}, \bibinfo {author} {\bibfnamefont {N.}~\bibnamefont {Laflorencie}},
  \ and\ \bibinfo {author} {\bibfnamefont {F.}~\bibnamefont {Alet}},\ }\href
  {\doibase 10.21468/SciPostPhys.6.4.050} {\bibfield  {journal} {\bibinfo
  {journal} {SciPost Phys.}\ }\textbf {\bibinfo {volume} {6}},\ \bibinfo
  {pages} {50} (\bibinfo {year} {2019})}\BibitemShut {NoStop}%
\bibitem [{\citenamefont {Mac\'e}\ \emph {et~al.}(2019)\citenamefont {Mac\'e},
  \citenamefont {Alet},\ and\ \citenamefont
  {Laflorencie}}]{mace_multifractal_2019}%
  \BibitemOpen
  \bibfield  {author} {\bibinfo {author} {\bibfnamefont {N.}~\bibnamefont
  {Mac\'e}}, \bibinfo {author} {\bibfnamefont {F.}~\bibnamefont {Alet}}, \ and\
  \bibinfo {author} {\bibfnamefont {N.}~\bibnamefont {Laflorencie}},\ }\href
  {\doibase 10.1103/PhysRevLett.123.180601} {\bibfield  {journal} {\bibinfo
  {journal} {Phys. Rev. Lett.}\ }\textbf {\bibinfo {volume} {123}},\ \bibinfo
  {pages} {180601} (\bibinfo {year} {2019})}\BibitemShut {NoStop}%
\bibitem [{\citenamefont {Nauts}\ and\ \citenamefont
  {Wyatt}(1983)}]{nauts_new_1983}%
  \BibitemOpen
  \bibfield  {author} {\bibinfo {author} {\bibfnamefont {A.}~\bibnamefont
  {Nauts}}\ and\ \bibinfo {author} {\bibfnamefont {R.~E.}\ \bibnamefont
  {Wyatt}},\ }\href {\doibase 10.1103/PhysRevLett.51.2238} {\bibfield
  {journal} {\bibinfo  {journal} {Phys. Rev. Lett.}\ }\textbf {\bibinfo
  {volume} {51}},\ \bibinfo {pages} {2238} (\bibinfo {year}
  {1983})}\BibitemShut {NoStop}%
\bibitem [{\citenamefont {Oganesyan}\ and\ \citenamefont
  {Huse}(2007)}]{oganesyan_localization_2007}%
  \BibitemOpen
  \bibfield  {author} {\bibinfo {author} {\bibfnamefont {V.}~\bibnamefont
  {Oganesyan}}\ and\ \bibinfo {author} {\bibfnamefont {D.~A.}\ \bibnamefont
  {Huse}},\ }\href {https://link.aps.org/doi/10.1103/PhysRevB.75.155111}
  {\bibfield  {journal} {\bibinfo  {journal} {Phys. Rev. B}\ }\textbf {\bibinfo
  {volume} {75}},\ \bibinfo {pages} {155111} (\bibinfo {year}
  {2007})}\BibitemShut {NoStop}%
\bibitem [{\citenamefont {Atas}\ \emph {et~al.}(2013)\citenamefont {Atas},
  \citenamefont {Bogomolny}, \citenamefont {Giraud},\ and\ \citenamefont
  {Roux}}]{atas_distribution_2013}%
  \BibitemOpen
  \bibfield  {author} {\bibinfo {author} {\bibfnamefont {Y.~Y.}\ \bibnamefont
  {Atas}}, \bibinfo {author} {\bibfnamefont {E.}~\bibnamefont {Bogomolny}},
  \bibinfo {author} {\bibfnamefont {O.}~\bibnamefont {Giraud}}, \ and\ \bibinfo
  {author} {\bibfnamefont {G.}~\bibnamefont {Roux}},\ }\href {\doibase
  10.1103/PhysRevLett.110.084101} {\bibfield  {journal} {\bibinfo  {journal}
  {Phys. Rev. Lett.}\ }\textbf {\bibinfo {volume} {110}},\ \bibinfo {pages}
  {084101} (\bibinfo {year} {2013})}\BibitemShut {NoStop}%
\bibitem [{Note2()}]{Note2}%
  \BibitemOpen
  \bibinfo {note} {Considering only the clusters with $L_x=4$ leads to a
  smaller critical disorder $V \sim 12$ for the one-dimensional geometry of
  this four-leg tube. Results on other observables are consistent with this
  estimate.}\BibitemShut {Stop}%
\bibitem [{\citenamefont {Papi\'c}\ \emph {et~al.}(2015)\citenamefont
  {Papi\'c}, \citenamefont {Stoudenmire},\ and\ \citenamefont
  {Abanin}}]{papic_many-body_2015}%
  \BibitemOpen
  \bibfield  {author} {\bibinfo {author} {\bibfnamefont {Z.}~\bibnamefont
  {Papi\'c}}, \bibinfo {author} {\bibfnamefont {E.~M.}\ \bibnamefont
  {Stoudenmire}}, \ and\ \bibinfo {author} {\bibfnamefont {D.~A.}\ \bibnamefont
  {Abanin}},\ }\href {\doibase 10.1016/j.aop.2015.08.024} {\bibfield  {journal}
  {\bibinfo  {journal} {Ann. Phys. (Berl.)}\ }\textbf {\bibinfo {volume}
  {362}},\ \bibinfo {pages} {714} (\bibinfo {year} {2015})}\BibitemShut
  {NoStop}%
\bibitem [{\citenamefont {Morampudi}\ \emph {et~al.}(2020)\citenamefont
  {Morampudi}, \citenamefont {Chandran},\ and\ \citenamefont
  {Laumann}}]{morampudi_universal_2020}%
  \BibitemOpen
  \bibfield  {author} {\bibinfo {author} {\bibfnamefont {S.~C.}\ \bibnamefont
  {Morampudi}}, \bibinfo {author} {\bibfnamefont {A.}~\bibnamefont {Chandran}},
  \ and\ \bibinfo {author} {\bibfnamefont {C.~R.}\ \bibnamefont {Laumann}},\
  }\href {\doibase 10.1103/PhysRevLett.124.050602} {\bibfield  {journal}
  {\bibinfo  {journal} {Phys. Rev. Lett.}\ }\textbf {\bibinfo {volume} {124}},\
  \bibinfo {pages} {050602} (\bibinfo {year} {2020})}\BibitemShut {NoStop}%
\bibitem [{\citenamefont {Khemani}\ \emph
  {et~al.}(2017{\natexlab{a}})\citenamefont {Khemani}, \citenamefont {Lim},
  \citenamefont {Sheng},\ and\ \citenamefont {Huse}}]{khemani_critical_2017}%
  \BibitemOpen
  \bibfield  {author} {\bibinfo {author} {\bibfnamefont {V.}~\bibnamefont
  {Khemani}}, \bibinfo {author} {\bibfnamefont {S.}~\bibnamefont {Lim}},
  \bibinfo {author} {\bibfnamefont {D.}~\bibnamefont {Sheng}}, \ and\ \bibinfo
  {author} {\bibfnamefont {D.~A.}\ \bibnamefont {Huse}},\ }\href {\doibase
  10.1103/PhysRevX.7.021013} {\bibfield  {journal} {\bibinfo  {journal} {Phys.
  Rev. X}\ }\textbf {\bibinfo {volume} {7}},\ \bibinfo {pages} {021013}
  (\bibinfo {year} {2017}{\natexlab{a}})}\BibitemShut {NoStop}%
\bibitem [{\citenamefont {Carrasquilla}\ and\ \citenamefont
  {Melko}(2017)}]{carrasquilla_machine_2017}%
  \BibitemOpen
  \bibfield  {author} {\bibinfo {author} {\bibfnamefont {J.}~\bibnamefont
  {Carrasquilla}}\ and\ \bibinfo {author} {\bibfnamefont {R.~G.}\ \bibnamefont
  {Melko}},\ }\href {\doibase 10.1038/nphys4035} {\bibfield  {journal}
  {\bibinfo  {journal} {Nat. Phys.}\ }\textbf {\bibinfo {volume} {13}},\
  \bibinfo {pages} {431} (\bibinfo {year} {2017})}\BibitemShut {NoStop}%
\bibitem [{\citenamefont {Schindler}\ \emph {et~al.}(2017)\citenamefont
  {Schindler}, \citenamefont {Regnault},\ and\ \citenamefont
  {Neupert}}]{schindler_probing_2017}%
  \BibitemOpen
  \bibfield  {author} {\bibinfo {author} {\bibfnamefont {F.}~\bibnamefont
  {Schindler}}, \bibinfo {author} {\bibfnamefont {N.}~\bibnamefont {Regnault}},
  \ and\ \bibinfo {author} {\bibfnamefont {T.}~\bibnamefont {Neupert}},\ }\href
  {\doibase 10.1103/PhysRevB.95.245134} {\bibfield  {journal} {\bibinfo
  {journal} {Phys. Rev. B}\ }\textbf {\bibinfo {volume} {95}},\ \bibinfo
  {pages} {245134} (\bibinfo {year} {2017})}\BibitemShut {NoStop}%
\bibitem [{\citenamefont {Li}\ \emph {et~al.}(2017)\citenamefont {Li},
  \citenamefont {Luo},\ and\ \citenamefont {Wan}}]{li_extracting_2017}%
  \BibitemOpen
  \bibfield  {author} {\bibinfo {author} {\bibfnamefont {Z.}~\bibnamefont
  {Li}}, \bibinfo {author} {\bibfnamefont {M.}~\bibnamefont {Luo}}, \ and\
  \bibinfo {author} {\bibfnamefont {X.}~\bibnamefont {Wan}},\ }\href
  {http://arxiv.org/abs/1711.04252} {\bibfield  {journal} {\bibinfo  {journal}
  {arXiv:1711.04252}\ } (\bibinfo {year} {2017})}\BibitemShut {NoStop}%
\bibitem [{\citenamefont {Venderley}\ \emph {et~al.}(2018)\citenamefont
  {Venderley}, \citenamefont {Khemani},\ and\ \citenamefont
  {Kim}}]{venderley_machine_2018}%
  \BibitemOpen
  \bibfield  {author} {\bibinfo {author} {\bibfnamefont {J.}~\bibnamefont
  {Venderley}}, \bibinfo {author} {\bibfnamefont {V.}~\bibnamefont {Khemani}},
  \ and\ \bibinfo {author} {\bibfnamefont {E.-A.}\ \bibnamefont {Kim}},\ }\href
  {\doibase 10.1103/PhysRevLett.120.257204} {\bibfield  {journal} {\bibinfo
  {journal} {Phys. Rev. Lett.}\ }\textbf {\bibinfo {volume} {120}},\ \bibinfo
  {pages} {257204} (\bibinfo {year} {2018})}\BibitemShut {NoStop}%
\bibitem [{\citenamefont {Hsu}\ \emph {et~al.}(2018)\citenamefont {Hsu},
  \citenamefont {Li}, \citenamefont {Deng},\ and\ \citenamefont
  {Das~Sarma}}]{hsu_machine_2018}%
  \BibitemOpen
  \bibfield  {author} {\bibinfo {author} {\bibfnamefont {Y.-T.}\ \bibnamefont
  {Hsu}}, \bibinfo {author} {\bibfnamefont {X.}~\bibnamefont {Li}}, \bibinfo
  {author} {\bibfnamefont {D.-L.}\ \bibnamefont {Deng}}, \ and\ \bibinfo
  {author} {\bibfnamefont {S.}~\bibnamefont {Das~Sarma}},\ }\href {\doibase
  10.1103/PhysRevLett.121.245701} {\bibfield  {journal} {\bibinfo  {journal}
  {Phys. Rev. Lett.}\ }\textbf {\bibinfo {volume} {121}},\ \bibinfo {pages}
  {245701} (\bibinfo {year} {2018})}\BibitemShut {NoStop}%
\bibitem [{\citenamefont {Zhang}\ \emph {et~al.}(2019)\citenamefont {Zhang},
  \citenamefont {Wang},\ and\ \citenamefont {Wang}}]{zhang_interpretable_2019}%
  \BibitemOpen
  \bibfield  {author} {\bibinfo {author} {\bibfnamefont {W.}~\bibnamefont
  {Zhang}}, \bibinfo {author} {\bibfnamefont {L.}~\bibnamefont {Wang}}, \ and\
  \bibinfo {author} {\bibfnamefont {Z.}~\bibnamefont {Wang}},\ }\href {\doibase
  10.1103/PhysRevB.99.054208} {\bibfield  {journal} {\bibinfo  {journal} {Phys.
  Rev. B}\ }\textbf {\bibinfo {volume} {99}},\ \bibinfo {pages} {054208}
  (\bibinfo {year} {2019})}\BibitemShut {NoStop}%
\bibitem [{\citenamefont {Huembeli}\ \emph {et~al.}(2019)\citenamefont
  {Huembeli}, \citenamefont {Dauphin}, \citenamefont {Wittek},\ and\
  \citenamefont {Gogolin}}]{huembeli_automated_2019}%
  \BibitemOpen
  \bibfield  {author} {\bibinfo {author} {\bibfnamefont {P.}~\bibnamefont
  {Huembeli}}, \bibinfo {author} {\bibfnamefont {A.}~\bibnamefont {Dauphin}},
  \bibinfo {author} {\bibfnamefont {P.}~\bibnamefont {Wittek}}, \ and\ \bibinfo
  {author} {\bibfnamefont {C.}~\bibnamefont {Gogolin}},\ }\href {\doibase
  10.1103/PhysRevB.99.104106} {\bibfield  {journal} {\bibinfo  {journal} {Phys.
  Rev. B}\ }\textbf {\bibinfo {volume} {99}},\ \bibinfo {pages} {104106}
  (\bibinfo {year} {2019})}\BibitemShut {NoStop}%
\bibitem [{\citenamefont {Serbyn}\ \emph {et~al.}(2016)\citenamefont {Serbyn},
  \citenamefont {Michailidis}, \citenamefont {Abanin},\ and\ \citenamefont
  {Papi\'c}}]{serbyn_power-law_2016}%
  \BibitemOpen
  \bibfield  {author} {\bibinfo {author} {\bibfnamefont {M.}~\bibnamefont
  {Serbyn}}, \bibinfo {author} {\bibfnamefont {A.~A.}\ \bibnamefont
  {Michailidis}}, \bibinfo {author} {\bibfnamefont {D.~A.}\ \bibnamefont
  {Abanin}}, \ and\ \bibinfo {author} {\bibfnamefont {Z.}~\bibnamefont
  {Papi\'c}},\ }\href {http://arxiv.org/abs/1605.05737} {\bibfield  {journal}
  {\bibinfo  {journal} {Physical Review Letters}\ }\textbf {\bibinfo {volume}
  {117}} (\bibinfo {year} {2016})}\BibitemShut {NoStop}%
\bibitem [{\citenamefont {Th\'eveniaut}\ and\ \citenamefont
  {Alet}(2019)}]{theveniaut_neural_2019}%
  \BibitemOpen
  \bibfield  {author} {\bibinfo {author} {\bibfnamefont {H.}~\bibnamefont
  {Th\'eveniaut}}\ and\ \bibinfo {author} {\bibfnamefont {F.}~\bibnamefont
  {Alet}},\ }\href {\doibase 10.1103/PhysRevB.100.224202} {\bibfield  {journal}
  {\bibinfo  {journal} {Phys. Rev. B}\ }\textbf {\bibinfo {volume} {100}},\
  \bibinfo {pages} {224202} (\bibinfo {year} {2019})}\BibitemShut {NoStop}%
\bibitem [{\citenamefont {Kim}\ and\ \citenamefont
  {Huse}(2013)}]{kim_ballistic_2013}%
  \BibitemOpen
  \bibfield  {author} {\bibinfo {author} {\bibfnamefont {H.}~\bibnamefont
  {Kim}}\ and\ \bibinfo {author} {\bibfnamefont {D.~A.}\ \bibnamefont {Huse}},\
  }\href {\doibase 10.1103/PhysRevLett.111.127205} {\bibfield  {journal}
  {\bibinfo  {journal} {Phys. Rev. Lett.}\ }\textbf {\bibinfo {volume} {111}},\
  \bibinfo {pages} {127205} (\bibinfo {year} {2013})}\BibitemShut {NoStop}%
\bibitem [{\citenamefont {Serbyn}\ \emph
  {et~al.}(2013{\natexlab{b}})\citenamefont {Serbyn}, \citenamefont
  {Papi{\'c}},\ and\ \citenamefont {Abanin}}]{serbyn_universal_2013}%
  \BibitemOpen
  \bibfield  {author} {\bibinfo {author} {\bibfnamefont {M.}~\bibnamefont
  {Serbyn}}, \bibinfo {author} {\bibfnamefont {Z.}~\bibnamefont {Papi{\'c}}}, \
  and\ \bibinfo {author} {\bibfnamefont {D.~A.}\ \bibnamefont {Abanin}},\
  }\href {\doibase 10.1103/PhysRevLett.110.260601} {\bibfield  {journal}
  {\bibinfo  {journal} {Phys. Rev. Lett.}\ }\textbf {\bibinfo {volume} {110}},\
  \bibinfo {pages} {260601} (\bibinfo {year} {2013}{\natexlab{b}})}\BibitemShut
  {NoStop}%
\bibitem [{\citenamefont {Nanduri}\ \emph {et~al.}(2014)\citenamefont
  {Nanduri}, \citenamefont {Kim},\ and\ \citenamefont
  {Huse}}]{nanduri_entanglement_2014}%
  \BibitemOpen
  \bibfield  {author} {\bibinfo {author} {\bibfnamefont {A.}~\bibnamefont
  {Nanduri}}, \bibinfo {author} {\bibfnamefont {H.}~\bibnamefont {Kim}}, \ and\
  \bibinfo {author} {\bibfnamefont {D.~A.}\ \bibnamefont {Huse}},\ }\href
  {\doibase 10.1103/PhysRevB.90.064201} {\bibfield  {journal} {\bibinfo
  {journal} {Phys. Rev. B}\ }\textbf {\bibinfo {volume} {90}},\ \bibinfo
  {pages} {064201} (\bibinfo {year} {2014})}\BibitemShut {NoStop}%
\bibitem [{\citenamefont {Khemani}\ \emph
  {et~al.}(2017{\natexlab{b}})\citenamefont {Khemani}, \citenamefont {Sheng},\
  and\ \citenamefont {Huse}}]{khemani_two_2017}%
  \BibitemOpen
  \bibfield  {author} {\bibinfo {author} {\bibfnamefont {V.}~\bibnamefont
  {Khemani}}, \bibinfo {author} {\bibfnamefont {D.~N.}\ \bibnamefont {Sheng}},
  \ and\ \bibinfo {author} {\bibfnamefont {D.~A.}\ \bibnamefont {Huse}},\
  }\href {\doibase 10.1103/PhysRevLett.119.075702} {\bibfield  {journal}
  {\bibinfo  {journal} {Phys. Rev. Lett.}\ }\textbf {\bibinfo {volume} {119}},\
  \bibinfo {pages} {075702} (\bibinfo {year} {2017}{\natexlab{b}})}\BibitemShut
  {NoStop}%
\bibitem [{\citenamefont {Glaetzle}\ \emph {et~al.}(2014)\citenamefont
  {Glaetzle}, \citenamefont {Dalmonte}, \citenamefont {Nath}, \citenamefont
  {Rousochatzakis}, \citenamefont {Moessner},\ and\ \citenamefont
  {Zoller}}]{glaetzle_quantum_2014}%
  \BibitemOpen
  \bibfield  {author} {\bibinfo {author} {\bibfnamefont {A.~W.}\ \bibnamefont
  {Glaetzle}}, \bibinfo {author} {\bibfnamefont {M.}~\bibnamefont {Dalmonte}},
  \bibinfo {author} {\bibfnamefont {R.}~\bibnamefont {Nath}}, \bibinfo {author}
  {\bibfnamefont {I.}~\bibnamefont {Rousochatzakis}}, \bibinfo {author}
  {\bibfnamefont {R.}~\bibnamefont {Moessner}}, \ and\ \bibinfo {author}
  {\bibfnamefont {P.}~\bibnamefont {Zoller}},\ }\href {\doibase
  10.1103/PhysRevX.4.041037} {\bibfield  {journal} {\bibinfo  {journal} {Phys.
  Rev. X}\ }\textbf {\bibinfo {volume} {4}},\ \bibinfo {pages} {041037}
  (\bibinfo {year} {2014})}\BibitemShut {NoStop}%
\bibitem [{\citenamefont {Celi}\ \emph {et~al.}(2020)\citenamefont {Celi},
  \citenamefont {Vermersch}, \citenamefont {Viyuela}, \citenamefont {Pichler},
  \citenamefont {Lukin},\ and\ \citenamefont {Zoller}}]{celi_emerging_2020}%
  \BibitemOpen
  \bibfield  {author} {\bibinfo {author} {\bibfnamefont {A.}~\bibnamefont
  {Celi}}, \bibinfo {author} {\bibfnamefont {B.}~\bibnamefont {Vermersch}},
  \bibinfo {author} {\bibfnamefont {O.}~\bibnamefont {Viyuela}}, \bibinfo
  {author} {\bibfnamefont {H.}~\bibnamefont {Pichler}}, \bibinfo {author}
  {\bibfnamefont {M.~D.}\ \bibnamefont {Lukin}}, \ and\ \bibinfo {author}
  {\bibfnamefont {P.}~\bibnamefont {Zoller}},\ }\href {\doibase
  10.1103/PhysRevX.10.021057} {\bibfield  {journal} {\bibinfo  {journal} {Phys.
  Rev. X}\ }\textbf {\bibinfo {volume} {10}},\ \bibinfo {pages} {021057}
  (\bibinfo {year} {2020})}\BibitemShut {NoStop}%
\bibitem [{\citenamefont {Nath}\ \emph {et~al.}(2015)\citenamefont {Nath},
  \citenamefont {Dalmonte}, \citenamefont {Glaetzle}, \citenamefont {Zoller},
  \citenamefont {Schmidt-Kaler},\ and\ \citenamefont
  {Gerritsma}}]{nath_hexagonal_2015}%
  \BibitemOpen
  \bibfield  {author} {\bibinfo {author} {\bibfnamefont {R.}~\bibnamefont
  {Nath}}, \bibinfo {author} {\bibfnamefont {M.}~\bibnamefont {Dalmonte}},
  \bibinfo {author} {\bibfnamefont {A.~W.}\ \bibnamefont {Glaetzle}}, \bibinfo
  {author} {\bibfnamefont {P.}~\bibnamefont {Zoller}}, \bibinfo {author}
  {\bibfnamefont {F.}~\bibnamefont {Schmidt-Kaler}}, \ and\ \bibinfo {author}
  {\bibfnamefont {R.}~\bibnamefont {Gerritsma}},\ }\href {\doibase
  10.1088/1367-2630/17/6/065018} {\bibfield  {journal} {\bibinfo  {journal}
  {New J. Phys.}\ }\textbf {\bibinfo {volume} {17}},\ \bibinfo {pages} {065018}
  (\bibinfo {year} {2015})}\BibitemShut {NoStop}%
\bibitem [{\citenamefont {Sundar}\ \emph {et~al.}(2019)\citenamefont {Sundar},
  \citenamefont {Rutkowski}, \citenamefont {Mueller},\ and\ \citenamefont
  {Lawler}}]{sundar_quantum_2019}%
  \BibitemOpen
  \bibfield  {author} {\bibinfo {author} {\bibfnamefont {B.}~\bibnamefont
  {Sundar}}, \bibinfo {author} {\bibfnamefont {T.~C.}\ \bibnamefont
  {Rutkowski}}, \bibinfo {author} {\bibfnamefont {E.~J.}\ \bibnamefont
  {Mueller}}, \ and\ \bibinfo {author} {\bibfnamefont {M.~J.}\ \bibnamefont
  {Lawler}},\ }\href {\doibase 10.1103/PhysRevA.99.043623} {\bibfield
  {journal} {\bibinfo  {journal} {Phys. Rev. A}\ }\textbf {\bibinfo {volume}
  {99}},\ \bibinfo {pages} {043623} (\bibinfo {year} {2019})}\BibitemShut
  {NoStop}%
\bibitem [{\citenamefont {Ioffe}\ \emph {et~al.}(2002)\citenamefont {Ioffe},
  \citenamefont {Feigel'man}, \citenamefont {Ioselevich}, \citenamefont
  {Ivanov}, \citenamefont {Troyer},\ and\ \citenamefont
  {Blatter}}]{ioffe_topologically_2002}%
  \BibitemOpen
  \bibfield  {author} {\bibinfo {author} {\bibfnamefont {L.~B.}\ \bibnamefont
  {Ioffe}}, \bibinfo {author} {\bibfnamefont {M.~V.}\ \bibnamefont
  {Feigel'man}}, \bibinfo {author} {\bibfnamefont {A.}~\bibnamefont
  {Ioselevich}}, \bibinfo {author} {\bibfnamefont {D.}~\bibnamefont {Ivanov}},
  \bibinfo {author} {\bibfnamefont {M.}~\bibnamefont {Troyer}}, \ and\ \bibinfo
  {author} {\bibfnamefont {G.}~\bibnamefont {Blatter}},\ }\href {\doibase
  10.1038/415503a} {\bibfield  {journal} {\bibinfo  {journal} {Nature}\
  }\textbf {\bibinfo {volume} {415}},\ \bibinfo {pages} {503} (\bibinfo {year}
  {2002})}\BibitemShut {NoStop}%
\bibitem [{\citenamefont {Marcos}\ \emph {et~al.}(2014)\citenamefont {Marcos},
  \citenamefont {Widmer}, \citenamefont {Rico}, \citenamefont {Hafezi},
  \citenamefont {Rabl}, \citenamefont {Wiese},\ and\ \citenamefont
  {Zoller}}]{marcos_two-dimensional_2014}%
  \BibitemOpen
  \bibfield  {author} {\bibinfo {author} {\bibfnamefont {D.}~\bibnamefont
  {Marcos}}, \bibinfo {author} {\bibfnamefont {P.}~\bibnamefont {Widmer}},
  \bibinfo {author} {\bibfnamefont {E.}~\bibnamefont {Rico}}, \bibinfo {author}
  {\bibfnamefont {M.}~\bibnamefont {Hafezi}}, \bibinfo {author} {\bibfnamefont
  {P.}~\bibnamefont {Rabl}}, \bibinfo {author} {\bibfnamefont {U.~J.}\
  \bibnamefont {Wiese}}, \ and\ \bibinfo {author} {\bibfnamefont
  {P.}~\bibnamefont {Zoller}},\ }\href {\doibase 10.1016/j.aop.2014.09.011}
  {\bibfield  {journal} {\bibinfo  {journal} {Ann. Phys. (Berl.)}\ }\textbf
  {\bibinfo {volume} {351}},\ \bibinfo {pages} {634} (\bibinfo {year}
  {2014})}\BibitemShut {NoStop}%
\bibitem [{\citenamefont {{Pietracaprina}}\ and\ \citenamefont
  {{Alet}}(2020)}]{pietracaprina_probing_2020}%
  \BibitemOpen
  \bibfield  {author} {\bibinfo {author} {\bibfnamefont {F.}~\bibnamefont
  {{Pietracaprina}}}\ and\ \bibinfo {author} {\bibfnamefont {F.}~\bibnamefont
  {{Alet}}},\ }\href {https://arxiv.org/abs/2005.10233} {\bibfield  {journal}
  {\bibinfo  {journal} {arXiv:2005.10233}\ } (\bibinfo {year}
  {2020})}\BibitemShut {NoStop}%
\bibitem [{\citenamefont {Agarwal}\ \emph {et~al.}(2015)\citenamefont
  {Agarwal}, \citenamefont {Gopalakrishnan}, \citenamefont {Knap},
  \citenamefont {M\"uller},\ and\ \citenamefont
  {Demler}}]{agarwal_anomalous_2015}%
  \BibitemOpen
  \bibfield  {author} {\bibinfo {author} {\bibfnamefont {K.}~\bibnamefont
  {Agarwal}}, \bibinfo {author} {\bibfnamefont {S.}~\bibnamefont
  {Gopalakrishnan}}, \bibinfo {author} {\bibfnamefont {M.}~\bibnamefont
  {Knap}}, \bibinfo {author} {\bibfnamefont {M.}~\bibnamefont {M\"uller}}, \
  and\ \bibinfo {author} {\bibfnamefont {E.}~\bibnamefont {Demler}},\ }\href
  {\doibase 10.1103/PhysRevLett.114.160401} {\bibfield  {journal} {\bibinfo
  {journal} {Phys. Rev. Lett.}\ }\textbf {\bibinfo {volume} {114}},\ \bibinfo
  {pages} {160401} (\bibinfo {year} {2015})}\BibitemShut {NoStop}%
\bibitem [{\citenamefont {Luitz}\ \emph {et~al.}(2016)\citenamefont {Luitz},
  \citenamefont {Laflorencie},\ and\ \citenamefont
  {Alet}}]{luitz_extended_2016}%
  \BibitemOpen
  \bibfield  {author} {\bibinfo {author} {\bibfnamefont {D.~J.}\ \bibnamefont
  {Luitz}}, \bibinfo {author} {\bibfnamefont {N.}~\bibnamefont {Laflorencie}},
  \ and\ \bibinfo {author} {\bibfnamefont {F.}~\bibnamefont {Alet}},\ }\href
  {\doibase 10.1103/PhysRevB.93.060201} {\bibfield  {journal} {\bibinfo
  {journal} {Phys. Rev. B}\ }\textbf {\bibinfo {volume} {93}},\ \bibinfo
  {pages} {060201} (\bibinfo {year} {2016})}\BibitemShut {NoStop}%
\bibitem [{\citenamefont {\ifmmode \check{Z}\else
  \v{Z}\fi{}nidari\ifmmode~\check{c}\else \v{c}\fi{}}\ \emph
  {et~al.}(2016)\citenamefont {\ifmmode \check{Z}\else
  \v{Z}\fi{}nidari\ifmmode~\check{c}\else \v{c}\fi{}}, \citenamefont
  {Scardicchio},\ and\ \citenamefont {Varma}}]{znidaric_diffusive_2016}%
  \BibitemOpen
  \bibfield  {author} {\bibinfo {author} {\bibfnamefont {M.}~\bibnamefont
  {\ifmmode \check{Z}\else \v{Z}\fi{}nidari\ifmmode~\check{c}\else
  \v{c}\fi{}}}, \bibinfo {author} {\bibfnamefont {A.}~\bibnamefont
  {Scardicchio}}, \ and\ \bibinfo {author} {\bibfnamefont {V.~K.}\ \bibnamefont
  {Varma}},\ }\href {\doibase 10.1103/PhysRevLett.117.040601} {\bibfield
  {journal} {\bibinfo  {journal} {Phys. Rev. Lett.}\ }\textbf {\bibinfo
  {volume} {117}},\ \bibinfo {pages} {040601} (\bibinfo {year}
  {2016})}\BibitemShut {NoStop}%
\bibitem [{\citenamefont {Blote}\ and\ \citenamefont
  {Hilhorst}(1982)}]{blote_roughening_1982}%
  \BibitemOpen
  \bibfield  {author} {\bibinfo {author} {\bibfnamefont {H.~W.~J.}\
  \bibnamefont {Blote}}\ and\ \bibinfo {author} {\bibfnamefont {H.~J.}\
  \bibnamefont {Hilhorst}},\ }\href {\doibase 10.1088/0305-4470/15/11/011}
  {\bibfield  {journal} {\bibinfo  {journal} {J. Phys. A}\ }\textbf {\bibinfo
  {volume} {15}},\ \bibinfo {pages} {L631} (\bibinfo {year}
  {1982})}\BibitemShut {NoStop}%
\bibitem [{\citenamefont {Alet}\ \emph {et~al.}(2005)\citenamefont {Alet},
  \citenamefont {Jacobsen}, \citenamefont {Misguich}, \citenamefont {Pasquier},
  \citenamefont {Mila},\ and\ \citenamefont {Troyer}}]{alet_interacting_2005}%
  \BibitemOpen
  \bibfield  {author} {\bibinfo {author} {\bibfnamefont {F.}~\bibnamefont
  {Alet}}, \bibinfo {author} {\bibfnamefont {J.~L.}\ \bibnamefont {Jacobsen}},
  \bibinfo {author} {\bibfnamefont {G.}~\bibnamefont {Misguich}}, \bibinfo
  {author} {\bibfnamefont {V.}~\bibnamefont {Pasquier}}, \bibinfo {author}
  {\bibfnamefont {F.}~\bibnamefont {Mila}}, \ and\ \bibinfo {author}
  {\bibfnamefont {M.}~\bibnamefont {Troyer}},\ }\href {\doibase
  10.1103/PhysRevLett.94.235702} {\bibfield  {journal} {\bibinfo  {journal}
  {Phys. Rev. Lett.}\ }\textbf {\bibinfo {volume} {94}},\ \bibinfo {pages}
  {235702} (\bibinfo {year} {2005})}\BibitemShut {NoStop}%
\bibitem [{\citenamefont {Balay}\ \emph {et~al.}(1997)\citenamefont {Balay},
  \citenamefont {Gropp}, \citenamefont {McInnes},\ and\ \citenamefont
  {Smith}}]{petsc-efficient}%
  \BibitemOpen
  \bibfield  {author} {\bibinfo {author} {\bibfnamefont {S.}~\bibnamefont
  {Balay}}, \bibinfo {author} {\bibfnamefont {W.~D.}\ \bibnamefont {Gropp}},
  \bibinfo {author} {\bibfnamefont {L.~C.}\ \bibnamefont {McInnes}}, \ and\
  \bibinfo {author} {\bibfnamefont {B.~F.}\ \bibnamefont {Smith}},\ }in\ \href
  {\doibase 10.1007/978-1-4612-1986-6_8} {\emph {\bibinfo {booktitle} {Modern
  Software Tools in Scientific Computing}}},\ \bibinfo {editor} {edited by\
  \bibinfo {editor} {\bibfnamefont {E.}~\bibnamefont {Arge}}, \bibinfo {editor}
  {\bibfnamefont {A.~M.}\ \bibnamefont {Bruaset}}, \ and\ \bibinfo {editor}
  {\bibfnamefont {H.~P.}\ \bibnamefont {Langtangen}}}\ (\bibinfo  {publisher}
  {Birkh{\"{a}}user Press},\ \bibinfo {year} {1997})\ pp.\ \bibinfo {pages}
  {163--202}\BibitemShut {NoStop}%
\bibitem [{\citenamefont {Balay}\ \emph {et~al.}(2017)\citenamefont {Balay},
  \citenamefont {Abhyankar}, \citenamefont {Adams}, \citenamefont {Brown},
  \citenamefont {Brune}, \citenamefont {Buschelman}, \citenamefont {Dalcin},
  \citenamefont {Eijkhout}, \citenamefont {Gropp}, \citenamefont {Kaushik},
  \citenamefont {Knepley}, \citenamefont {McInnes}, \citenamefont {Rupp},
  \citenamefont {Smith}, \citenamefont {Zampini}, \citenamefont {Zhang},\ and\
  \citenamefont {Zhang}}]{petsc-user-ref}%
  \BibitemOpen
  \bibfield  {author} {\bibinfo {author} {\bibfnamefont {S.}~\bibnamefont
  {Balay}}, \bibinfo {author} {\bibfnamefont {S.}~\bibnamefont {Abhyankar}},
  \bibinfo {author} {\bibfnamefont {M.~F.}\ \bibnamefont {Adams}}, \bibinfo
  {author} {\bibfnamefont {J.}~\bibnamefont {Brown}}, \bibinfo {author}
  {\bibfnamefont {P.}~\bibnamefont {Brune}}, \bibinfo {author} {\bibfnamefont
  {K.}~\bibnamefont {Buschelman}}, \bibinfo {author} {\bibfnamefont
  {L.}~\bibnamefont {Dalcin}}, \bibinfo {author} {\bibfnamefont
  {V.}~\bibnamefont {Eijkhout}}, \bibinfo {author} {\bibfnamefont {W.~D.}\
  \bibnamefont {Gropp}}, \bibinfo {author} {\bibfnamefont {D.}~\bibnamefont
  {Kaushik}}, \bibinfo {author} {\bibfnamefont {M.~G.}\ \bibnamefont
  {Knepley}}, \bibinfo {author} {\bibfnamefont {L.~C.}\ \bibnamefont
  {McInnes}}, \bibinfo {author} {\bibfnamefont {K.}~\bibnamefont {Rupp}},
  \bibinfo {author} {\bibfnamefont {B.~F.}\ \bibnamefont {Smith}}, \bibinfo
  {author} {\bibfnamefont {S.}~\bibnamefont {Zampini}}, \bibinfo {author}
  {\bibfnamefont {H.}~\bibnamefont {Zhang}}, \ and\ \bibinfo {author}
  {\bibfnamefont {H.}~\bibnamefont {Zhang}},\ }\href@noop {} {\emph {\bibinfo
  {title} {{PETS}c Users Manual}}},\ \bibinfo {type} {Tech. Rep.}\ \bibinfo
  {number} {ANL-95/11 - Revision 3.8}\ (\bibinfo  {institution} {Argonne
  National Laboratory},\ \bibinfo {year} {2017})\BibitemShut {NoStop}%
\bibitem [{\citenamefont {Hernandez}\ \emph {et~al.}(2005)\citenamefont
  {Hernandez}, \citenamefont {Roman},\ and\ \citenamefont
  {Vidal}}]{slepc-toms}%
  \BibitemOpen
  \bibfield  {author} {\bibinfo {author} {\bibfnamefont {V.}~\bibnamefont
  {Hernandez}}, \bibinfo {author} {\bibfnamefont {J.~E.}\ \bibnamefont
  {Roman}}, \ and\ \bibinfo {author} {\bibfnamefont {V.}~\bibnamefont
  {Vidal}},\ }\href {\doibase 10.1145/1089014.1089019} {\bibfield  {journal}
  {\bibinfo  {journal} {{ACM} Trans. Math. Software}\ }\textbf {\bibinfo
  {volume} {31}},\ \bibinfo {pages} {351} (\bibinfo {year} {2005})}\BibitemShut
  {NoStop}%
\bibitem [{\citenamefont {Amestoy}\ \emph {et~al.}(2001)\citenamefont
  {Amestoy}, \citenamefont {Duff}, \citenamefont {L'Excellent},\ and\
  \citenamefont {Koster}}]{amestoy_fully_2001}%
  \BibitemOpen
  \bibfield  {author} {\bibinfo {author} {\bibfnamefont {P.}~\bibnamefont
  {Amestoy}}, \bibinfo {author} {\bibfnamefont {I.}~\bibnamefont {Duff}},
  \bibinfo {author} {\bibfnamefont {J.}~\bibnamefont {L'Excellent}}, \ and\
  \bibinfo {author} {\bibfnamefont {J.}~\bibnamefont {Koster}},\ }\href
  {\doibase 10.1137/S0895479899358194} {\bibfield  {journal} {\bibinfo
  {journal} {SIAM J. Matrix Anal. \& Appl.}\ }\textbf {\bibinfo {volume}
  {23}},\ \bibinfo {pages} {15} (\bibinfo {year} {2001})}\BibitemShut {NoStop}%
\bibitem [{\citenamefont {Amestoy}\ \emph {et~al.}(2006)\citenamefont
  {Amestoy}, \citenamefont {Guermouche}, \citenamefont {L’Excellent},\ and\
  \citenamefont {Pralet}}]{amestoy_hybrid_2006}%
  \BibitemOpen
  \bibfield  {author} {\bibinfo {author} {\bibfnamefont {P.~R.}\ \bibnamefont
  {Amestoy}}, \bibinfo {author} {\bibfnamefont {A.}~\bibnamefont {Guermouche}},
  \bibinfo {author} {\bibfnamefont {J.-Y.}\ \bibnamefont {L’Excellent}}, \
  and\ \bibinfo {author} {\bibfnamefont {S.}~\bibnamefont {Pralet}},\ }\href
  {\doibase 10.1016/j.parco.2005.07.004} {\bibfield  {journal} {\bibinfo
  {journal} {Parallel Computing}\ }\bibinfo {series} {Parallel {Matrix}
  {Algorithms} and {Applications} ({PMAA}’04)},\ \textbf {\bibinfo {volume}
  {32}},\ \bibinfo {pages} {136} (\bibinfo {year} {2006})}\BibitemShut
  {NoStop}%
\bibitem [{\citenamefont {Ghysels}\ \emph {et~al.}(2016)\citenamefont
  {Ghysels}, \citenamefont {Li}, \citenamefont {Rouet}, \citenamefont
  {Williams},\ and\ \citenamefont {Napov}}]{ghysels_efficient_2016}%
  \BibitemOpen
  \bibfield  {author} {\bibinfo {author} {\bibfnamefont {P.}~\bibnamefont
  {Ghysels}}, \bibinfo {author} {\bibfnamefont {X.}~\bibnamefont {Li}},
  \bibinfo {author} {\bibfnamefont {F.}~\bibnamefont {Rouet}}, \bibinfo
  {author} {\bibfnamefont {S.}~\bibnamefont {Williams}}, \ and\ \bibinfo
  {author} {\bibfnamefont {A.}~\bibnamefont {Napov}},\ }\href {\doibase
  10.1137/15M1010117} {\bibfield  {journal} {\bibinfo  {journal} {SIAM J. Sci.
  Comput.}\ }\textbf {\bibinfo {volume} {38}},\ \bibinfo {pages} {S358}
  (\bibinfo {year} {2016})}\BibitemShut {NoStop}%
\bibitem [{\citenamefont {Ghysels}\ \emph {et~al.}(2017)\citenamefont
  {Ghysels}, \citenamefont {Li}, \citenamefont {Gorman},\ and\ \citenamefont
  {Rouet}}]{ghysels_robust_2017}%
  \BibitemOpen
  \bibfield  {author} {\bibinfo {author} {\bibfnamefont {P.}~\bibnamefont
  {Ghysels}}, \bibinfo {author} {\bibfnamefont {X.~S.}\ \bibnamefont {Li}},
  \bibinfo {author} {\bibfnamefont {C.}~\bibnamefont {Gorman}}, \ and\ \bibinfo
  {author} {\bibfnamefont {F.~H.}\ \bibnamefont {Rouet}},\ }in\ \href {\doibase
  10.1109/IPDPS.2017.21} {\emph {\bibinfo {booktitle} {2017 {IEEE}
  {International} {Parallel} and {Distributed} {Processing} {Symposium}
  ({IPDPS})}}}\ (\bibinfo {year} {2017})\ pp.\ \bibinfo {pages}
  {897--906}\BibitemShut {NoStop}%
\bibitem [{\citenamefont {Abadi}\ \emph {et~al.}(2015)\citenamefont {Abadi},
  \citenamefont {Agarwal}, \citenamefont {Barham}, \citenamefont {Brevdo},
  \citenamefont {Chen}, \citenamefont {Citro}, \citenamefont {Corrado},
  \citenamefont {Davis}, \citenamefont {Dean}, \citenamefont {Devin},
  \citenamefont {Ghemawat}, \citenamefont {Goodfellow}, \citenamefont {Harp},
  \citenamefont {Irving}, \citenamefont {Isard}, \citenamefont {Jia},
  \citenamefont {Jozefowicz}, \citenamefont {Kaiser}, \citenamefont {Kudlur},
  \citenamefont {Levenberg}, \citenamefont {Man\'{e}}, \citenamefont {Monga},
  \citenamefont {Moore}, \citenamefont {Murray}, \citenamefont {Olah},
  \citenamefont {Schuster}, \citenamefont {Shlens}, \citenamefont {Steiner},
  \citenamefont {Sutskever}, \citenamefont {Talwar}, \citenamefont {Tucker},
  \citenamefont {Vanhoucke}, \citenamefont {Vasudevan}, \citenamefont
  {Vi\'{e}gas}, \citenamefont {Vinyals}, \citenamefont {Warden}, \citenamefont
  {Wattenberg}, \citenamefont {Wicke}, \citenamefont {Yu},\ and\ \citenamefont
  {Zheng}}]{tensorflow2015-whitepaper}%
  \BibitemOpen
  \bibfield  {author} {\bibinfo {author} {\bibfnamefont {M.}~\bibnamefont
  {Abadi}}, \bibinfo {author} {\bibfnamefont {A.}~\bibnamefont {Agarwal}},
  \bibinfo {author} {\bibfnamefont {P.}~\bibnamefont {Barham}}, \bibinfo
  {author} {\bibfnamefont {E.}~\bibnamefont {Brevdo}}, \bibinfo {author}
  {\bibfnamefont {Z.}~\bibnamefont {Chen}}, \bibinfo {author} {\bibfnamefont
  {C.}~\bibnamefont {Citro}}, \bibinfo {author} {\bibfnamefont {G.~S.}\
  \bibnamefont {Corrado}}, \bibinfo {author} {\bibfnamefont {A.}~\bibnamefont
  {Davis}}, \bibinfo {author} {\bibfnamefont {J.}~\bibnamefont {Dean}},
  \bibinfo {author} {\bibfnamefont {M.}~\bibnamefont {Devin}}, \bibinfo
  {author} {\bibfnamefont {S.}~\bibnamefont {Ghemawat}}, \bibinfo {author}
  {\bibfnamefont {I.}~\bibnamefont {Goodfellow}}, \bibinfo {author}
  {\bibfnamefont {A.}~\bibnamefont {Harp}}, \bibinfo {author} {\bibfnamefont
  {G.}~\bibnamefont {Irving}}, \bibinfo {author} {\bibfnamefont
  {M.}~\bibnamefont {Isard}}, \bibinfo {author} {\bibfnamefont
  {Y.}~\bibnamefont {Jia}}, \bibinfo {author} {\bibfnamefont {R.}~\bibnamefont
  {Jozefowicz}}, \bibinfo {author} {\bibfnamefont {L.}~\bibnamefont {Kaiser}},
  \bibinfo {author} {\bibfnamefont {M.}~\bibnamefont {Kudlur}}, \bibinfo
  {author} {\bibfnamefont {J.}~\bibnamefont {Levenberg}}, \bibinfo {author}
  {\bibfnamefont {D.}~\bibnamefont {Man\'{e}}}, \bibinfo {author}
  {\bibfnamefont {R.}~\bibnamefont {Monga}}, \bibinfo {author} {\bibfnamefont
  {S.}~\bibnamefont {Moore}}, \bibinfo {author} {\bibfnamefont
  {D.}~\bibnamefont {Murray}}, \bibinfo {author} {\bibfnamefont
  {C.}~\bibnamefont {Olah}}, \bibinfo {author} {\bibfnamefont {M.}~\bibnamefont
  {Schuster}}, \bibinfo {author} {\bibfnamefont {J.}~\bibnamefont {Shlens}},
  \bibinfo {author} {\bibfnamefont {B.}~\bibnamefont {Steiner}}, \bibinfo
  {author} {\bibfnamefont {I.}~\bibnamefont {Sutskever}}, \bibinfo {author}
  {\bibfnamefont {K.}~\bibnamefont {Talwar}}, \bibinfo {author} {\bibfnamefont
  {P.}~\bibnamefont {Tucker}}, \bibinfo {author} {\bibfnamefont
  {V.}~\bibnamefont {Vanhoucke}}, \bibinfo {author} {\bibfnamefont
  {V.}~\bibnamefont {Vasudevan}}, \bibinfo {author} {\bibfnamefont
  {F.}~\bibnamefont {Vi\'{e}gas}}, \bibinfo {author} {\bibfnamefont
  {O.}~\bibnamefont {Vinyals}}, \bibinfo {author} {\bibfnamefont
  {P.}~\bibnamefont {Warden}}, \bibinfo {author} {\bibfnamefont
  {M.}~\bibnamefont {Wattenberg}}, \bibinfo {author} {\bibfnamefont
  {M.}~\bibnamefont {Wicke}}, \bibinfo {author} {\bibfnamefont
  {Y.}~\bibnamefont {Yu}}, \ and\ \bibinfo {author} {\bibfnamefont
  {X.}~\bibnamefont {Zheng}},\ }\href {https://www.tensorflow.org/} {\enquote
  {\bibinfo {title} {{TensorFlow}: Large-scale machine learning on
  heterogeneous systems},}\ } (\bibinfo {year} {2015}),\ \bibinfo {note}
  {software available from tensorflow.org}\BibitemShut {NoStop}%
\bibitem [{\citenamefont {Vidmar}\ and\ \citenamefont
  {Rigol}(2017)}]{vidmar_entanglement_2017}%
  \BibitemOpen
  \bibfield  {author} {\bibinfo {author} {\bibfnamefont {L.}~\bibnamefont
  {Vidmar}}\ and\ \bibinfo {author} {\bibfnamefont {M.}~\bibnamefont {Rigol}},\
  }\href {\doibase 10.1103/PhysRevLett.119.220603} {\bibfield  {journal}
  {\bibinfo  {journal} {Phys. Rev. Lett.}\ }\textbf {\bibinfo {volume} {119}},\
  \bibinfo {pages} {220603} (\bibinfo {year} {2017})}\BibitemShut {NoStop}%
\bibitem [{\citenamefont {Garrison}\ and\ \citenamefont
  {Grover}(2018)}]{garrison_does_2018}%
  \BibitemOpen
  \bibfield  {author} {\bibinfo {author} {\bibfnamefont {J.~R.}\ \bibnamefont
  {Garrison}}\ and\ \bibinfo {author} {\bibfnamefont {T.}~\bibnamefont
  {Grover}},\ }\href {\doibase 10.1103/PhysRevX.8.021026} {\bibfield  {journal}
  {\bibinfo  {journal} {Phys. Rev. X}\ }\textbf {\bibinfo {volume} {8}},\
  \bibinfo {pages} {021026} (\bibinfo {year} {2018})}\BibitemShut {NoStop}%
\bibitem [{\citenamefont {Zhou}\ and\ \citenamefont
  {Luitz}(2017)}]{zhou_operator_2017}%
  \BibitemOpen
  \bibfield  {author} {\bibinfo {author} {\bibfnamefont {T.}~\bibnamefont
  {Zhou}}\ and\ \bibinfo {author} {\bibfnamefont {D.~J.}\ \bibnamefont
  {Luitz}},\ }\href {\doibase 10.1103/PhysRevB.95.094206} {\bibfield  {journal}
  {\bibinfo  {journal} {Phys. Rev. B}\ }\textbf {\bibinfo {volume} {95}},\
  \bibinfo {pages} {094206} (\bibinfo {year} {2017})}\BibitemShut {NoStop}%
\end{thebibliography}%

\end{document}